\documentclass[12pt]{article}
\pdfoutput=1
\usepackage{amsmath,amssymb,amsthm,mathtools,bm,graphicx,float,array,multirow,multicol,rotfloat,caption,subcaption,enumerate,geometry,mathdots,adjustbox,booktabs,parskip,mathtools,tikz,tikz-cd,csquotes,rotating,empheq,mathrsfs,xfrac}
\usepackage{pdflscape}
\usepackage{dsfont}
\usepackage{pifont}
\usepackage[all]{xy}
\usepackage[shortlabels]{enumitem}
\usepackage[normalem]{ulem}
\usepackage[vcentermath]{youngtab}
\usepackage[boxsize=.8em]{ytableau}
\usepackage[numbers,sort&compress]{natbib}
\usepackage{pgfplots}
\usepgfplotslibrary{statistics}
\usepackage{pgfplotstable}
\usepackage{xcolor}
\usepackage[
  hidelinks,
linktocpage]{hyperref}
\usepackage[capitalise,sort]{cleveref}
\usepackage{longtable}
\usepackage{hhline}
\usepackage{subcaption}
\usepackage{nowidow}
\usepackage{makecell}
\usepackage{algorithm}
\usepackage{algpseudocode}

\pgfplotsset{compat=1.18}
\numberwithin{equation}{section}
\graphicspath{{figs/}}
\makeatletter
\def\myrotate{\ifodd\c@page\else+\fi 90}
\g@addto@macro{\landscape}{\PLS@Rotate{\myrotate}}
\usetikzlibrary{arrows,shapes.misc,positioning,decorations.pathmorphing,decorations.markings,decorations.pathreplacing,matrix,patterns,backgrounds,calc,fit}
\tikzset{
scale cd/.style={every label/.append style={scale=#1}, cells={nodes={scale=#1}}},
gauge/.style={rounded rectangle, draw=black!100, thick, minimum size=2mm},
gaugeD/.style={rounded rectangle, draw=black!100,double,thick,minimum size=2mm},
empty/.style={rounded rectangle, draw=white!100, thick, minimum size=2mm},
flavor/.style={rectangle, draw=black!100, thick, minimum size=2mm},
flavorD/.style={rectangle, draw=black!100, double,thick, minimum size=2mm},
node/.style={circle, thick, draw=black!100,fill=white!100,  minimum size=3mm, inner sep=0pt},
sqnode/.style={rectangle
, thick, draw=black!100,fill=white!100,  minimum size=2mm, inner sep=0pt
},
sonode/.style={circle, thick, draw=black!100,fill=red!100,  minimum size=2mm, inner sep=0pt},
spnode/.style={circle, thick, draw=black!100,fill=blue!100,  minimum size=2mm, inner sep=0pt},
fnode/.style={rectangle, thick, draw=black!100,fill=white!100,  minimum size=2mm, inner sep=0pt},
tnode/.style={rounded rectangle, outer sep=0pt, thick, minimum size=2mm},
brace/.style={decoration={brace, mirror},decorate},
arrow/.style={->, >=stealth, thick}
}

\theoremstyle{plain}
\newtheorem*{thm*}{Theorem}

\DeclareMathOperator{\tr}{Tr}
\makeatother

\DeclareSymbolFont{largesymbolsA}{U}{jkpexa}{m}{n}
\SetSymbolFont{largesymbolsA}{bold}{U}{jkpexa}{bx}{n}
\DeclareMathSymbol{\varprod}{\mathop}{largesymbolsA}{16}

\DeclareMathOperator{\SU}{SU}

\newcommand{\CC}{\mathbb{C}}

\newcommand{\ZZ}{\mathbb{Z}}

\newcommand{\coma}{\, , \quad}
\newcommand{\fstop}{\, .}

\newcommand{\mailref}[1]{\href{mailto:#1}{\color{black}\nolinkurl{#1}}}

\newcommand{\Lpagenumber}{\ifdim\textwidth=\linewidth\else\bgroup
	\dimendef\margin=0 
	\ifodd\value{page}\margin=\oddsidemargin
	\else\margin=\evensidemargin
	\fi
	\raisebox{\dimexpr -3\topmargin-\headheight-\headsep-0.5\linewidth}[0pt][0pt]{%
		\rlap{\hspace{\dimexpr \margin+\textheight+3\footskip}%
			\llap{\rotatebox{90}{\hspace{-4.5cm}\thepage\hfill}}}}%
	\egroup\fi}
\AddToHook{shipout/background}{\Lpagenumber}%

\tikzset{
    ncbar angle/.initial=90,
    ncbar/.style={
        to path=(\tikztostart)
        -- ($(\tikztostart)!#1!\pgfkeysvalueof{/tikz/ncbar angle}:(\tikztotarget)$)
        -- ($(\tikztotarget)!($(\tikztostart)!#1!\pgfkeysvalueof{/tikz/ncbar angle}:(\tikztotarget)$)!\pgfkeysvalueof{/tikz/ncbar angle}:(\tikztostart)$)
        -- (\tikztotarget)
    },
    ncbar/.default=0.5cm,
}

\tikzset{square left brace/.style={ncbar=0.2cm}}
\tikzset{square right brace/.style={ncbar=-0.2cm}}

\tikzset{snake it/.style={decorate, decoration={snake, amplitude=.4mm, segment length=2mm,
                       post length=0mm,pre length=0mm}}}

\hypersetup{
	pdftitle={Learning to Trace Seiberg Dualities},    
	pdfauthor={\textcopyright\ Jonathan J. Heckman,  Shani Nadir Meynet, Alessandro Mininno, Gary Shiu},     
	pdfsubject={hep-th},   
	pdfcreator={pdfLaTex},   
	pdfproducer={LaTex}, 
	pdfkeywords={},
	colorlinks=true
  ,urlcolor=blue
  ,anchorcolor=blue
  ,citecolor=blue
  ,filecolor=blue
  ,linkcolor=blue
  ,menucolor=blue
  ,linktocpage=true
}

\crefname{figure}{Figure}{Figures}
\crefname{table}{Table}{Tables}
\crefname{definition}{Definition}{Definitions}
\crefname{proposition}{Proposition}{Propositions}
\crefname{claim}{Claim}{Claims}
\crefname{conjecture}{Conjecture}{Conjectures}

\allowdisplaybreaks[1]

\begin{document}

\begin{titlepage}

\begin{center}
\vspace{1.7cm}
{\LARGE\bfseries Learning to Trace Seiberg Dualities }
\vspace{0.4cm}

{\large Jonathan J. Heckman,$^{1,2}$  Shani Nadir Meynet,$^{1}$ \\[4mm] Alessandro Mininno,$^{3}$ and Gary Shiu$^3$}\\
\vspace{.6cm}

{$^1$Department of Physics and Astronomy, University of Pennsylvania,\\
209 South 33rd Street, Philadelphia, PA 19104, USA}\\
\vspace{.1cm}
{$^2$Department of Mathematics, University of Pennsylvania, \\
209 South 33rd Street, Philadelphia, PA 19104, USA}\\
\vspace{.1cm}
{$^3$Department of Physics, University of Wisconsin--Madison,\\1150 University Avenue, Madison, WI 53706, USA}\par
\vspace{.2cm}

\scalebox{0.8}{\tt \mailref{jheckman@sas.upenn.edu}, \mailref{smeynet@sas.upenn.edu},} \\
\scalebox{0.8}{\tt \mailref{mininno@physics.wisc.edu}, \mailref{shiu@physics.wisc.edu}}
\vspace{0.3cm}

\begin{abstract}


Dualities play an important role in establishing both microscopic and emergent
phenomena in a wide range of physical systems. In practice, though, it can often be computationally
challenging to establish when two systems are dual, even when all of the ``rules of the game'' are well-known.
Said differently, when confronted with two systems, how can one efficiently establish that they are in fact dual?
In this paper we use machine learning methods to address this question for Seiberg dualities of supersymmetric quiver gauge theories.
Mathematically, this involves establishing mutations of quivers, which is in turn a variation on the theme of ``learning to unknot''.
On the one hand, this leads us to a practical tool for establishing the computational complexity of different dualities.
On the other hand, it also allows us to study how different network architectures learn how to trace Seiberg dualities.
We find that for quivers with a modest number of quiver nodes (of order $10$), different network architectures consisting of transformers and multi-layer perceptrons tend to outperform deterministic algorithms. Supplementing the network by well-established pathfinder algorithms (essentially ``Google Maps for quivers'') leads to an additional improvement in the efficiency and accuracy of the search strategy. We anticipate that this class of questions can serve as a useful benchmark for frontier AI models applied to theoretical physics. 

\end{abstract}

\end{center}

\vspace{1cm}
\vfill
\end{titlepage}

\tableofcontents
\bigskip\medskip
\hrule
\bigskip\bigskip
\newpage

\section{Introduction}

Dualities provide important complementary perspectives in a wide range of physical systems. Celebrated examples include the Kramers--Wannier duality of the 2D Ising model, the electric-magnetic duality of Maxwell theory, and many modern examples ranging from interacting quantum field theories to string / M- / F-theory to holography.\footnote{The literature is vast, so rather than providing a partial and biased list of references, we trust the reader to search the original literature for themselves.} Indeed, dualities provide a window into the structure of many physical systems at strong coupling.

More precisely, a ``duality'' can often connect more than two presentations of a theory. Indeed, in many well-known examples,
successive dualities lead to a wild proliferation of different possible descriptions. Turning the question around, if one is confronted with two theories, how does one establish that the pair is connected to one another via a chain of dualities, and if so, how many simple duality operations does it take to connect the pair?

In this paper, we study this question for Seiberg dualities \cite{Seiberg:1994pq} applied to quiver gauge theories, i.e., mutations of quivers in the sense of \cite{Fomin:2001mwn, Fomin:2002nsg}. In the original examples of supersymmetric QCD, this duality relates an ``electric'' non-abelian gauge theory with a ``magnetic'' dual non-abelian gauge theory. This can be extended to a wide range of supersymmetric gauge theories with multiple simple gauge group factors, where the basic operation of duality involves acting on one gauge group factor at a time. A celebrated example of precisely this sort is the duality cascade of reference \cite{Klebanov:2000hb}, in which there is a pair of gauge groups that share matter in bifundamental representations. This example also prominently features in many string theory applications since it characterizes fractional branes near the warped throat specified by the conifold \cite{Klebanov:1998hh, Klebanov:2000nc, Klebanov:2000hb}. Physically, the flow of gauge couplings in these examples dictates how duality proceeds (i.e., which gauge group is running to strong coupling).

In many cases, however, the structure of dualities is far more involved, and there can often be more than one prescribed answer depending on the parameters of the quiver. For example, \cite{Franco:2004jz} showed that the far more generic situation is that there is an infinite family of dual theories. The specific trajectory also exhibits a high level of sensitivity to the initial conditions of gauge couplings, resulting in a huge chaotic range of possible duality sequences.

From this perspective, there is already a practical question: Given two supersymmetric quiver gauge theories, are they in fact related via a sequence of Seiberg dualities? If so, what are the most efficient/optimal simple duality operations available to realize these transformations? In physical terms, this helps to establish the level of computational complexity in reaching dual phases of quiver gauge theories. From a holographic standpoint, this amounts to understanding how far into the bulk a given flow proceeds down a throat.\footnote{It is also of interest in the context of statistical inference in the sense of \cite{Heckman:2013kza, Balasubramanian:2014bfa, Fowler:2021oje}: what is the level of distinguishability between different string compactifications?}

Treating this question in purely analytic terms is a daunting task, and so we shall instead seek a more practical, artificial intelligence/machine learning (AI/ML)-based approach.\footnote{One way to narrow the search is to also look for simple numerical invariants of the quiver that could be different, e.g., the defect group associated with a quiver, or its matrix Hilbert series. See e.g., \cite{DelZotto:2022fnw, Chakrabhavi:2026iku} for some recent examples of invariants along these lines.} For earlier work on AI/ML approaches to Seiberg dualities and quiver gauge theories, see, e.g., \cite{Krefl:2017yox, Bao:2020nbi, Dechant:2022ccf, Choi:2023rqg, Seong:2023njx, Seong:2024wkt, Capuozzo:2024vdw, Armstrong-Williams:2024nzy,Carta:2025asr}.  We comment that in the context of \textit{brane tilings} \cite{Franco:2005rj, Hanany:2005ve, Franco:2005sm,2003math.....10326K,kasteleyn1967graph} Seiberg duality is also referred to as \textit{toric duality} \cite{Feng:2000mi, Feng:2001bn, Feng:2001xr, Feng:2002zw}. AI/ML techniques are expected to be especially helpful in the case of infinite mutation type quivers, i.e., each successive duality dramatically increases the combinatorial possibilities.

To focus on the essential points, in this paper we suppress some of the physical data of a quiver gauge theory. Along these lines, we shall specify a ``simplified quiver'' by a directed graph, with each node of the graph decorated with a choice of a non-negative integer (i.e., the integer $N$ of an $\SU(N)$ gauge theory). The quivers are dictated by the condition that the gauge theory is free of anomalies; namely, there is a weighted balancing condition on the arrows of the graph. A Seiberg duality on a node corresponds to a transformation of the gauge theory that alters the dualized node and changes the local connectivity of arrows on the graph (see Section \ref{sec:seiberg_duality} for details). A sequence of dualities can thus be viewed as specifying an alphabet, with words formed by concatenating all of these mutations. The underlying mathematical structure thus shares some similarities with the process of determining whether a given knot is related to an unknot \cite{Gukov:2020qaj} (see also \cite{Davies:2021mxf, Gukov:2023, Applebaum:2024, 2025arXiv251006284D,Dranowski:2026}).

Much as in \cite{Gukov:2020qaj}, our aim here will be to compare a few AI methodologies to understand which neural network (NN) architectures are most effective at learning how to trace Seiberg dualities.\footnote{See also \cite{gromov2023grokkingmodulararithmetic} for work on the study of how NNs learn the structure of arithmetic and group laws.} Our purpose here is twofold: On the one hand, we use these different architectures for the practical goal of establishing the existence of dualities and obtaining an estimate of the overall computational complexity of duality transformations. On the other hand, there is also the question of performance, namely how efficient and accurate different approaches are in establishing a path to dualization. Indeed, while we focus our analysis on Seiberg dualities in quiver gauge theories, the same sort of combinatorial methods shows up in a broad class of other computational problems, ranging, e.g., from the mathematics of cluster algebras \cite{Fomin:2001mwn, Fomin:2002nsg} to scattering amplitudes \cite{Arkani-Hamed:2012zlh}. 
As such, we anticipate that the analysis we provide will have far broader applications.

\subsubsection*{Summary of Methodology and Results}

Let us now turn to a brief summary of our methodology. Additional details are presented later on. As in most machine learning tasks, we first collect/generate a suitable batch of training data. With this in place, we then train different neural network architectures (detailed below). We then consider the post-training phase, comparing their relative performance against some pre-specified performance metrics.

In our case, the quivers used to generate the training data consist of D-branes probing a toric Calabi--Yau (CY) threefold singularity. This 
leads to a rather rich class of quiver gauge theories that have been extensively studied in the literature (see e.g., \cite{He:2004rn, Aspinwall:2004jr, Franco:2017jeo} for reviews). 
Starting from these ``canonical presentations'' of a quiver, we proceed to generate our training data by Seiberg dualizing these quivers multiple times. We primarily focus on quivers with $13$ or fewer nodes, with the training set built from up to $12$ Seiberg dualities. We note that after dualizing, one often reaches a non-toric phase (i.e., after flops of the geometry). In practice, generating the full dataset was accomplished with rather modest computational resources on personal laptops and the UPenn cluster over two to three days, with 256 GB of RAM across 8 CPUs.

We then used this data to train different neural network architectures. To keep a broad perspective, we considered a few different choices of graph neural networks (GNNs) with transformer layers \cite{Vaswani:2017}. The first NN we trained is a GNN that estimates the distance, i.e., the number of minimal mutations, between a pair of theories. We call this network a \textit{Distance GNN} (DGNN) because, conceptually, it compares the two graphs and attempts to estimate their distance/difference. The accumulated difference is used to estimate how far apart the two quivers are. We show a schematic representation of the DGNN architecture in Figure \ref{sfig:DGNN}, while the detailed description of each layer is given in Section \ref{sec:siamese}.

The second NN we trained is another GNN that aims to estimate, in a pair of theories, which vertices are most likely to be dualized to transition from one theory to the other. We call this network an \textit{Adviser GNN} (AGNN), where the GNN functions as a policy adviser, i.e., it takes the input quiver and provides a probability that will be used by the pathfinder algorithm to make the choice of which node to dualize at each step. We show a schematic representation of the AGNN architecture that we will describe in Section \ref{sec:ar_gnn}, in Figure \ref{sfig:AGNN}.

We find it most effective to use the NN as a policy guide for heuristic-driven pathfinders that search over possible ways of dualizing a quiver gauge theory. We considered two kinds of pathfinders: one is a bidirectional A$^*$ search pathfinder \cite{Hart:1968}, and the other is a beam search pathfinder. For both, we used various definitions of cost and heuristic functions\footnote{Search algorithms seek a path that minimizes specific criteria. These are formalized using a \textit{cost function} that computes the accumulated penalty of the route taken so far and a \textit{heuristic function} that estimates the remaining effort to reach the goal. In this sense, a beam search pathfinder is an A$^*$ search without a heuristic function and with a reduced search frontier.} which guides the search for a path connecting a pair of quivers. The output of the DGNN will be used as the heuristic function, while the AGNN acts as the cost function. A schematic representation of the A$^*$ search is shown in Figure \ref{sfig:schematicbidirA*search}, with pseudocode in Algorithm \ref{alg:bidirectional_astar}, while a beam search \cite{Lowerre:1976} is shown in Figure \ref{sfig:schematicbeamsearch} and described in Algorithm \ref{alg:beam_search}. This approach of mixing a GNN with a heuristic search algorithm has been proven to be quite efficient in the data science literature (see e.g. \cite{Wang:2024}), but as far as we are aware, the present work constitutes the first application to a problem in theoretical physics.

Moreover, we designed a physics-informed pathfinder, which we call the \textit{Lowest Common Ancestor} (LCA) pathfinder, that is also an A$^*$ search pathfinder, with the policy determined by attempting to find the common quiver between a pair of theories with minimal ranks. This LCA pathfinder is inspired by the way (human) physicists approach the problem of finding the path between two theories related by Seiberg dualities, and it is similar to how we generated the database to train our NNs. The last pathfinder we studied is essentially a hybrid approach that unifies all the previous policies into a single pathfinder, guided both by the output of the NNs and the LCA policies. The policies for the pathfinders we designed are summarized in Table \ref{tab:summaryPaths}.

We tested the performance of these pathfinders by comparing them first with a Breadth-First Search (BFS) algorithm. In this way, we defined the efficiency ratio as  the average number of nodes expected to be explored by a BFS baseline over the number of nodes explored by the heuristic pathfinder. We also kept track of their success rate, namely the percentage of paths found before exhausting the maximum number of steps. Each pathfinder has its pros and cons based on maximizing either the efficiency ratios or their success rates. We analyzed the performance over the dataset used to train the NNs, i.e., the in-distribution (ID) dataset, and over out-of-distribution (OOD) theories. See Table \ref{tab:EER+SR_summary} for a summary of these results.

We then compared the performance of the pathfinders among themselves, using the LCA pathfinder as the baseline, as summarized in Table \ref{tab:EER+SR_LCA_summary}.\footnote{A careful comparison of \cref{tab:EER+SR_summary,tab:EER+SR_LCA_summary} reveals an apparent paradox when comparing the ERs of the OOD theories: the average EERs of the LCA relative to the BFS are superior to those of the Hybrid model. However, when evaluating the ER normalized by the LCA baseline, the Hybrid model outperforms the LCA. The reason is that the BFS baseline grows exponentially with the distances, whereas the LCA is efficient in finding the paths. On the other hand, the LCA pathfinder becomes trapped when the distance is not too large, whereas the Hybrid pathfinder is more efficient. We will discuss these performance tests further in \cref{sec:LCA-performance,sec:hyb_det_performance}.} In that table, one can also find the performance of the Hybrid LCA pathfinder, which is the final pathfinder we considered, with both the Hybrid and LCA policies added to guide the A$^*$ search. This pathfinder has the best overall performance of all the pathfinders we considered if we insist on having a $100\%$ success rate.

The NNs (with our best checkpoints) and the pathfinders are available at our \href{https://github.com/alexmininno/GNN-Pathfinders}{GitHub repository}. In order to study the breaking point of the NN-guided pathfinders, we also introduce the concept of ``complexity'' defined as $C = D\,\log_{10}K$, where $D$ is the distance between two quivers and $K$ is the number of nodes of the quivers. The NNs for which we provide the checkpoints have been trained up to complexity $\sim 11.5$, as we considered theories with up to $13$ nodes and distances of up to $12$ mutations. In the conclusion of the analysis of the breaking point of the NN-guided pathfinders, we estimate that the Hybrid or Hybrid LCA pathfinders will perform better than pure LCA up to a complexity $\sim 1.5/3\times$ of the largest complexity seen during the training of the NNs.

\subsubsection*{Structure of the Paper}

The rest of this paper is organized as follows. In Section \ref{sec:seiberg_duality}, we discuss some combinatorial aspects of Seiberg duality and pose the general question: What is the computational complexity of establishing whether two quiver gauge theories are Seiberg dual? With this in place, in Section \ref{sec:dataset_generation}, we provide further details on our method for generating training data for our neural networks. Section \ref{sec:GNNforSD} discusses in greater detail the different GNN architectures, and Section \ref{sec:pathfindersforSD} discusses pathfinder algorithms and their hybrid implementation with GNNs. In Section \ref{sec:RESULTS}, we summarize the results from the various tests. Moreover, in Section \ref{sec:breakpoint}, we analyze the breaking point of the pathfinders. Finally, we present our conclusions and potential future directions in Section \ref{sec:CONC}.  Additional details of GNNs are reviewed in Appendix \ref{app:gnn_appendix}; while details on the different pathfinder algorithms used in the main body are summarized in Appendix \ref{app:pathfindersgeneralities}. The seed theories used for our training dataset are given in Appendix \ref{sec:basictheories}.

\textbf{Note Added:} As this paper was being finalized, reference \cite{Yu:2026ogu} appeared, which also considers the task of identifying Seiberg dual quiver gauge theories. Our approach is centered on understanding the physics, computational complexity, and inference underlying such duality operations, whereas \cite{Yu:2026ogu} focuses on evaluating AI models, specifically testing their ability to use a verifier to debug incorrect duality claims. 

\section{Tracing Seiberg Dualities} 
\label{sec:seiberg_duality}

In this section, we discuss in more detail the central physics goal of the present work: determining how many ``simple'' Seiberg duality operations are necessary to connect two quiver gauge theories. To frame the discussion to follow, we briefly review the relevant physics and some motivating questions.

\subsection{Mutating Quiver Gauge Theories}

We shall be interested in 4D $\mathcal{N} = 1$ quiver gauge theories with $\SU$-type gauge groups.\footnote{We will not be dealing with global properties of the gauge group, so we will conflate the usage of gauge group and gauge algebra since it affects nothing in our discussion.} In physical terms, we introduce a collection of 4D $\mathcal{N} = 1$ vector multiplets for the $\SU(N_i)$ gauge groups, which we denote by circles. Flavor symmetry factors can also be included, being formally viewed as gauge symmetry nodes at zero coupling.\footnote{One does not need to enforce 4D anomaly cancellation constraints on such nodes.} In the quiver gauge theory literature, these are typically presented as squares. In what follows, we focus on quivers without such flavor symmetry factors.

The ``matter'' of the theory is dictated by a choice of adjacency matrix $\mathbf{A}$. While in top-down constructions $\mathbf{A}$ arises from a signed intersection pairing matrix, for our purposes it will suffice to consider a presentation in which all entries are non-negative i.e., $A_{ij} \in \mathbb{Z}_{\geq 0}$. This entry specifies the number of 4D $\mathcal{N} = 1$ chiral multiplets in the bifundamental representation $(\mathbf{N}_i, \overline{\mathbf{N}}_j)$ of $\SU(N_i) \times \SU(N_j)$. We draw this as a directed arrow that emanates from node $i$ to node $j$. We also allow for arrows $i \rightarrow j$ as well as $j \rightarrow i$, which in the physics literature are often referred to as ``vector-like pairs''. We note that the adjacency matrix is different from its anti-symmetrization $\mathbf{B} = \mathbf{A} - \mathbf{A}^{T}$. This quiver data amounts to specifying a directed graph along with the decoration provided by rank assignments $\{ N_i \}_{i \in \mathrm{Nodes}}$ (the rank of each $\SU(N_i)$ being $N_i - 1$). The full specification of the theory is supplemented by additional data such as gauge-invariant superpotential interaction terms, gauge couplings, and other (possibly irrelevant) interaction terms of the QFT. While this additional physical data is quite important in fully specifying a theory, in what follows we shall primarily focus on the defining data of the directed graph (i.e., the adjacency matrix $\mathbf{A}$) and the rank assignments. We denote this data as $(\mathbf{A}, \mathbf{N})$ in the obvious notation. This should be viewed as necessary but not sufficient data in fully specifying the 4D QFT.

Now, for 4D quiver gauge theories, there are a few basic consistency conditions we need to maintain. First of all, the absence of $\SU(N_i)^3$ gauge anomalies means that the rank assignments are partially constrained. Consistency requires that for each node $i$, the following condition holds:\footnote{Recall that in our conventions, all entries of $A_{ij}$ are non-negative. Additionally, we are specifically excluding quivers with flavor symmetry factors. These can be included by extending the sum over nodes, but one does not need to impose an anomaly cancellation condition on the non-abelian flavor symmetry itself.}
\begin{equation}
\mathrm{No\, \SU(N_i)^3 \,\,Anomalies} \Rightarrow \underset{j}{\sum} A_{ij} N_j - \underset{j}{\sum} N_j A_{ji} = 0\fstop
\end{equation}
We do not impose any further constraints aside from the requirement that any duality operation produces another theory with rank assignments $N_i \geq 0$. In special cases such as superconformal field theories, one often imposes additional constraints on the $N_i$. We do not impose such conditions here.\footnote{It is worth noting that even the condition of anomaly cancellation is a special feature of dealing with a 4D QFT. For example, in closely related 1D quiver quantum mechanics theories, no such condition needs to be imposed.}

Given a 4D $\mathcal{N} = 1$ quiver gauge theory, we can consider a Seiberg dual presentation of the IR theory. This is an IR equivalence among theories, whose prototypical example is given by Supersymmetric Quantum Chromodynamics (SQCD), i.e., $\SU(N_c)$ with $N_f$ flavors, which is dual to $\SU(N_f-N_c)$ with $N_f$ flavors and a set of meson fields interacting via a cubic superpotential. This operation naturally extends to quiver gauge theories, where one considers the dualization of a specified node. We refer to this as a simple duality operation.

Chaining together multiple such simple duality operations then relates many different quiver gauge theories. We refer to dualization of a quiver at node $k$ by $D_{k}(\mathbf{A}, \mathbf{N})$. Generalizing the SQCD example above, the mutation maps $N_k$ to $N_f^k-N_k$, where $N_f^k$ is the number of flavors at node $k$,  given by the sum of the ranks of all nodes $j$ connecting to $k$, weighted by the number of incoming arrows $A_{jk}$. On the other hand, all the ranks of the other nodes in the quiver remain unchanged. In formulae, the dual theory has a rank vector
\begin{equation}
    N'_i = \begin{dcases}
      \sum_{j=1}^K N_j A_{jk} - N_k & \text{if } i = k, \\
      N_i & \text{if } i \neq k\fstop
   \end{dcases}
\end{equation}
This operation is well-defined so long as the new rank $N'_k$ remains non-negative.\footnote{When the rank is too low compared to the number of flavors, a non-perturbative superpotential can be generated (see e.g, \cite{Affleck:1983mk, Intriligator:2005aw}). This can often be diagnosed simply by dualizing to a theory with formally negative ranks, clearly a pathological situation. For our purposes, it suffices to work at the level of the directed graph and its ranks assignments.} Note also that duality preserves the condition of anomaly cancellation, i.e., we have an equal number of incoming and outgoing arrows (weighted by the ranks of the neighboring nodes).\footnote{We will also consider theories that are not necessarily anomaly free, but we will not use them as a training set for the neural networks; instead, they will only be used for our out-of-distribution tests. As already mentioned, such quivers also arise in physical systems such as quiver quantum mechanics. We will still impose that the resulting theory after dualization is given by a connected graph.}

\begin{figure}[!htp]
    \centering
     \begin{tikzpicture}[font=\footnotesize]
    \begin{scope}[local bounding box=SD1]
        \node[node,label={above:$N_k$},fill=red] (N1_a) {};
        \node[node,label={above:$N_j$}] (N2_a) [right=2cm of N1_a] {};
        \node[node,label={above:$N_i$}] (N3_a) [left=2cm of N1_a] {};
        \draw[-Triangle] (N3_a) -- node[fill=white,inner sep=0.5pt]{\tiny $A_{ik}$} (N1_a);
        \draw[-Triangle] (N1_a) -- node[fill=white,inner sep=0.5pt]{\tiny $A_{kj}$} (N2_a);
    \end{scope}
    \begin{scope}[local bounding box=SD2, xshift=7cm] 
        \node[node,label={above:$\sum_l N_lA_{lk}-N_k$},fill=red] (N1_b) {};
        \node[node,label={above:$N_j$}] (N2_b) [right=2cm of N1_b] {};
        \node[node,label={above:$N_i$}] (N3_b) [left=2cm of N1_b] {};
        \draw[Triangle-] (N3_b) --  node[fill=white,inner sep=0.5pt]{\tiny $A_{ki}$} (N1_b);
        \draw[Triangle-] (N1_b) -- node[fill=white,inner sep=0.5pt]{\tiny $A_{jk}$} (N2_b);
        \draw[-Triangle] (N3_b) to[bend right=30]  node[fill=white,inner sep=0.5pt,pos=0.5]{\tiny $A_{ik}A_{kj}$}(N2_b);
    \end{scope}
    \draw[very thick, ->, shorten >=2mm, shorten <=2mm] (SD1.east |- N1_a.center) -- (SD2.west |- N1_b.center);
\end{tikzpicture}
    \caption{Schematic representation of Seiberg duality at the level of quiver theories. The labels on the arrows indicate the multiplicities of the fields in the various bifundamental representations.}
    \label{fig:SDonquiver}
\end{figure}
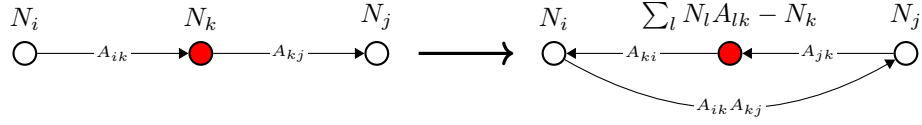

The mutation of the adjacency matrix modifies the matter content. The effects of dualizing node $k$ include the reversal of the chiralities of the fundamental and antifundamental matter fields coupled to $k$, and the generation of composite meson fields corresponding to paths of length two that pass through $k$. The mesonic fields add new arrows between adjacent nodes $i$ and $j$. We schematically show the action of Seiberg duality on node $k$ in Figure \ref{fig:SDonquiver}.

In addition to the changes to the directed graph and rank vectors, there is an accompanying change to the superpotential, i.e., we add a new cubic interaction term that naturally generalizes the SQCD case. As already mentioned, we neglect such superpotential data in what follows, focusing exclusively on the data $(\mathbf{A}, \mathbf{N})$. The only way in which the physics of a superpotential enters is that any vector-like pairs \textit{generated} by Seiberg duality/quiver mutation are automatically deleted. This naturally happens because vector-like pairs generated in this way also introduce a mass term that eliminates these contributions anyway. We note that this does not exclude the possibility of vector-like pairs in the original seed theory. 

Under this assumption, the Fomin--Zelevinsky mutation rule \cite{Fomin:2001mwn, Fomin:2002nsg} now becomes:
\begin{equation}
    A'_{ij} = \begin{dcases}
    A_{ji} & \text{if } i = k \text{ or } j = k, \\
    \max(0, \Delta_{ij}) & \text{otherwise,}
    \end{dcases}
\end{equation}
where the net flow of arrows $\Delta_{ij}$ from node $i$ to node $j$ is computed by taking the existing arrows $A_{ij}$, adding the new mesonic arrows $A_{ik} A_{kj}$, and subtracting any oppositely oriented arrows:
\begin{equation}
    \Delta_{ij} = (A_{ij} + A_{ik}A_{kj}) - (A_{ji} + A_{jk}A_{ki})\fstop
\end{equation}

\subsection{Mutation Trees}

Having specified the defining data for mutating a quiver gauge theory, we can now proceed to introduce a sequence of dualities, i.e., we pick a node of the quiver gauge theory $Q = (\mathbf{A}, \mathbf{N})$, mutate some node $k$ to reach $D_k \,Q$, and then mutate again. Starting from a seed theory, we thus obtain a sequence of dualities, reaching a new quiver:
\begin{equation}
Q^{(n)} = D_{j_n} \circ \cdots \circ D_{j_1} Q\fstop
\end{equation}
By inspection, each quiver obtained in this way amounts to a sequence of letters forming a mutation word. There can, in principle, be non-trivial relations between these words (i.e., a braid algebra), and in this sense, the structure of mutations shares many formal similarities with the problem of unknotting \cite{Gukov:2020qaj}.

\begin{figure}[!htp]
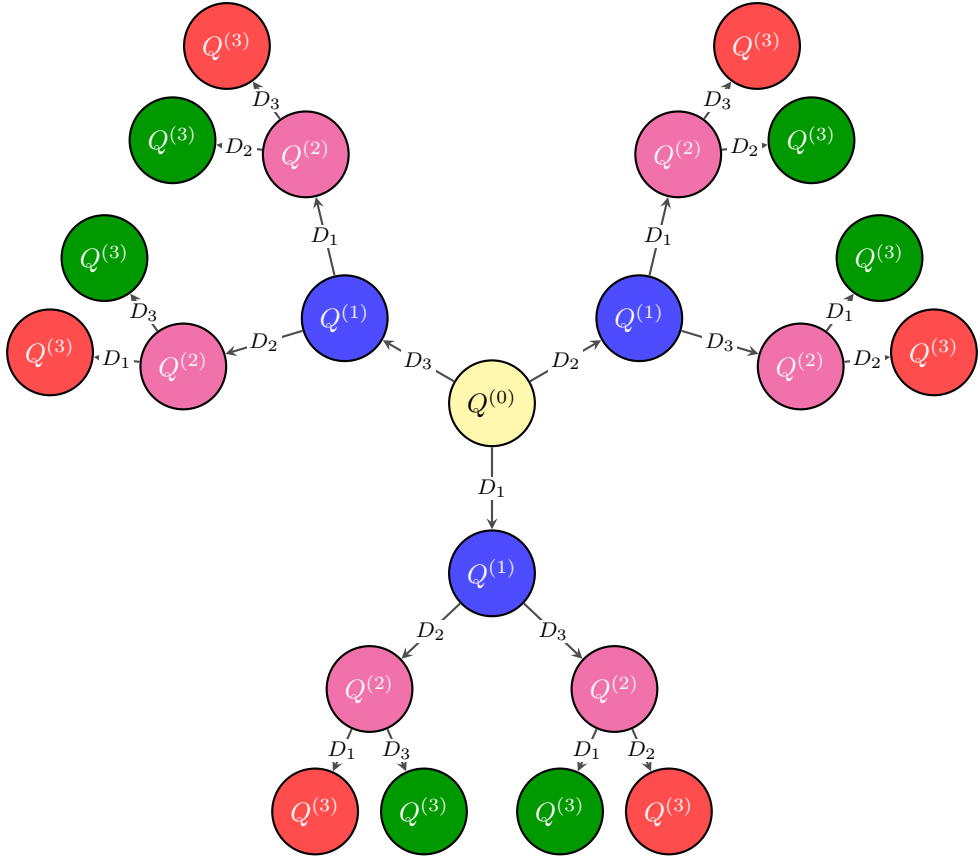

    \centering

    \caption{Seiberg Duality Tree for $\CC^3/\ZZ_3$ inspired by \cite[Figure 2]{Franco:2003ja}. Each node in the tree is a gauge theory $Q^{(n)}$ with $n$ being the number of non-trivial mutations $D_{j}$ done from $Q^{(0)}$. Generally, the action of $D_j$ leads to different gauge theories. However, $\CC^3/\ZZ_3$ has a permutation symmetry such that each node of the same color corresponds to the same theory up to relabeling of the gauge nodes.}
    \label{fig:Sdualitytree}
\end{figure}

\begin{figure}[!htp]
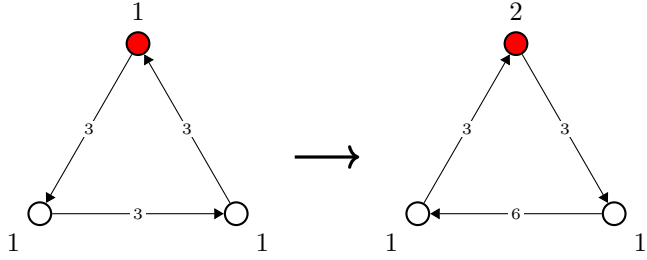

    \centering
    %
    \caption{Example of dualization of $\CC^3/\ZZ_3^{(1,1,1)}$ with respect to the {\color{red}{red}} node (top quiver node).}
    \label{fig:C3Z3examples}
\end{figure}

In a physical quiver gauge theory where we also include the data of the gauge couplings and superpotential couplings, the trajectory of an RG flow can often dictate a preferred sequence of dualities. That being said, there is also a great deal of sensitivity to these boundary conditions, i.e., the RG flow trajectory is chaotic \cite{Franco:2004jz}. From this perspective, one can, in principle, entertain \textit{any} sequence of mutations. The set of quiver gauge theories that can mutate into each other, therefore, forms a treelike structure. Indeed, at each stage, we can opt to mutate any of the nodes.\footnote{In the context of quiver gauge theories realized by D-branes probing a CY singularity, we comment that the braiding relations involve left- and right-mutation on an exceptional collection of sheaves \cite{Berenstein:2002fi,Herzog:2003zc, Herzog:2004qw, Aspinwall:2004vm}.} Taking into account relations in the braid algebra, the number of distinct quivers reached in this way can be smaller. For example, finite-type quivers are precisely those that always return to the same starting point after a finite number of moves. On the other hand, the far more generic cas is that one can mutate endlessly. A simple example of this sort is the quiver obtained from $N$ D3-branes probing the orbifold $\mathbb{C}^{3}/\mathbb{Z}_3^{(1,1,1)}$ (see \cite{Franco:2004jz}), as schematically depicted in Figure \ref{fig:Sdualitytree}. We also show the example of dualizing one of the nodes of $\mathbb{C}^{3}/\mathbb{Z}_3^{(1,1,1)}$ in Figure \ref{fig:C3Z3examples}.

In the context of string based constructions, RG flows/duality moves correspond to motion along a warped throat. Moving further into the bulk of a string compactification thus has a direct bearing on how energy scales impact the computational complexity of a given collection of quivers.

\subsection{Seed and Validation Theories}

A priori, there are an infinite number of possible quiver gauge theories one could introduce. To keep things concrete, we primarily 
focus on theories where the underlying quiver is dual to a well-known string theory construction, e.g., a D3-brane probing a toric Calabi--Yau singularity. At a small number of gauge group nodes, the entire list of such quiver gauge theories is completely known, and there are well-known algorithms for extracting more intricate quivers (see, e.g., \cite{Franco:2017jeo} for an overview). In practice, then, we start with well-known toric Calabi--Yau geometries where the corresponding quiver gauge theory consists of up to $12$ gauge group nodes. From this starting point, we then generate our training dataset by mutating many times. We comment that after mutation, the resulting geometry probed by the D3-brane may be in a non-toric phase.

To test the efficacy of our method, we compare against a number of different kinds of quivers. First of all, we consider validation on Seiberg-dualized quivers derived from D3-branes probing toric and non-toric Calabi--Yau singularities not included in the original training set. 
Additionally, we also consider more abstract quivers which are not known to be realized by any D3-brane probing a Calabi--Yau singularity. This case is especially interesting to consider because it directly points to another implicit question: when does a quiver gauge theory arise from a D-brane probing a singular geometry?

Having stated the physics motivation and rules, we now build a machine to check for Seiberg dualities.

\section{Dataset Generation} 
\label{sec:dataset_generation}

In this section we describe the theories we have used to generate the database for the training of the NNs we considered in our work. We also explain the method used to generate the database via BFS over the duality tree of the theories in Appendix \ref{sec:basictheories}.

\subsection{Breadth-First Search over Duality Trees}
\label{sec:BFS}

We now explain how we generated the pairs of theories that we used to train and test the NNs and the pathfinders. 

We considered the theories described in Appendix \ref{sec:basictheories}, which we will refer to as \textit{seed theories} from this moment on. Let us call $\mathcal{Q}$ the set of quivers that can be generated by applying Seiberg duality to any of the seed theories $Q^{(0)}\in \mathcal{Q}$. Every quiver $Q\in \mathcal{Q}$ is defined by its rank vector $\mathbf{N}$ and adjacency matrix $\mathbf{A}$. We then define $D_k:  \mathcal{Q} \to \mathcal{Q} \cup \{\emptyset\}$ to be the map of a Seiberg duality transformation at node $k$. The operator maps a quiver to its dual, $D_k\,Q^{(n)} = Q^{(n+1)}$, according to the rules defined in Section \ref{sec:seiberg_duality}. In the following, if mutating a node makes the rank of the dualized node non-positive, or if the quiver breaks into subquivers, we prune that branch; the operator returns $\emptyset$. We also introduce the concept of \textit{depth} $\mathfrak{d}$ of the duality tree, which is given by the number of non-trivial applications of $D_k$ to $Q^{(0)}$.

\begin{figure}[!htp]
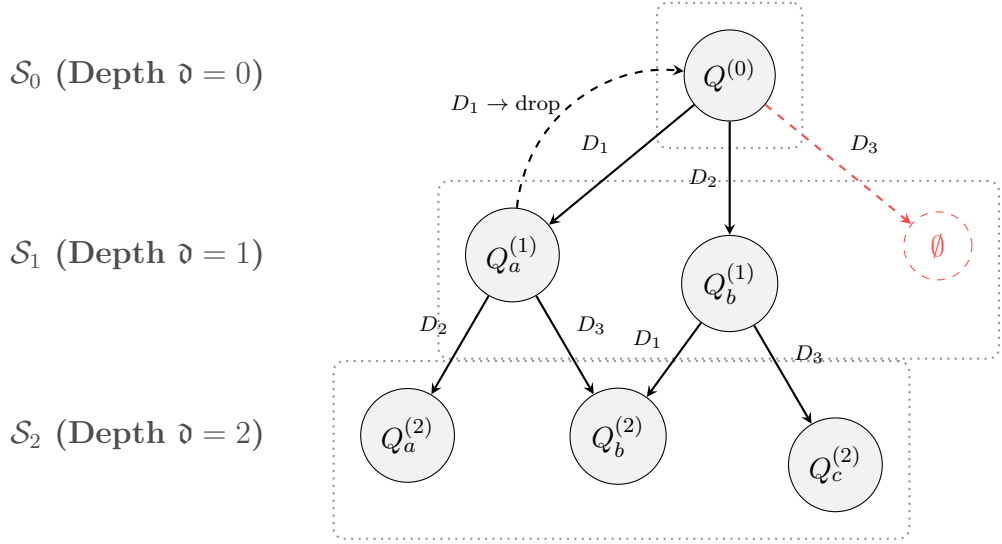

\centering

\caption{Schematic representation of the BFS generated by the mutation $D_k$.}
\label{fig:bfs_diagram}
\end{figure}

In order to generate the pairs, we start from $Q^{(0)}$ at depth $\mathfrak{d}=0$. We then make a BFS exploration, layer by layer, with $\mathcal{S}_{\mathfrak{d}}$ denoting the set of quivers $Q$ discovered by mutating $Q^{(0)}$ $\mathfrak{d}$ times. Moreover, we keep track of all visited quivers to avoid repetition in the database, and we call $\mathcal{V}_{\mathfrak{d}}$ the set of quivers found up to depth $\mathfrak{d}$.\footnote{Introducing $\mathcal{V}_{\mathfrak{d}}$ allows us to avoid artificially inflating the depth by applying $D_k$ twice consecutively to the same node, which returns the original quiver $Q$. This ensures only ``non-trivial" applications of $D_k$ are considered.} Starting from $$\mathcal{S}_0=\{Q^{(0)}\}\coma \mathcal{V}_0=\{Q^{(0)}\}\coma$$ we proceed up to depth $\mathfrak{d}$, via the recursive definition:
\begin{align}
    \mathcal{S}_{\mathfrak{d}} &= \left\{ Q^{(\mathfrak{d})} = D_k\,Q^{(\mathfrak{d}-1)} \;\middle|\; Q^{(\mathfrak{d}-1)} \in \mathcal{S}_{\mathfrak{d}-1}, \; k \in \{1, \dots, N\}, \; Q^{(\mathfrak{d})} \neq \emptyset, \; Q^{(\mathfrak{d})} \notin \mathcal{V}_{\mathfrak{d}-1} \right\}\coma \\
    \mathcal{V}_{\mathfrak{d}} &= \mathcal{V}_{\mathfrak{d}-1} \cup \mathcal{S}_{\mathfrak{d}}\fstop
\end{align}
A schematic representation of the generation is shown in Figure \ref{fig:bfs_diagram} and the pseudocode is in Algorithm \ref{alg:bfs_duality}.

\begin{figure}[!htp]
\centering
\begin{tikzpicture}[
    node distance=1.5cm,
    state/.style={circle, draw, minimum size=0.6cm, fill=gray!10, font=\scriptsize},
    target/.style={circle, draw, minimum size=0.6cm, fill=blue!20, font=\scriptsize, thick},
    edge/.style={-, draw=black!50, thick},
    path_edge/.style={-, draw=red!80, line width=1.5pt}
]

\node[target] (Ga) {$Q_A$};
\node[state, right=1.5cm of Ga, yshift=0.8cm] (n1) {};
\node[state, right=1.5cm of Ga, yshift=-0.8cm] (n2) {};
\node[state, right=3cm of Ga, yshift=1.2cm] (n3) {};
\node[state, right=3cm of Ga, yshift=0cm] (n4) {};
\node[state, right=3cm of Ga, yshift=-1.2cm] (n5) {};
\node[state, right=4.5cm of Ga, yshift=0.8cm] (n6) {};
\node[state, right=4.5cm of Ga, yshift=-0.8cm] (n7) {};
\node[target, right=6cm of Ga] (Gb) {$Q_B$};

\draw[edge] (Ga) -- (n1);
\draw[edge] (n1) -- (n3);
\draw[edge] (n3) -- (n6);
\draw[edge] (n6) -- (Gb);
\draw[edge] (n1) -- (n4);
\draw[edge] (n2) -- (n5);
\draw[edge] (n5) -- (n7);
\draw[edge] (n4) -- (n7);
\draw[edge] (n4) -- (n6);
\draw[edge] (n3) -- (n4);
\draw[edge] (n7) -- (Gb);

\draw[edge] (Ga) -- (n2);
\draw[edge] (n2) -- (n4);

\draw[path_edge] (Ga) -- node[above, font=\scriptsize, text=red!80] {$k_i$} (n2);
\draw[path_edge] (n2) -- node[above, font=\scriptsize, text=red!80] {$k_j$} (n4);
\draw[path_edge] (n4) -- node[above, font=\scriptsize, text=red!80] {$k_m$} (Gb);

\end{tikzpicture}
\caption{Schematic representation of the duality distance $d(Q_A, Q_B) = 3$, defined as the shortest path (red) corresponding to the dualizations $D_{k_m} \circ D_{k_j} \circ D_{k_i}\,Q_A$.}
\label{fig:state_space_graph}
\end{figure}
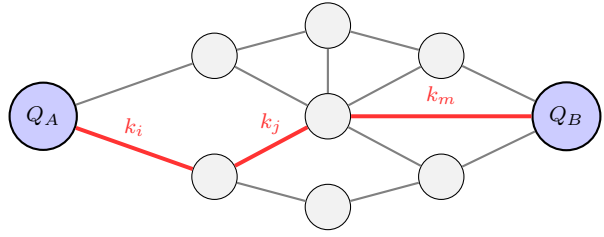

Once we have generated the full duality tree for the quiver $Q^{(0)}$ up to $\mathfrak{d}_\text{max}$, we need to construct the pairs $(Q_A,Q_B)$ of theories divided by their distance $d(Q_A,Q_B)$, i.e., the minimal number of simple Seiberg duality operations that are necessary to go from $Q_A$ to $Q_B$. This can be obtained by considering the graph obtained by placing each mutant of the quiver on the nodes, with the edges representing the mutations relating them; we will denote this graph as $\mathcal{Q}^*=(\mathcal{V}^*, \mathcal{E}^*)$. We define the duality distance between $(Q_A,Q_B)$ as the length of the shortest sequence of edges connecting them in this graph:
\begin{equation}
    d(Q_A, Q_B) = \min \left\{ p : p \text{ number of edges between } Q_A \text{ and } Q_B \text{ in } \mathcal{Q}^* \right\}\fstop
\end{equation}
This distance is illustrated in Figure \ref{fig:state_space_graph}. 

The distance is computed during the BFS generation. Every discovered state $Q_i \in \mathcal{V}^*$ is associated with a specific mutation path sequence $P_i = (k_{i,1}, k_{i,2}, \dots, k_{i, \mathfrak{d}_i})$ originating from $Q^{(0)}$. Since all the quivers $Q\in \mathcal{Q}$ are generated up to depth $\mathfrak{d}_\text{max}$, the distance between any two quivers $Q_A$ and $Q_B$ is obtained from their path vectors. We define the LCA of the two theories as the state corresponding to the longest shared prefix between their path sequences:
\begin{equation}
    P_{\text{LCA}} = P_A \cap P_B\fstop
\end{equation}
The shortest path connecting $Q_A$ to $Q_B$ is constructed by taking the union of their path sequences and removing the shared intersection (the path from the root to the LCA). Expressed as a formula, this is given by:
\begin{equation}\label{eq:dQAQB}
    d(Q_A, Q_B) = |P_A| + |P_B| - 2|P_{\text{LCA}}|\fstop
\end{equation}
In generating the dataset, we keep track of the distance as well as the specific path we used to compute it. However, as is clear from Figure \ref{fig:state_space_graph}, this path is, in general, not unique. Moreover, if the quiver has permutation symmetries, the distance \eqref{eq:dQAQB} becomes an upper bound. As we will explain in Section \ref{sec:trainingevalDGNN}, we trained the NNs using \eqref{eq:dQAQB} as the target distance to be learned, without identifying quivers related by symmetries. This allows us to validate the training using the mean absolute error (MAE) between the predicted distance and \eqref{eq:dQAQB}. However, as we will see later, the DGNN loses accuracy at large distances, predicting shorter distances compared to the database distance \eqref{eq:dQAQB}. Once the permutations are taken into account, the accuracy of the network improves, meaning that the model was already accounting for these symmetries.

\begin{algorithm}[!htp]
\caption{BFS over Duality Trees}
\label{alg:bfs_duality}
\footnotesize
\begin{algorithmic}[1]
\State \textbf{Input:} Initial quiver $Q^{(0)}$, maximum depth $\mathfrak{d}_\text{max}$
\State \textbf{Output:} Set of visited quivers $\mathcal{V}$
\State $\mathcal{S}_0 \gets \{ Q^{(0)} \}$
\State $\mathcal{V} \gets \{ Q^{(0)} \}$
\For{$\mathfrak{d} \gets 1$ to $\mathfrak{d}_\text{max}$}
    \State $\mathcal{S}_{\mathfrak{d}} \gets \emptyset$
    \For{$Q^{(\mathfrak{d}-1)} \in \mathcal{S}_{\mathfrak{d}-1}$}
        \For{$k \gets 1$ to $K$} \Comment{Iterate over all nodes}
            \State $Q^{(\mathfrak{d})} \gets D_k\,Q^{(\mathfrak{d}-1)}$
            \If{$Q^{(\mathfrak{d})} \neq \emptyset$ \textbf{and} $Q^{(\mathfrak{d})} \notin \mathcal{V}$} \Comment{Check validity and uniqueness}
                \State $\mathcal{S}_{\mathfrak{d}} \gets \mathcal{S}_{\mathfrak{d}} \cup \{ Q^{(\mathfrak{d})} \}$
                \State $\mathcal{V} \gets \mathcal{V} \cup \{ Q^{(\mathfrak{d})} \}$
            \EndIf
        \EndFor
    \EndFor
    \If{$\mathcal{S}_{\mathfrak{d}} = \emptyset$}
        \State \textbf{break} \Comment{Terminate if no new states are generated}
    \EndIf
\EndFor
\State \Return $\mathcal{V}$
\end{algorithmic}
\end{algorithm}

\section{Graph Neural Networks for Seiberg Dualities} 
\label{sec:GNNforSD}

In this section, we describe the specific Graph Neural Networks (GNNs) we designed for this work. See Appendix \ref{app:gnn_appendix} for a review of terminology as well as a general introduction to the Graph Neural Network paradigm. We primarily consider two types of GNN architectures, though they will serve two different purposes. The first one, called the Distance GNN (DGNN), is trained to estimate the distance $d(Q_A,Q_B)$ between two theories $Q_A$ and $Q_B$ related by Seiberg duality. The second GNN, called the Adviser GNN (AGNN), predicts the first node to mutate in the sequence of Seiberg dualities that brings $Q_A$ to $Q_B$.

Both our architectures share the building blocks of standard MPNN \cite{Gilmer:2017} and Graph Transformers (similar to GraphGPS models, where GPS stands for General, Powerful, and Scalable \cite{Rampavek:2022}), but we adapt them to evaluate duality operations rather than classification tasks. Specifically, they use a message-passing tokenizer followed by a Transformer encoder. The tokenizer captures the structure of each node in the quiver, e.g., the rank of its gauge group or the number of edges connecting to it, while the global attention mechanism allows the network to process the totality of the graph. The differences lie in how the architectures handle the outputs.

\begin{figure}[!htp]
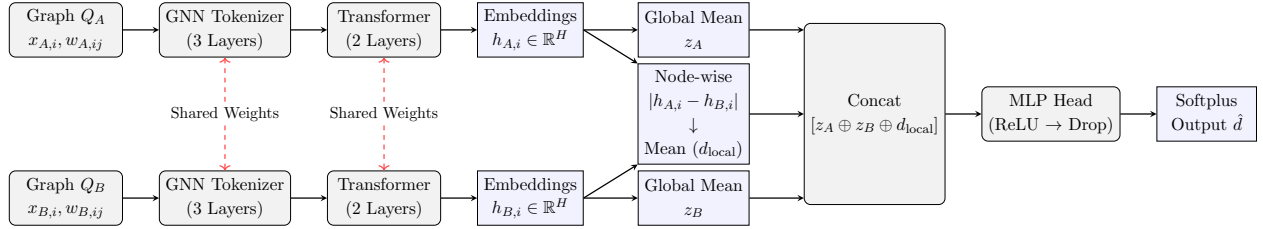
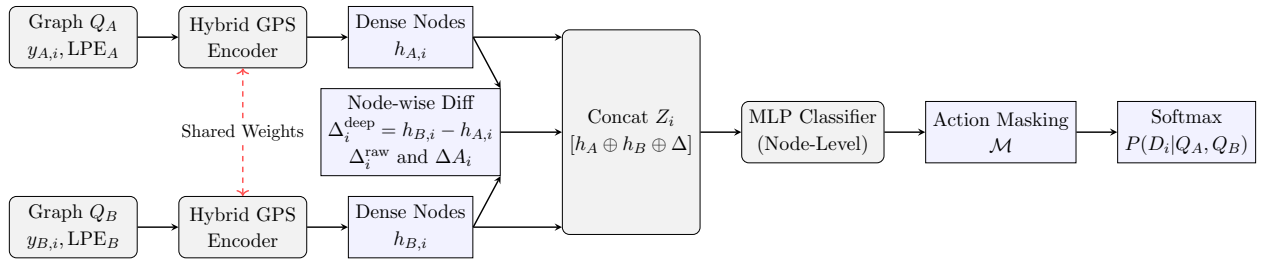

    \centering
\begin{subfigure}[b]{\textwidth}
\centering
\resizebox{\textwidth}{!}{%

}
\caption{DGNN Architecture.}
\label{sfig:DGNN}
\end{subfigure}

\vspace{1em}

\begin{subfigure}[b]{\textwidth}
\centering
\resizebox{\textwidth}{!}{%
%
}
\caption{AGNN Architecture.}
\label{sfig:AGNN}
\end{subfigure}
    \caption{Schematic representations of DGNN and AGNN architectures.}
    \label{fig:DGNNandAGNN}
\end{figure}

\subsection{Distance Graph Neural Network Architecture}
\label{sec:siamese}

We now explain how the DGNN architecture is constructed. Its purpose is to estimate the distance $d(Q_A,Q_B)$ between two quivers $Q_A$ and $Q_B$ related by mutations. The DGNN architecture is depicted in Figure \ref{sfig:DGNN}, showing how it processes a pair $(Q_A, Q_B)$ through weight-sharing encoding branches to map them into a latent space. Processing the latent space leads to the prediction of a continuous distance $\hat{d} \in \mathbb{R}^+$.

First of all, every quiver $Q$ is represented by its ranks and adjacency matrix $(\mathbf{N},\mathbf{A})$. These inputs are element-wise normalized  as:
\begin{equation}
    x_i = \log(1 + N_i)\coma  w_{ij} = \log(1 + A_{ij})\fstop
\end{equation}

The normalized inputs are then given to an encoder consisting of a 3-layer GNN tokenizer. The primary role of this module is to construct locally-aware ``tokens", so that starting from the first layer encoding the information about the ranks of the nodes $x_i$, the NN can process more and more information about the edges connecting each node.

Let $h_i^{(\ell)} \in \mathbb{R}^H$ represent the hidden feature vector of node $i$ at layer $\ell$,\footnote{In our case, $H=64$.} where the initial input is the 1-dimensional normalized rank $h_i^{(0)} \equiv x_i \in \mathbb{R}^1$. The convolution layers are parametrized by two distinct learnable weight matrices each: $W_{\text{self}}^{(\ell)}$ for self-loops and $W_{\text{msg}}^{(\ell)}$ for neighbor messages.\footnote{For the first layer $\ell=0$, the weight matrices project from the 1-dimensional input to the hidden dimension $H$, i.e., $W_{\text{self}}^{(0)}, W_{\text{msg}}^{(0)} \in \mathbb{R}^{H \times 1}$. For subsequent layers $\ell>0$, the matrices are square, $W_{\text{self}}^{(\ell)}, W_{\text{msg}}^{(\ell)} \in \mathbb{R}^{H \times H}$.} At each step, the unactivated update rule aggregates information from the immediate neighborhood:
\begin{equation}
    m_i^{(\ell+1)} = W_{\text{self}}^{(\ell)} h_i^{(\ell)} + \sum_{j \in \mathcal{N}(i)} W_{\text{msg}}^{(\ell)} \left( h_j^{(\ell)} \cdot w_{ji} \right)\fstop
\end{equation}
The weights $W_{\text{self}}^{(\ell)}$ and $W_{\text{msg}}^{(\ell)}$ are shared between $Q_A$ and $Q_B$. These vectors $m_i^{(\ell+1)}$ are normalized via LayerNorm and passed through a LeakyReLU activation function with a negative slope $\alpha = 0.2$.\footnote{The precise definitions of LayerNorm and LeakyReLU are discussed in Appendix \ref{app:MessagePassing}.}
The fully activated node representation is therefore:
\begin{equation}
    \tilde{h}_i^{(\ell+1)} = \text{LeakyReLU}\left( \text{LayerNorm}\left( m_i^{(\ell+1)} \right) \right)\fstop
\end{equation}
Crucially, residual skip connections are enforced for all layers $\ell > 0$ such that:
\begin{equation}
    h_i^{(\ell+1)} = h_i^{(\ell)} + \tilde{h}_i^{(\ell+1)}\fstop
\end{equation}
Standard convolutional GNNs replace the node state at each step. This architecture differs by using residual connections to accumulate the outputs, adding the new features $\tilde{h}_i^{(\ell+1)}$ directly to the prior state $h_i^{(\ell)}$. We accumulate these features to ensure that the network retains some of the initial information of the quiver extracted in previous layers. Consequently, the final token representation $h_i^{(3)}$ is a sum of the information gathered by all the convolutional layers, and it will be the token passed through the transformer we describe next.

While the GNN tokenizer has captured the local information of each node and its surroundings up to $\ell=3$, when comparing $(Q_A,Q_B)$, we need a way to understand the global changes in the quiver that distinguish them after dualization. To achieve this, we pad $h_i^{(3)}$ into $X \in \mathbb{R}^{K \times H}$ and process it using a 2-layer Transformer encoder via the Scaled Dot-Product Self-Attention mechanism \cite{vaswani2017attention}. We use Multi-Head Attention (MHA) with $h=4$ independent heads and a projection dimension $d_k = H/h = 16$, followed by a position-wise Feed-Forward Network (FFN). The progression of the node state $X$ through a single encoder layer uses standard residual connections and LayerNorm:
\begin{equation}
\begin{split}
    \tilde{X} &= \text{LayerNorm}(X + \text{MHA}(X))\coma \\
    X_{\text{out}} &= \text{LayerNorm}(\tilde{X} + \text{FFN}(\tilde{X}))\fstop
    \end{split}
\end{equation}
After passing through the 2 distinct Transformer layers, the model yields $h_{A,i}$ and $h_{B,i}$ for the paired quivers $Q_A$ and $Q_B$, respectively. From these vectors, we define a node-wise difference vector $d_{\text{local}} \in \mathbb{R}^H$:
\begin{equation}
    d_{\text{local}} = \frac{1}{K} \sum_{i=1}^K \left| h_{A,i} - h_{B,i} \right|\coma
\end{equation}
which explicitly records the local discrepancies caused by the sequence of Seiberg dualities.

Finally, we define the average of $h_{A,i}$ and $h_{B,i}$ over the nodes of the quiver:
\begin{equation}
    z_A = \frac{1}{K} \sum_{i=1}^K h_{A,i}\coma z_B = \frac{1}{K} \sum_{i=1}^K h_{B,i}\coma
\end{equation}
which provide global representations of each theory to the distance regressor. The averaging operation smooths out the individual node features, giving the macroscopic difference $d_\text{global}=|z_A - z_B|$. For our purposes, we need the combined information of $d_\text{global}$ and $d_\text{local}$. Indeed, unlike standard GNNs that typically pool node features to process the graph data (e.g., for classification purposes), the DGNN computes the absolute difference between corresponding nodes of two graphs, and it needs to retain finer information. If this discrepancy were evaluated solely after global pooling (yielding only $d_{\text{global}}$), localized node variations would be smoothed out. Incorporating the pre-pooling difference $d_{\text{local}}$ ensures that the regressor retains fine-grained structures that global averaging alone would discard.

After the various encoding layers, the continuous distance is processed through a Multi-Layer Perceptron (MLP) head, with  $[z_A \oplus z_B \oplus d_{\text{local}}] \in \mathbb{R}^{3H}$ as input. The purpose of this part of the NN is to reduce the $3H$-dimensional vector down to a single scalar. To achieve this, we use standard linear projection with a ReLU activation function and a Dropout regularization layer ($p=0.2$) to mitigate overfitting. The output of this projection is passed through a Softplus function, i.e.
\begin{equation}
    \hat{d} = \text{Softplus}(s) = \log(1 + e^s)\coma
\end{equation}
leading to the predicted distance $\hat{d}$.

\subsubsection{Training Evaluation}
\label{sec:trainingevalDGNN}

\begin{figure}[!htp]
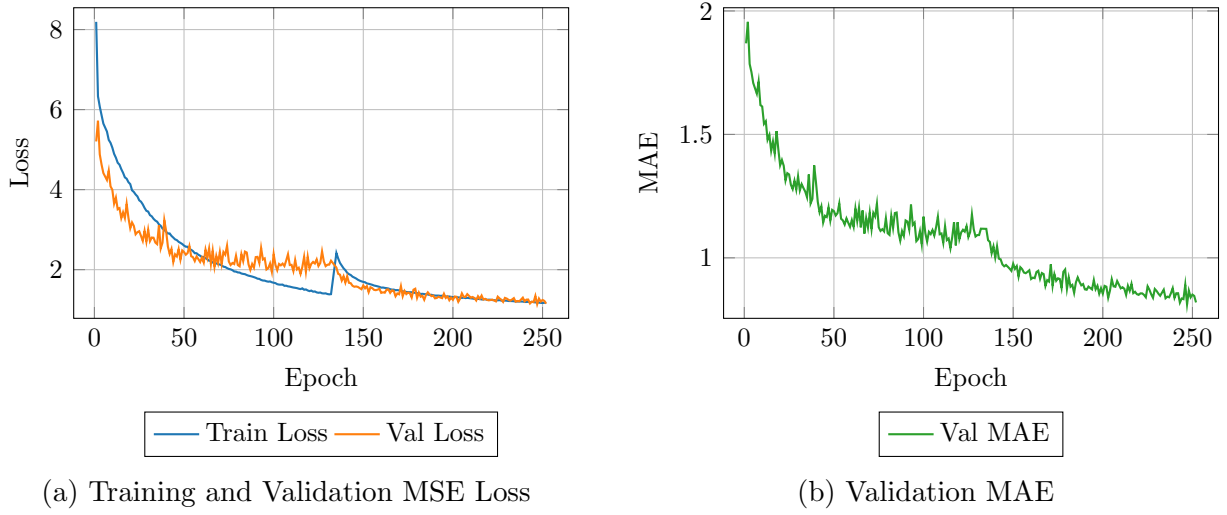

    \begin{subfigure}[b]{0.49\textwidth}
    \centering

        \caption{Training and Validation MSE Loss}
    \end{subfigure}\hfill
    \begin{subfigure}[b]{0.49\textwidth}
        \centering
        %
        \caption{Validation MAE}
    \end{subfigure}
    \caption{Training logs for the DGNN model}
    \label{fig:siamese_training_logs}
\end{figure}

The network parameters are optimized by minimizing the Mean Squared Error (MSE) between the prediction $\hat{d}$ and the path distance $d$ computed at the generation of the database. For a given batch of size $B$, the loss function is defined as:
\begin{equation}
    \mathcal{L}_{\text{MSE}} = \frac{1}{B} \sum_{k=1}^{B} (\hat{d}_k - d_k)^2\fstop
\end{equation}
While MSE is utilized for gradient descent to penalize outlier predictions, the accuracy of the model is tracked using the Mean Absolute Error (MAE):
\begin{equation}
    \text{MAE} = \frac{1}{B} \sum_{k=1}^{B} |\hat{d}_k - d_k|\fstop
\end{equation}
In the context of distance regression, MAE provides a direct measure of the average step deviation.

The DGNN was trained for approximately 250 epochs to estimate the distance between quivers. Figure \ref{fig:siamese_training_logs} displays the training trajectory. Checkpoints for the final evaluation were selected based on the epoch minimizing the MAE. Specifically, the checkpoint used for the pathfinder was reached at epoch 247 with a final MSE loss of $1.137$.

\begin{figure}[!htp]
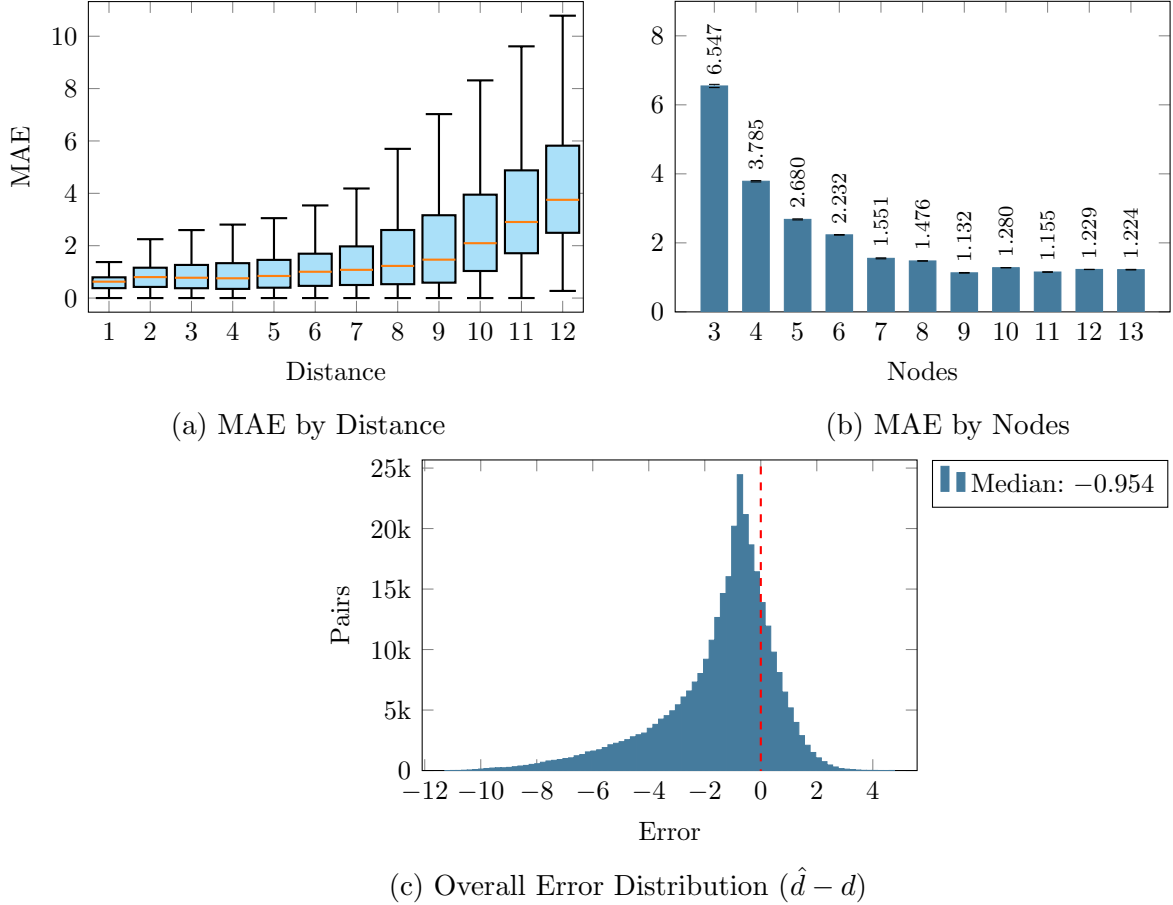

    \centering
    \begin{subfigure}[b]{0.49\textwidth}

        \caption{MAE by Distance}
    \end{subfigure}\hfill
     \begin{subfigure}[b]{0.49\textwidth}
     %
        \caption{MAE by Nodes}
    \end{subfigure}
    
    \begin{subfigure}[b]{0.49\textwidth}
%
        \caption{Overall Error Distribution ($\hat{d} - d$)}
    \end{subfigure}
    \caption{Benchmark results for DGNN.}
    \label{fig:siamese_benchmarks}
   \end{figure}

There can be multiple reasons why the MAE could not go lower. One possibility is that whenever the distance between two quivers is larger than the number of nodes, the graph difference taken by the DGNN of the hidden layers of the NN becomes less reliable, leading to an inability to estimate larger distances than the number of nodes of the quivers. This is corroborated by looking at the MAE by the number of nodes in Figure \ref{fig:siamese_benchmarks}, where it is clear that the NN struggled to predict larger distances for smaller quivers while being extremely accurate for larger quivers. 

\begin{figure}[!htp]
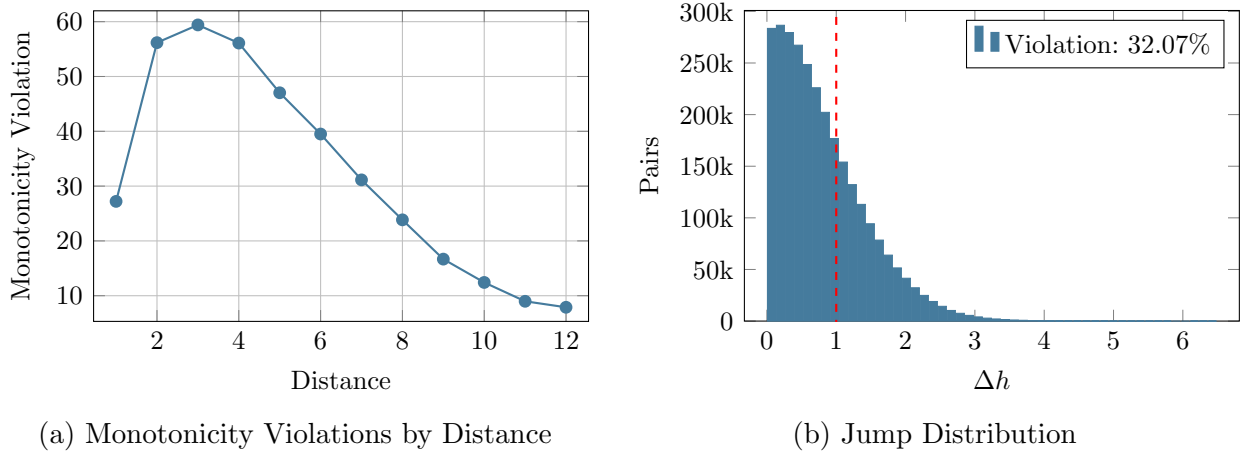

    \begin{subfigure}[b]{0.49\textwidth}\centering

        \caption{Monotonicity Violations by Distance}
    \end{subfigure}\hfill
     \begin{subfigure}[b]{0.49\textwidth}\centering
    %
        \caption{Jump Distribution}
    \end{subfigure}
    \caption{Evaluation of DGNN monotonicity. The non-monotonicity of the heuristic results in excessive node expansions and timeouts.}
    \label{fig:siamese_monotonicity}
\end{figure}

Because we apply the DGNN as the heuristic function of an A$^*$ search, as we will explain in Section \ref{sec:PathforSeibergDualities}, we have decided to perform further benchmark tests on the kind of heuristic function that the DGNN can provide. One test is to check if the heuristic function is monotonic, so that we reduce the possibility that the A$^*$ search requires processing the same node multiple times. However, as we show in Figure \ref{fig:siamese_monotonicity}, the DGNN output exhibits non-monotonicity. More precisely, we note that the DGNN is non-monotonic approximately $32.07\%$ of the time. Consequently, during a pure A$^*$ search, the priority queue can be delayed by repeatedly exploring uninformative branches. We will see that, despite the high success of the A$^*$ search guided by the DGNN for both in distribution (ID) and out of distribution (OOD) theories, it is precisely the absence of monotonicity in the DGNN that causes the algorithm's failure.

\subsection{Adviser Graph Neural Network Architecture}
\label{sec:ar_gnn}

We now proceed to describe the AGNN architecture. The objective of the AGNN is to predict the most probable node that should be mutated first in the sequence of Seiberg dualities from quiver $Q_A$ to quiver $Q_B$. Unlike the DGNN, which predicts a distance, the AGNN acts as a localized policy function, yielding a probability distribution over the $K$ nodes of $Q_A$. The AGNN architecture is described in Figure \ref{sfig:AGNN}.

First of all, much as in the DGNN case, we want to encode the information of the rank and adjacency matrix of the quiver via a suitable ``tokenization". The initial feature vector is 5-dimensional:
\begin{equation}
    y_i = \left( \log(1+N_i), \, \log(1+A_i^{\text{in}}), \, \log(1+A_i^{\text{out}}), \, \log(1+N_i^{\text{in}}), \, \log(1+N_i^{\text{out}}) \right)
\end{equation}
where 
\begin{equation}
    A_i^{\text{in}} =  \sum_{j=1}^K A_{ji}\coma A_i^{\text{out}} =  \sum_{j=1}^K A_{ij} \coma N_i^{\text{in}} = \sum_j A_{ji} N_j \coma N_i^{\text{out}} = \sum_j A_{ij} N_j\fstop
\end{equation}

To allow the network to distinguish nodes, we evaluate the Laplacian Positional Encodings (LPE).\footnote{Let $\mathbf{A}$ be the adjacency matrix of the directed quiver. We define the symmetric adjacency matrix of the underlying undirected graph as $\mathbf{A}_{\text{sym}} = \mathbf{A} + \mathbf{A}^T$. The degree $d_i$ of node $i$ is, then, the total number of edges connected to the node, i.e. $d_i = \sum_j (\mathbf{A}_{\text{sym}})_{ij}$. The corresponding graph Laplacian is $L = \mathcal{D} - \mathbf{A}_{\text{sym}}$, where $\mathcal{D}$ is the diagonal degree matrix with entries $\mathcal{D}_{ii} = d_i$ and $\mathcal{D}_{i\neq j}=0$.} Let us denote as $\text{LPE}_i \in \mathbb{R}^{d_{\text{LPE}}}$ the Laplacian eigenvectors. These LPE$_i$ vectors and the features $\mathfrak{x}_i$ are projected into the hidden layer of dimension $H$, summed, and normalized to form the initial embeddings $h_{A,i}^{(0)}$ and $h_{B,i}^{(0)}$:
\begin{equation}
    h_i^{(0)} = \text{LayerNorm}\left( W_{\text{node}} y_i + W_{\text{LPE}} \text{LPE}_i \right)\coma
\end{equation}
where $W_{\text{node}} \in \mathbb{R}^{H \times 5}$ and $W_{\text{LPE}} \in \mathbb{R}^{H \times d_{\text{LPE}}}$ are learnable projection matrices, shared by both quivers $Q_A$ and $Q_B$.

These initial embeddings are then passed through a Hybrid GPS Encoder, identical in structure to the combined GNN Tokenizer and Transformer used in the DGNN, from which we extract $h_{A,i}$ and $h_{B,i}$, as before. However, despite what one might expect, we find that these more refined inputs empirically under-perform when used in the training of the DGNN. Indeed, the introduction of the LPE$_i$ vectors as an input degraded training performance, most likely because it made it harder to extract macroscopic properties of the graph, resulting in higher error rates when predicting distances.

Similarly to the DGNN case, we also compute node-wise differences:
\begin{equation}
    \Delta_i^{\text{deep}} = h_{B,i} - h_{A,i}\coma \quad \Delta_i^{\text{raw}} = h_{B,i}^{(0)} - h_{A,i}^{(0)} \coma 
\end{equation}
as well as the element-wise absolute difference of the adjacency matrices:
\begin{equation}
    \Delta A_i = \sum_j |A_{B,ij} - A_{A,ij}|\fstop
\end{equation}
In this way, we keep track of all possible local differences between the two quivers. The dense nodes and these local differences are concatenated into $Z_i \in \mathbb{R}^{4H+6}$ for each node:
\begin{equation}
    Z_i = \left[ h_{A,i} \oplus h_{B,i} \oplus \Delta_i^{\text{deep}} \oplus \Delta_i^{\text{raw}} \oplus y_{A,i} \oplus \Delta A_i \right]\fstop
\end{equation}
The concatenated vector $Z_i$ is then processed by an MLP classification head, which projects the features down to a single scalar value for each node $i$. This scalar, called \textit{logit}, represents the network's preliminary preference for mutating node $i$.

However, not all nodes can, in general, be dualized since, for instance, mutating certain nodes might yield negative gauge ranks or disconnect the quivers. We generate a binary action mask $\mathcal{M}$ prior to evaluation, which takes the value $\mathcal{M}_i = 1$ if the mutation at node $i$ is permissible, and $\mathcal{M}_i = 0$ otherwise. In practice, the network checks if $N_i' = \sum_j A_{ji} N_j - N_i$ is positive. Before converting the logits into probabilities, we apply this mask by manually overriding the scores of all invalid nodes to $-\infty$. 

Finally, we apply a Softmax function across the $N$ nodes. Because the exponential of $-\infty$ is exactly zero, this masking guarantees that the network assigns a $0\%$ probability to any mutation that is not possible, forcing the remaining valid nodes to absorb the entirety of the probability distribution. The final policy is thus given by: 
\begin{equation} 
P(D_i | Q_A, Q_B) = \text{Softmax}(\tilde{y})_i \coma \quad \text{where } \tilde{y}_i = \begin{dcases} \mathfrak{y}_i & \text{if } \mathcal{M}_i = 1 \coma \\ -\infty & \text{if } \mathcal{M}_i = 0 \coma \end{dcases} 
\end{equation} 
where $\mathfrak{y}_i$ denotes the logit output of the MLP.

Once this valid probability distribution is obtained, we sort the probabilities $P(D_i | Q_A, Q_B)$ in descending order and extract the top-$k$ nodes.

\subsubsection{Training Evaluation}
\label{sec:trainingevalAGNN}

\begin{figure}[!htp]
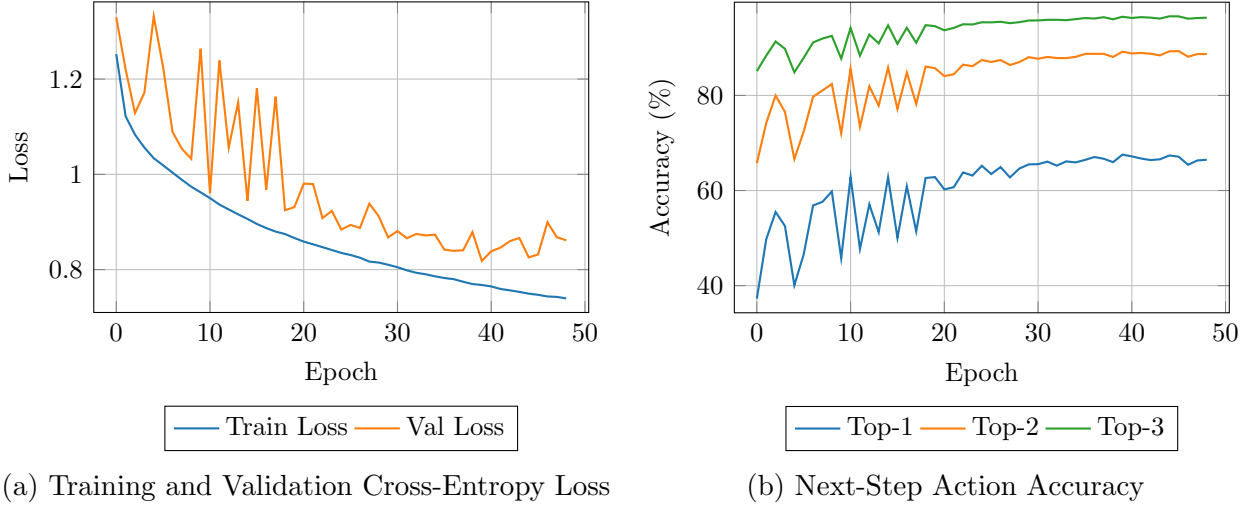

    \centering
    \begin{subfigure}[b]{0.49\textwidth}

        \caption{Training and Validation Cross-Entropy Loss}
    \end{subfigure}\hfill
    \begin{subfigure}[b]{0.49\textwidth}
        %
        \caption{Next-Step Action Accuracy}
    \end{subfigure}
    \caption{Training logs for the Autoregressive model.}
    \label{fig:ar_training_logs}
\end{figure}

The AGNN is optimized by minimizing the Cross Entropy (CE) loss between the predicted probability distribution and the exact first-step mutation extracted from the pre-computed shortest path sequence. For a given mini-batch of size $B$, let $t_q \in \{1, \dots, K\}$ denote the index of the node that must be dualized in the $q$-th quiver $Q_{A,q}$. The loss function penalizes the network based on the predicted log-probability assigned to these actions and is defined as:
\begin{equation}
    \mathcal{L}_{\text{CE}} = - \frac{1}{B} \sum_{q=1}^{B} \log P(D_{t_q} | Q_{A,q}, Q_{B,q})\fstop
\end{equation}
During training, we check the prediction accuracy of the top-$k$ nodes, which monitor whether the node we predict to be the one to dualize, i.e., $t_q$, appears within the highest probability choices predicted by the network. However, it is crucial to note that this metric does not perfectly reflect the true success rate of the AGNN. Due to quiver symmetries and the non-uniqueness of shortest paths in the duality tree, multiple nodes may represent equally optimal first-step mutations. The Cross Entropy loss penalizes the network for deviating from $t_q$ in our dataset, even if the AGNN can, in principle, propose a perfectly valid alternative node. Consequently, the top-$k$ accuracy serves as a lower bound on the true success rate of the policy.

The AGNN was trained to predict the next-step Seiberg duality operation, with the NN outputting the probabilities over the nodes. Figure \ref{fig:ar_training_logs} shows the cross-entropy loss and top-1 next-step accuracy across the 49 training epochs, at which point the training was stopped to prevent overfitting the data, since the training loss kept decreasing while the validation loss started increasing. However, the accuracy converges rapidly, with the optimal validation loss achieved at epoch 39. 

\begin{figure}[!htp]
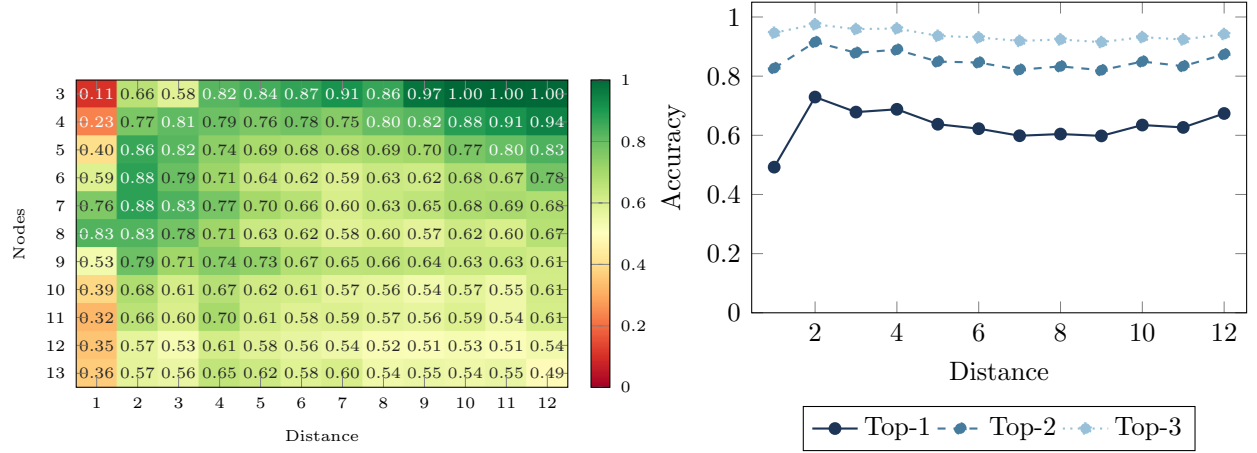
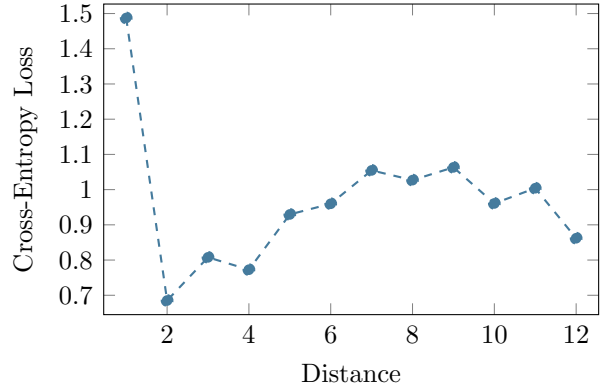

\centering
    \begin{subfigure}[b]{0.49\textwidth}\centering

        \caption{Top-1 Accuracy by Nodes and Distance}
    \end{subfigure}\hfill
    \begin{subfigure}[b]{0.49\textwidth}\centering
%
        \caption{Top-$k$ Accuracy Decay over Distance}
    \end{subfigure}
    
  \vspace{0.2cm}

    \begin{subfigure}[b]{0.49\textwidth}
        %
        \caption{Mean Cross-Entropy Loss by Distance}
    \end{subfigure}
    
    \caption{Benchmark results for the AGNN. While next-step accuracy degrades at larger duality distances, the validity rate remains high, indicating that the network has implicitly learned the rules of Seiberg dualities.}
    \label{fig:ar_benchmarks}
\end{figure}

Due to the symmetry of the quivers, the top-1 accuracy does not exceed $70\%$, meaning that for a better policy, we need to rely on more than one probability. In fact, top-3 accuracy is, on average, around 96\%, resulting in the correct node being dualized most of the time. This can be seen from Figure \ref{fig:ar_benchmarks}. We note that the accuracy actually \textit{increases} with distance, resulting in a better oracle whenever two quivers are far apart, while the accuracy drops for smaller distances. However, unlike the DGNN, there is not much difference in accuracy for the different theories used for training.

\begin{figure}[!htp]
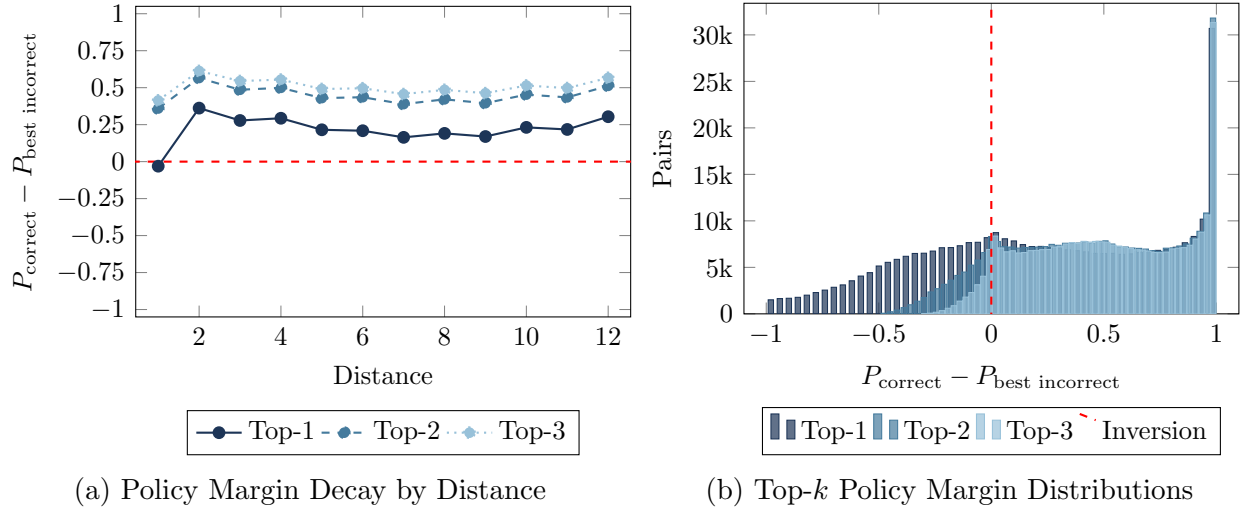

    \begin{subfigure}[b]{0.49\textwidth}

        \caption{Policy Margin Decay by Distance}
    \end{subfigure}\hfill
    \begin{subfigure}[b]{0.49\textwidth}
%
        \caption{Top-$k$ Policy Margin Distributions}
    \end{subfigure}
    \caption{Policy Margin analysis. While Top-1 inversions cause some paths to fail, the high Top-3 margins enable pruning of a large portion of the duality tree.}
    \label{fig:ar_policy_margin}
\end{figure}

The AGNN will also be utilized to guide a pathfinding search. Initially, we evaluate it as a unidirectional pathfinder, where the trajectory is constructed by iteratively selecting the actions with the top-$k$ probabilities predicted by the AGNN. To benchmark the reliability of the AGNN in this guiding role, we analyze the \textit{policy margin}, defined as the difference between the probability assigned to the correct node ($P_{\text{correct}}$) and the highest probability assigned to an incorrect node ($P_{\text{best incorrect}}$).\footnote{Because multiple valid paths can connect two theories, we define an `incorrect' node as any node that does not belong to the specific shortest path recorded during our dataset generation.} A negative policy margin indicates an \textit{inversion}, meaning the network prefers an incorrect operation over the correct one. As shown in Figure \ref{fig:ar_policy_margin}, the model exhibits such inversions $20\%$ to $40\%$ of the time. However, despite this inversion rate, the average policy margin remains positive across all target distances; this occurs because the network is highly confident when it is correct (yielding large positive margins), easily outweighing the margins of the incorrect predictions, which tend to be only slightly negative.

From the discussion above, we can conclude that, considering the three most probable nodes as branching, the AGNN is correct over $90\%$ of the time. This insight is critical: while the AGNN is not an infallible oracle and cannot navigate the duality tree on its own, it suppresses the vast majority of incorrect branches. This confidence motivates the last kind of pathfinder that will use the AGNN policy to restrict the branching factors, allowing the DGNN to evaluate only the most promising paths.

Furthermore, analyzing the tail of the margin distribution in Figure \ref{fig:ar_policy_margin} reveals that for a substantial subset of states (approximately $5.7 \times 10^4$ theories), the network exhibits an extreme policy margin of $P_\text{correct}-P_{\text{best incorrect}} \approx 1.0$. This signifies that the AGNN model is highly polarized: it appears to be uncertain between two options for only a small subset of nodes. Large regions of the duality tree are instead considered trivial and are solved with high confidence. From an algorithmic perspective, this is beneficial, as a guided search experiences a zero effective branching factor in these states. However, this same polarization means that when the network does make an error (inversions where the margin is $< 0$), it also often does so with high confidence. This extreme confidence in both correct and incorrect directions further highlights why a pure greedy search with AGNN is brittle, necessitating the safety net of the DGNN to backtrack out of confident dead-ends.

\section{Pathfinder for Seiberg Dualities} 
\label{sec:pathfindersforSD}

In this section, we discuss the other component of our inference model, namely the guided pathfinders that are used to determine the chain of dualities that connect two theories $Q_A$ and $Q_B$. The pathfinders are of two kinds: a Bidirectional A$^*$ Search Pathfinder and a Beam Search Pathfinder, as generally reviewed in Appendix \ref{app:pathfindersgeneralities}. For the latter kind, the beam is guided by the output of the AGNN model. On the other hand, we propose different kinds of cost and heuristic functions for the A$^*$ Search Pathfinder, as summarized in Table \ref{tab:summaryPaths}. 

\subsection{Brute-Force Search Baseline and Branching Factor}
\label{sec:branching_factor}

To benchmark the efficiency of guided search, we first formalize the complexity of unguided exploration. In standard graph search algorithms, the branching factor $b$ characterizes the number of direct children a node possesses. In the present framework, the branching factor of a specific quiver $Q = (\mathbf{N}, \mathbf{A})$ corresponds to the number of valid Seiberg duality transformations it admits. 
Let $K$ be the total number of nodes in the quiver. A priori, one might attempt to dualize any of the $K$ nodes, implying a maximum branching factor of $K$. However, as established in Section \ref{sec:dataset_generation}, a mutation $D_k\,Q$ is admissible if and only if it satisfies two conditions:
\begin{enumerate}
    \item \textit{Rank Positivity}: The mutated gauge rank must remain positive.
    \item \textit{Graph Connectivity}: The resulting adjacency matrix $\mathbf{A}'$ still describes a connected quiver.
\end{enumerate}
Therefore, the valid branching factor $b_Q$ for any state $Q$ is defined as the number of valid mutations:
\begin{equation}
    b_Q = \left| \left\{ k \in \{1, \dots, K\} \;\middle|\; D_k\,Q \neq \emptyset \right\} \right| \leq K\fstop
\end{equation}
For a bidirectional search between $Q_A$ and $Q_B$, $b_A$ and $b_B$ designate the valid branching factors:
\begin{itemize}
    \item $b_A \equiv b_{Q_A}$: The number of valid dualizations originating from $Q_A$.
    \item $b_B \equiv b_{Q_B}$: The number of valid dualizations originating from $Q_B$.
\end{itemize}

To establish a baseline for the expected number of mutations performed by an unguided bidirectional search ($N_{\text{BFS}}$), we observe the following fact: since Seiberg duality is an involution, i.e. $D_k \circ D_k\,Q = Q$,\footnote{Strictly speaking, this requires a left- and right-mutation on the exceptional collection of sheaves rather than two mutations of the same sort (see e.g., \cite{Heckman:2006sk}). Such distinctions will not matter in the present analysis but would be interesting to study further.} the branching factor can be estimated to be, at depth $i \geq 1$, at most $b_{\text{avg.}}(b_{\text{avg.}} - 1)^{i-1}$. This means that starting from $Q$, we reach distance $d$ with $N_{\text{dir}}(d)$ dualities, which can be expressed as\footnote{If $b_Q = 2$, $N_\text{dir}(d) = 1+2d$.}
\begin{equation}
    N_{\text{dir}}(d) = 1 + b_{Q} \sum_{i=1}^{d} (b_{Q} - 1)^{i-1} = 1+ b_{Q} \frac{(b_{Q}-1)^d-1}{(b_{Q}-1)-1}  \fstop
\end{equation}
If we start a bidirectional search between two quivers at distance $d$, to guarantee finding an intersection, the unguided pathfinder that explores must reach depth $d/2$ from each side. To be precise, one of the two sides will reach $\lfloor d/2 \rfloor$ depth, and the other $\lceil d/2 \rceil$. The expectation is the sum of these two expansions:
\begin{equation}
\scalebox{0.96}{$\displaystyle
    N_{\text{BFS}} \approx N_{\text{dir}}(\lfloor d/2 \rfloor) + N_{\text{dir}}(\lceil d/2 \rceil) = 2+ \frac{b_{\text{avg.}}}{b_{\text{avg.}}-2}\left((b_{\text{avg.}}-1)^{\lfloor d/2 \rfloor}+(b_{\text{avg.}}-1)^{\lceil d/2 \rceil}-2\right)\fstop$}
\end{equation}
This formulation implies asymmetric growth. When $d$ changes from an even to an odd distance, one tree must expand by an entire extra depth level, causing a spike in the total number of explored nodes. Conversely, when $d$ goes from odd to even, the second tree catches up to the same depth, resulting in a smaller multiplicative increase. The exponential growth of $N_{\text{BFS}}$ therefore oscillates depending on the parity of $d$, an artifact that will be visible in the efficiency metrics defined in Section \ref{sec:RESULTS}.

\subsection{Pathfinders for Seiberg Dualities}
\label{sec:PathforSeibergDualities}

We now proceed to present the NN-informed A$^*$ pathfinder we developed in detail. A summary of the algorithms is given in Appendix \ref{app:pathfindersgeneralities}, while their evaluations, costs, and objective functions are given in Table \ref{tab:summaryPaths}.

Since quivers are defined up to node permutations, at each step of the pathfinder, we compare a canonicalized version of the quivers via a Weisfeiler--Lehman (WL) hashing of the graphs.\footnote{This avoids missing the pair because of a ``wrong" ordering of the nodes.} Once a pair is identified, the search terminates, and the duality sequence is reconstructed. 

\begin{figure}[!htp]
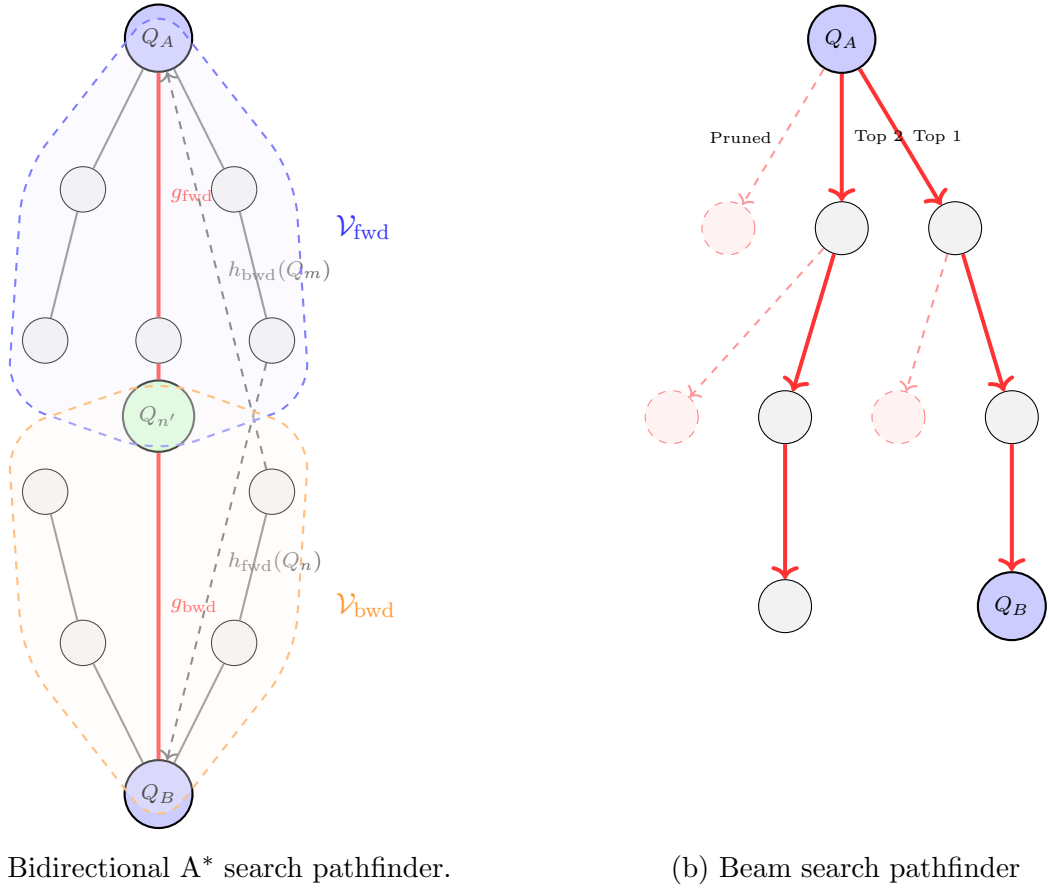

    \centering
    \begin{subfigure}[b]{0.49\textwidth}
        \centering

    \caption{Bidirectional A$^*$ search pathfinder.}
    \label{sfig:schematicbidirA*search}
    \end{subfigure}\hfill
    \begin{subfigure}[b]{0.49\textwidth}
        \centering
    %
    \caption{Beam search pathfinder}
    \label{sfig:schematicbeamsearch}
    \end{subfigure}
    \caption{Schematic representations of the two kinds of pathfinders we consider in this work.}
    \label{fig:schematicpathfinders}
\end{figure}

In contrast to the unguided expansion $N_{\text{BFS}}$, the total computational cost of the guided pathfinder is the sum of the unique quivers explored:
\begin{equation}\label{eq:Nexp}
    N_{\text{exp}} = |\mathcal{V}_{\text{fwd}}| + |\mathcal{V}_{\text{bwd}}|\fstop
\end{equation}
If the pathfinder is unidirectional, as will be the case for the AGNN pathfinder, then $\mathcal{V}_{\text{bwd}}=\emptyset$. If the search exceeds a predefined threshold of steps without finding an intersection, execution halts. This outcome indicates either that the two theories are not related by Seiberg dualities, or that their separation exceeds the maximum search depth. 

\begin{table}[!htp]
    \centering
    \renewcommand{\arraystretch}{1.2}
    \begin{tabular}{c||c|c|c}
        Pathfinder  & Search & Cost $g(.)$ & Heuristic $h(.)$\\
        \hhline{=#=|=|=}
        DGNN  & Bidir. A$^*$ & $1$ & $\text{DGNN}(Q_n,Q_\text{target})$\\\hline
        AGNN & $B=3$ Beam & $- \log P(D_i | Q_n,Q_\text{target})$ & $0$\\\hline
        Hybrid & Bidir. A$^*$ & $1-\lambda_\text{AR}\log P(D_i | Q_n,Q_\text{target})$ & $\lambda_{\text{DGNN}}\text{DGNN}(Q_n,Q_\text{target})$ \\\hline
        LCA & Bidir. A$^*$ & $c(D_i)$ & $0$\\\hline
        \makecell{Hybrid\\ LCA} & Bidir. A$^*$ &
        \makecell{$(1 - \lambda_{\text{det\_cost}}) + \lambda_{\text{det\_cost}} c_{\text{det}}(D_k)$ \\ $- \lambda_{\text{AR}} \log P(D_k | Q_n, Q_{\text{target}})$} &
        \makecell{$\lambda_{\text{DGNN}} \, \text{DGNN}(Q_n, Q_{\text{target}})$ \\ $+ \lambda_{\text{LCA}} \frac{\sum \text{rank}(Q_n)}{\sum \text{rank}(Q_{\text{root}})}$}
    \end{tabular}
    \caption{Summary of Pathfinders and their cost and heuristic functions.}
    \label{tab:summaryPaths}
\end{table}

\paragraph{BFS Pathfinder:} As a fundamental baseline for evaluating our neural network-guided models, we implement an unguided Breadth-First Search (BFS) pathfinder. This algorithm operates by systematically exploring the Seiberg duality graph level by level, generating all admissible mutations at depth $t$ before evaluating any transition at depth $t+1$. Structurally, it represents an unguided A$^*$ search where the heuristic function is identically zero ($h(Q) \equiv 0$) and the cumulative path cost simply tracks the traversal depth ($g(Q_{n+1}) = g(Q_n) + 1$), assigning an equal step cost to every edge. Consequently, the BFS pathfinder is topologically agnostic, treating every valid mutation as equally promising without any predictive physical intuition. To optimize execution and match the structural framework of the DGNN and Hybrid models, the BFS pathfinder is implemented as a bidirectional search originating simultaneously from both the root theory $Q_A$ and the target theory $Q_B$.

\paragraph{DGNN Pathfinder:} It consists of a bidirectional A$^*$ search algorithm with a heuristic function $h(\cdot)$ given by the output of the DGNN. This means that for any intermediate quiver states $Q_n$ and $Q_m$ originating from mutating $Q_A$ and $Q_B$, respectively, the evaluation functions for the forward and backward frontiers are defined as:
\begin{equation}
    \begin{split}
        f_{\text{fwd}}(Q_n) &= g_{\text{fwd}}(Q_n) + \text{DGNN}(Q_n, Q_B) 
        \\
    f_{\text{bwd}}(Q_m) &= g_{\text{bwd}}(Q_m) + \text{DGNN}(Q_m, Q_A)
    \coma
    \end{split}
\end{equation}
with $g_{\text{fwd}}(Q_A)=g_{\text{bwd}}(Q_B)=0$. The dequeued element at each step is determined by which evaluation function is smaller. This represents the first of a set of pathfinders in which the NN prediction provides the policy to guide the pathfinder. As we will discuss in Section \ref{sec:RESULTS}, despite being quite efficient and successful, the main problem with using the DGNN as a heuristic is its non-monotonicity; i.e., the distance between the two quivers is not monotonically decreasing. In fact, sometimes the heuristic delays the search queue, eventually degrading to a BFS algorithm.

\paragraph{AGNN Pathfinder:} Unlike the distance-guided approach, this implementation utilizes the probability distributions output by the Adviser GNN (AGNN) to parameterize the cost function $g(Q)$ of a beam search with beam width $B$.\footnote{Because beam search permanently discards alternative branches without a backtracking mechanism, the probability that a forward beam originating from $Q_A$ and a backward beam from $Q_B$ intersect in the vast intermediate mutation space is vanishingly small. Therefore, unlike the A$^*$ pathfinder, beam search is strictly implemented as a unidirectional search originating from $Q_A$.} Given an initial theory $Q_A$ and a target theory $Q_B$, the algorithm explores the space of valid Seiberg dualities starting from $Q_A$. Following our empirical evaluation of the network's classification accuracy in Section \ref{sec:ar_gnn}, we set the fixed beam width to $B=3$. 

At each expansion step, the algorithm processes all quivers currently residing in the beam. For a given state $Q_n \in \mathcal{B}_t$, the AGNN predicts the mutation probability distribution $P(D_k | Q_n, Q_B)$ across all admissible nodes. The algorithm selects the $B$ actions with the highest predicted probabilities and generates the corresponding mutant quivers $Q_{n+1} = D_k Q_n$. To formulate the trajectory evaluation as a cost minimization problem, each valid successor state $Q_{n+1}$ is assigned a cumulative trajectory cost $g(Q_{n+1})$, defined as the accumulated negative log-probability of the mutation sequence:
\begin{equation}
    g(Q_{n+1}) = g(Q_n) - \log P(D_k | Q_n, Q_B)\coma
\end{equation}
where the root theory is initialized at $g(Q_A) = 0$. Since $\log P \leq 0$, highly probable mutations contribute minimally to the accumulated cost, while unlikely ones heavily increase the cost. 

To prevent cyclic exploration, the algorithm checks the best candidates against the WL hashes of all previously evaluated quivers. If the candidate's structure has already been visited, the candidate is discarded.\footnote{Recall that the beam search algorithm simultaneously tracks up to $B$ active trajectories at each step. In this sense, it can be conceptualized as an A$^*$ search with a truncated frontier, trading the guarantee of completeness for memory efficiency.} Once possible loop branches have been pruned, the algorithm collects all surviving candidates generated across the current beam, sorts them in ascending order based on their cumulative trajectory cost $g(Q_{n+1})$, and retains only the top $B$ quivers to form the new beam $\mathcal{B}_{t+1}$.

The search procedure terminates successfully the moment a generated candidate is confirmed to be isomorphic to the target theory $Q_B$. Conversely, the search returns a failure status if the exploration reaches a predefined maximum depth or if the beam becomes empty due to dead-end mutations.\footnote{A dead-end occurs when all active trajectories in the beam are pruned. This scenario arises if the AGNN policy repeatedly favors mutations that lead to previously visited states. Because the algorithm systematically filters out historical structures to prevent infinite loops, an overconfident policy trapped in a cycle will cause all candidate branches to be discarded, leaving $\mathcal{B}_{t+1} = \emptyset$.}

\paragraph{Hybrid Pathfinder:} To overcome the limitations of the DGNN A$^*$ search and the AGNN beam search, we developed a \textit{Hybrid Pathfinder}. This approach combines the distance estimates of the DGNN with the branch probabilities of the AGNN. Specifically, the AGNN acts as a soft pruning mechanism, discouraging exploration along improbable branches and preventing the algorithm from stalling in flat or non-monotonic regions of the DGNN landscape. Retaining the A$^*$ priority queue provides a mechanism to backtrack if the AGNN predictions diverge. 

The Hybrid Pathfinder inherits the bidirectional exploration structure of the DGNN A$^*$ search but modifies the step-cost accumulation using the policy output of the AGNN. For any valid mutation $Q_{n+1} = D_k(Q_n)$, the evaluation function $f(Q_{n+1})$ is defined again as:
\begin{equation}
    f(Q_{n+1}) = g(Q_{n+1}) + \lambda_{\text{DGNN}} \, \text{DGNN}(Q_{n+1})\coma
\end{equation}
where $\lambda_{\text{DGNN}}$ is a scaling hyperparameter balancing the heuristic weight, and the cumulative trajectory cost $g(Q_{n+1})$ updates recursively via:
\begin{equation}
    g(Q_{n+1}) = g(Q_n) + 1 - \lambda_{\text{AR}} \log P(D_k \mid Q_n, Q_{\text{target}})\fstop
\end{equation}
Here, the base step cost of $1$ ensures admissibility by guaranteeing strictly increasing path costs, while the parameter $\lambda_{\text{AR}} \geq 0$ controls the penalty imposed on actions deviating from the optimal AGNN policy. When $\lambda_{\text{AR}} = 0$, the expansion reverts to the standard DGNN A$^*$ search; conversely, as $\lambda_{\text{AR}} \to \infty$ with $\lambda_{\text{DGNN}} = 0$, the trajectory cost dominates, recovering the behavior of the AGNN pathfinder.

\paragraph{LCA Pathfinder:} This algorithm works as an A$^*$ search without a heuristic function, where at each iteration, the pathfinder expands the active frontier in the direction that has the minimum path cost determined by hard-coded rules, not learned from training. At every state $Q_n$, the cost of mutating it with $D_k$ is based on the change in the rank of the dualized node $k$. The cost function $g(D_k)$ is determined by three weights, defined as $c_{\text{dec}} : c_{\text{eq}} : c_{\text{inc}} \sim 1 : 10 : 100$, such that the cost of every step is
\begin{equation}\label{eq:costLCA}
    g(D_k) = \begin{cases} 
      c_{\text{dec}} & \text{if } \text{rank}(k) \text{ decreases}\coma \\
      c_{\text{eq}} & \text{if } \text{rank}(k) \text{ is unchanged}\coma \\
      c_{\text{inc}} & \text{if } \text{rank}(k) \text{ increases}\fstop
   \end{cases}
\end{equation}
Because we penalize the increase of rank for both frontiers, the search does not attempt to find the direct path between two theories. Instead, the cost function acts as a ``gravity well" that funnels both search trees towards the model that has the lowest rank.

The motivation behind the LCA is to construct an algorithm that mimics the process that (human) physicists follow when guessing if two theories are related by Seiberg dualities: one tries to obtain a ``simpler theory" from both quivers, reducing ranks and arrows between nodes, hoping to reach a common ancestor that generates the pair of quivers. This heuristic serves as a more robust baseline than standard BFS for benchmarking our neural pathfinders. By explicitly incorporating the structural ``complexity'' of the quiver, ranks, and total number of arrows, this algorithm effectively prioritizes structural simplification. This contrasts with a standard BFS, which is topologically agnostic and blindly treats every valid mutation as equally promising, regardless of its impact on the theory's complexity.

\paragraph{Hybrid LCA Pathfinder:} Finally, a natural possibility is the integration of the LCA and Hybrid pathfinders. The idea is to further help the Hybrid pathfinder overcome the non-monotonicity and inversion problems: if the guidance of the NNs is unsure, the LCA can move the search towards the minimal quiver.

The LCA cost function enters into modifying both the cost and heuristic functions of the Hybrid pathfinder. To this end, we modify the total edge cost for the priority queue, which accumulates as:
\begin{equation}
    g(Q_{n+1}) = g(Q_n) + (1 - \lambda_{\text{det\_cost}}) + \lambda_{\text{det\_cost}}c_{\text{det}}(D_k) - \lambda_{\text{AR}} \log P(D_k | Q_n, Q_{\text{target}})\coma
\end{equation}
where $c_{\text{det}}(D_k)$ is defined similarly to \eqref{eq:costLCA}, i.e.
\begin{equation}
    c_{\text{det}}(D_k) = \begin{cases} 
      c_{\text{det\_dec}} & \text{if } \text{rank}(k) \text{ decreases}\coma \\
      c_{\text{det\_eq}} & \text{if } \text{rank}(k) \text{ is unchanged}\coma \\
      c_{\text{det\_inc}} & \text{if } \text{rank}(k) \text{ increases}\fstop
   \end{cases}
\end{equation}
Similarly, the heuristic incorporates a penalty proportional to the total rank of the quiver to encourage moving along quivers with smaller ranks. The modified function is defined as:
\begin{equation}
    h(Q_{n}) = \lambda_{\text{DGNN}} \, \text{DGNN}(Q_n, Q_{\text{target}}) + \lambda_{\text{LCA}} \frac{\sum \text{rank}(Q_n)}{\sum \text{rank}(Q_{\text{root}})}\coma
\end{equation}
where $Q_{\text{root}}$ denotes the initial graph $Q_A$ or $Q_B$ depending on the active frontier.

By adjusting the weights $\lambda_{\text{det\_cost}}$, $\lambda_{\text{AR}}$, $\lambda_{\text{DGNN}}$, and $\lambda_{\text{LCA}}$, this pathfinder averages between a purely NN-guided framework and the LCA pathfinder. However, this integration can cause conflicting search signals between the two pathfinders. The Hybrid pathfinder uses the guidance of DGNN and AGNN, which have learned to find the shortest path between two theories. This means that such a path might necessitate moving along trajectories where the quivers increase their ranks, completely avoiding the LCA. On the other hand, the LCA pathfinder works like an effective potential, moving precisely to find the theories with lower ranks. If the LCA is integrated into the Hybrid with the same hierarchy of weights defined for the LCA pathfinder, the resulting algorithm would simply reduce to the LCA pathfinder without any guidance given by the NNs. 

To balance the two guiding policies of the pathfinder, we introduce hyperparameters designed to maximize the efficiency ratio while maintaining a $100\%$ success rate. The optimizer converged to $c_{\text{det\_dec}} = 0.3$, $c_{\text{det\_eq}} = 2.7$, and $c_{\text{det\_inc}} = 3.1$, alongside a global weight $\lambda_{\text{det\_cost}} = 1.3$. Substituting these into the definition of $c_{\text{base}}$ yields effective base step costs of $c_{\text{base\_dec}} = 0.09$, $c_{\text{base\_eq}} = 3.21$, and $c_{\text{base\_inc}} = 3.73$ for the priority queue. 

This calibration imposes a penalty for rank-increasing mutations that is of the same order of magnitude as the neural network coefficients while contributing negligibly to the cost of rank-reducing mutations. Consequently, the LCA penalty does not dominate the Hybrid cost function. Instead, it biases search toward smaller-rank theories when network confidence is low while allowing high-probability shortcuts to proceed uninterrupted.

\subsection{Complexity Comparison}

We, now, compare the computational complexity of the NN-guided pathfinders against the unguided bidirectional BFS baseline established in Section \ref{sec:branching_factor}. In an unguided expansion, the number of explored states scales exponentially as $\mathcal{O}\left(b_{\text{avg.}}^{d/2}\right)$, where $b_{\text{avg.}}$ is the average valid branching factor and $d$ is the minimum mutation distance between the theories.

In the ideal limit of a monotonic heuristic, an A$^*$ search expands only the states along the optimal trajectory, achieving linear complexity $\mathcal{O}(d)$. However, as shown in Section \ref{sec:trainingevalDGNN}, the distance estimates generated by the DGNN exhibit non-monotonic fluctuations across mutation sequences, violating the formal conditions for algorithmic consistency. Because A$^*$ maintains an exhaustive frontier of unexplored branches, a locally misleading heuristic can attract the search into non-optimal trajectories. Consequently, in the worst-case scenario, the computational cost not only degrades to the exponential scaling of the BFS baseline but can even exceed it, exhausting the step budget before recovering the true path. The same worst-case bound applies to the LCA baseline, which operates as an A$^*$ search guided purely by a deterministic topological cost function. Because the LCA heuristic strictly penalizes increases in gauge ranks and arrow counts, the algorithm can easily become trapped in a local minimum of topological ``complexity''. In this case, the search might stagnate in minimal-rank configurations, eventually exhausting its computational budget.

Conversely, the AGNN beam search strictly confines the expansion tree to a maximum of $B$ candidate quivers per level. The number of evaluated states is strictly bounded by $\mathcal{O}(B \cdot b_{\text{avg.}} \cdot d)$, thereby guaranteeing predictable linear scaling with respect to distance at the cost of a decrease in NN accuracy. Indeed, this massive reduction in memory and execution time sacrifices algorithmic completeness: as discussed in Section \ref{sec:trainingevalAGNN}, if an incorrect AGNN prediction prunes the true path from the active beam, the algorithm permanently loses the ability to backtrack and fails to find a solution.

The Hybrid Pathfinder overcomes the limitations of both by integrating localized pruning policies with a backtracking memory. First, by applying a Top-$k$ filter based on the AGNN policy distributions, the effective branching factor is bounded by $b_{\text{eff.}} \leq B \ll b_{\text{avg.}}$. Second, embedding the negative log-likelihood penalty $-\log P(D_k | Q_n, Q_{\text{target}})$ into the step-cost $g(Q)$ counteracts the non-monotonicity of the DGNN heuristic.

Furthermore, because $P \leq 1$, the modified step cost is bounded from below by $g(Q_{n+1}) - g(Q_n) \geq 1$. This positivity guarantees that, even under conditions where both NNs produce inaccurate predictions, the Hybrid Pathfinder will degrade to a BFS-like exploration. Finally, the Hybrid LCA variant preserves these theoretical complexity bounds while introducing a bias that favors rank-reducing mutations when the NN-guidance is uncertain.

\subsection{Pathfinders Evaluation and Efficiency Metrics}
\label{sec:pathfinder_evaluation}

To quantify the performance and efficiency of the pathfinders, we compare them to baseline algorithms, namely the BFS search and the LCA pathfinder. Although the AGNN pathfinder performs a unidirectional search, we nevertheless decided to compare it with the bidirectional baselines to facilitate comparisons with the other models. 

\paragraph{Success Rate (SR):} We define, for each pathfinder, the success rate as the fraction of test pairs for which the model successfully finds a path, i.e., $N_\text{succ.}$, out of the total number of test pairs evaluated, i.e., $N_\text{tot.}$, obtaining
\begin{equation}
    \text{SR} = \frac{N_\text{succ.}}{N_\text{tot.}}\fstop
\end{equation}

\paragraph{Efficiency Ratio  (ER):} This indicates how successful the guidance coming from the cost and heuristic functions is at pruning the search space compared to a given baseline search. An $\text{ER} > 1$ indicates an active reduction in search complexity. We define two different baselines: the BFS search, representing an uninformed baseline, and the LCA search, serving as a physics-informed baseline. We thus compute the ERs against either the BFS explored nodes $N_{\text{BFS}}$ or the LCA ones $N_{\text{LCA}}$. 
 The ratio is evaluated against the baseline expectation:
    \begin{equation}
        \text{ER}_{\text{bidir}} = \frac{N_{\text{baseline}}}{N_{\text{exp}}}\coma
    \end{equation}
    where $N_{\text{exp}}$ is the total number of nodes explored by search algorithms as in \eqref{eq:Nexp}, and $N_{\text{baseline}}$ corresponds to $N_{\text{BFS, bidir}}$ or $N_{\text{LCA}}$ depending on the comparison.\footnote{For the LCA pathfinder, the comparison is obviously done over the BFS baseline.} Note that when $N_{\text{baseline}} = N_{\text{BFS}}$, the ratio inherits the oscillations originating from the asymmetric bidirectional BFS expansion. These spikes at odd distances are an artifact of the unguided baseline and do not reflect the underlying behavior of the pathfinders, as confirmed by the absence of such oscillations when evaluating the efficiency against the LCA pathfinder.
    
\paragraph{Guidance Efficiency (GE):} This metric quantifies the precision of the neural network policy by measuring the search relative to the theoretical minimum number of steps required to discover a path of distance $d$. Because it is defined as a ratio of observed evaluations to the ideal minimum, it is strictly bounded by $\text{GE} \geq 1$, where values approaching $1$ indicate a near-perfect trajectory with minimal branching redundancy.
    An optimal search would never expand branches outside the shortest path. Accounting for the immediate neighbors evaluated to confirm each step along the trajectory, the theoretical minimum number of explored states is $b_{\text{avg.}}$ at the root level and $b_{\text{avg.}} - 1$ for the remaining $d - 1$ transition steps. The bidirectional guidance efficiency is thus defined as:
        \begin{equation}\label{eq:GEbidir}
            \text{GE}_{\text{bidir}} = \frac{N_{\text{exp}}}{b_{\text{avg.}} + (d - 1)(b_{\text{avg.}} - 1)}\fstop
        \end{equation}
    For the AGNN pathfinder, however, the search is performed by only keeping a beam width $B$. 
     Under perfect guidance, the AGNN evaluates only the single root quiver at depth $0$ and at most $\min(B, b_{\text{avg.}})$ candidate branches at each subsequent depth level. The unidirectional guidance efficiency is therefore defined as 
        \begin{equation}
            \text{GE}_{\text{unidir}} = \frac{N_{\text{exp}}}{1 + (d - 1) \min(B, b_{\text{avg.}})}\fstop
        \end{equation}
  
\paragraph{Effective Efficiency Ratio (EER):} Because the evaluated models exhibit varying SR across different duality distances and quiver complexities, and since the underlying ERs are formulated differently for A$^*$ and beam search architectures, a unified comparison requires a composite metric. We, therefore, define the Effective Efficiency Ratio as the direct product of efficiency and accuracy:
\begin{equation}\label{eq:EERdef}
    \text{EER} = \text{ER} \times \text{SR}\fstop
\end{equation}
This formulation naturally balances the trade-off between search speed and algorithmic completeness, strictly penalizing pathfinders that fail to converge by assigning zero effective efficiency to unresolved pairs.

The benchmarking evaluations described above are initially conducted on an in-distribution (ID) test set drawn from the theory families detailed in Appendix \ref{sec:IDTheories}. This assesses pathfinder performance when the neural networks operate within familiar topological domains encountered during training. Furthermore, because the search algorithms actively transform theory structures through sequential Seiberg dualities, it is important to evaluate their robustness against unfamiliar architectures. Consequently, we extend our benchmarks to the out-of-distribution (OOD) theories cataloged in Appendix \ref{sec:OODTheories} to test the generalization capabilities of the learned policies.

\section{Results} 
\label{sec:RESULTS}

In this section, we evaluate the computational efficiency and routing performance of the GNN pathfinders introduced in Section \ref{sec:pathfindersforSD}. We begin our analysis by benchmarking the models on the in-distribution (ID) test set detailed in Appendix \ref{sec:IDTheories}, evaluating their search capabilities within familiar domains. In Section \ref{sec:Performance-OODTh}, we investigate their generalization properties when applied to the out-of-distribution (OOD) architectures listed in Appendix \ref{sec:OODTheories}. To ensure statistical consistency across all evaluations, each experimental benchmark is performed on a randomized sample of 500 quiver pairs, systematically categorized by node count and Seiberg duality distance.

\begin{table}[!htp]
    \centering
    \begin{subtable}[t]{\textwidth}
        \centering
        \caption{Performance against BFS Baseline}
        \label{tab:EER+SR_summary}
\begin{tabular}{c|c|c|c|c}
            \hhline{=|=|=|=|=}
            {Dataset} & {Pathfinder} & {$\overline{\text{SR}}$ (\%)} & {$\overline{\text{ER}}$} & {$\overline{\text{EER}}$} \\
            \hhline{=|=|=|=|=}
            \multirow{5}{*}{ID dataset} & DGNN & $99.94$ & $430.25$ & $429.99$ \\\cline{2-5}
            & AGNN & $76.12$ & $928.59$ & $706.84$ \\\cline{2-5}
            & Hybrid & $100.00$ & $790.08$ & $790.07$ \\\cline{2-5}
            & LCA & $100.00$ & $763.66$ & $763.66$ \\
            \hhline{=|=|=|=|=}
            \multirow{5}{*}{OOD dataset in Figure \ref{fig:Baoquivers}} & DGNN & $99.51$ & $134.38$ & $133.72$ \\\cline{2-5}
            & AGNN & $50.91$ & $305.15$ & $155.35$ \\\cline{2-5}
            & Hybrid & $99.84$ & $221.99$ & $221.63$ \\\cline{2-5}
            & LCA & $99.97$ & $176.13$ & $176.08$ \\
            \hhline{=|=|=|=|=}
            \multirow{5}{*}{OOD dataset in Figure \ref{fig:newquivers}} & DGNN & $99.34$ & $714.42$ & $709.74$ \\\cline{2-5}
            & AGNN & $78.64$ & $1175.18$ & $924.21$ \\\cline{2-5}
            & Hybrid & $99.92$ & $881.67$ & $880.96$ \\\cline{2-5}
            & LCA & $100.00$ & $568.25$ & $568.25$ \\%
            \hhline{=|=|=|=|=}
        \end{tabular}
    \end{subtable}

    \vspace{1em}
    
    \begin{subtable}[t]{\textwidth}
        \centering
        \caption{Performance against LCA Baseline}
        \label{tab:EER+SR_LCA_summary}
        \begin{tabular}{c|c|c|c|c}
            \hhline{=|=|=|=|=}
            {Dataset} & {Pathfinder} & {$\overline{\text{SR}}$ (\%)} & {$\overline{\text{ER}}$} & {$\overline{\text{EER}}$} \\
            \hhline{=|=|=|=|=}
            \multirow{5}{*}{ID dataset} & DGNN & $99.94$ & $0.83$ & $0.83$ \\\cline{2-5}
            & AGNN & $76.12$ & $2.04$ & $1.55$ \\\cline{2-5}
            & Hybrid & $100.00$ & $1.10$ & $1.10$ \\\cline{2-5}
            & Hybrid LCA & $100.00$ & $1.14$ & $1.14$ \\
            \hhline{=|=|=|=|=}
            \multirow{5}{*}{OOD dataset in Figure \ref{fig:Baoquivers}} & DGNN & $99.51$ & $0.54$ & $0.53$ \\\cline{2-5}
            & AGNN & $50.91$ & $1.03$ & $0.52$ \\\cline{2-5}
            & Hybrid & $99.84$ & $0.83$ & $0.83$ \\\cline{2-5}
            & Hybrid LCA & $100.00$ & $0.98$ & $0.98$ \\
            \hhline{=|=|=|=|=}
            \multirow{5}{*}{OOD dataset in Figure \ref{fig:newquivers}} & DGNN & $99.34$ & $0.83$ & $0.83$ \\\cline{2-5}
            & AGNN & $78.64$ & $1.91$ & $1.50$ \\\cline{2-5}
            & Hybrid & $99.92$ & $1.16$ & $1.16$ \\\cline{2-5}
            & Hybrid LCA & $100.00$ & $1.18$ & $1.18$ \\
            \hhline{=|=|=|=|=}
        \end{tabular}
    \end{subtable}
    \caption{Summary of pathfinder performances evaluated as the average across all distances and node counts in each dataset. The EER is the effective efficiency ratio obtained by multiplying the success rate by the corresponding efficiency ratio as in \eqref{eq:EERdef}. Table \ref{tab:EER+SR_summary} shows the comparison of the pathfinders against the BFS baseline, while Table \ref{tab:EER+SR_LCA_summary} shows the comparison of the pathfinders against the LCA baseline.}
    \label{tab:overall_performance}
\end{table}

\subsection{In-Distribution Theories Analysis}
\label{sec:Performance-IDTh}

\subsubsection{DGNN Pathfinder Performance}
\label{sec:DGNNPathPerformance}

We first examine the baseline performance of the A$^*$ search, guided purely by the DGNN distance heuristic, as formulated in Section \ref{sec:PathforSeibergDualities}. As illustrated in \cref{sfig:SH_siamese_ID}, the DGNN pathfinder demonstrates exceptional reliability across the ID evaluation set, achieving an overall SR of 99.94\%. The algorithm successfully resolves the vast majority of duality trajectories, exhibiting only minor performance degradation when traversing large duality distances.

\begin{figure}[!htp]
    \centering

\caption{Success Heatmap}
        \label{sfig:SH_siamese_ID}
\end{figure}

\begin{figure}[!htp]
    \centering
   
    \begin{subfigure}[b]{0.49\textwidth}
    \centering
%
\caption{Efficiency Ratio}
        \label{sfig:ER_siamese_ID}
    \end{subfigure}\hfill
    \begin{subfigure}[b]{0.49\textwidth}
    \centering
%
\caption{Guidance Efficiency}
        \label{sfig:GE_siamese_ID}
    \end{subfigure}
    \caption{Global performance of the DGNN pathfinder tested on theories in the testing set of DGNN.}
    \label{fig:siamese_global_ID}
\end{figure}

As illustrated in Figure \ref{sfig:ER_siamese_ID}, the search efficiency of the DGNN pathfinder remains stable and does not degrade significantly at large distances. However, the guidance of the DGNN model in Figure \ref{sfig:GE_siamese_ID} reduces when considering pairs of quivers separated by larger distances. This means that the pathfinder is slowly degrading to an unguided pathfinder. Despite this, the EER does not show signs of decreasing with distance; hence, it still outperforms a pure BFS algorithm. 

However, if we had the possibility to improve the guidance of the NN, the performance of the pathfinder could, in principle, improve. One of the reasons why the DGNN pathfinder becomes unguided at large distances is due to the non-monotonicity of the heuristic function given by the output of the DGNN. One way to compensate for this is to try to modify the cost function so that the pathfinder can prioritize certain directions when exploring the possible nodes to dualize. The AGNN pathfinder, discussed below, addresses this issue by modifying the cost function.

\subsubsection{AGNN Pathfinder Performance}
\label{sec:AGNNPathPerformance}

We next evaluate the unidirectional beam search guided by the AGNN policy. The main difference of the beam search compared to the A$^*$ search is the limited frontier that, at each step of the path, is maintained. However, limiting the frontier, if the policy is not perfect (as anticipated in Section \ref{sec:trainingevalAGNN}), may result in the pathfinder being unable to find a path at all. For this reason, the AGNN pathfinder will have a success rate lower than that of the DGNN A$^*$ pathfinder. 

\begin{figure}[!htp]
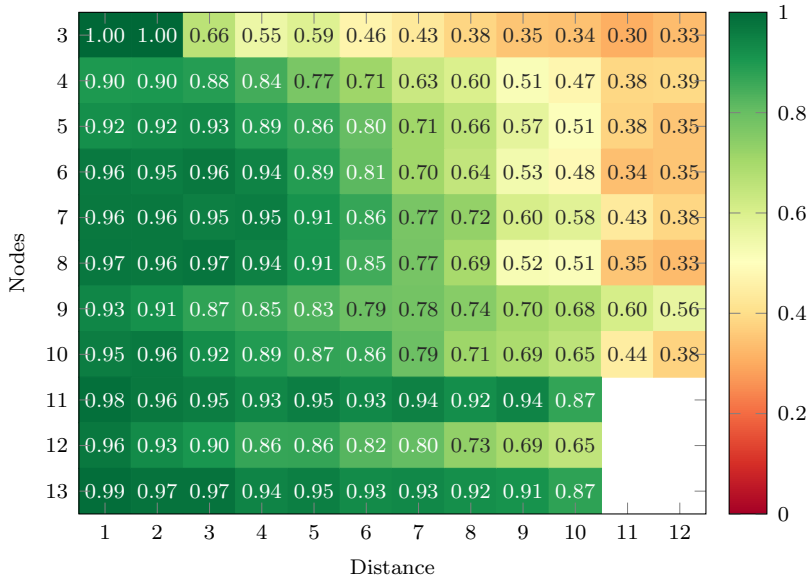

    \centering

\caption{Success Heatmap}
        \label{sfig:SH_ar_ID}
\end{figure}

As a benchmark, we evaluate the AGNN pathfinder across the same test set used in the previous section. As illustrated in \cref{sfig:SH_ar_ID}, the overall SR for the AGNN pathfinder converges to approximately 76.12\%. Nevertheless, the AGNN pathfinder achieves high search efficiency.

\begin{figure}[!htp]
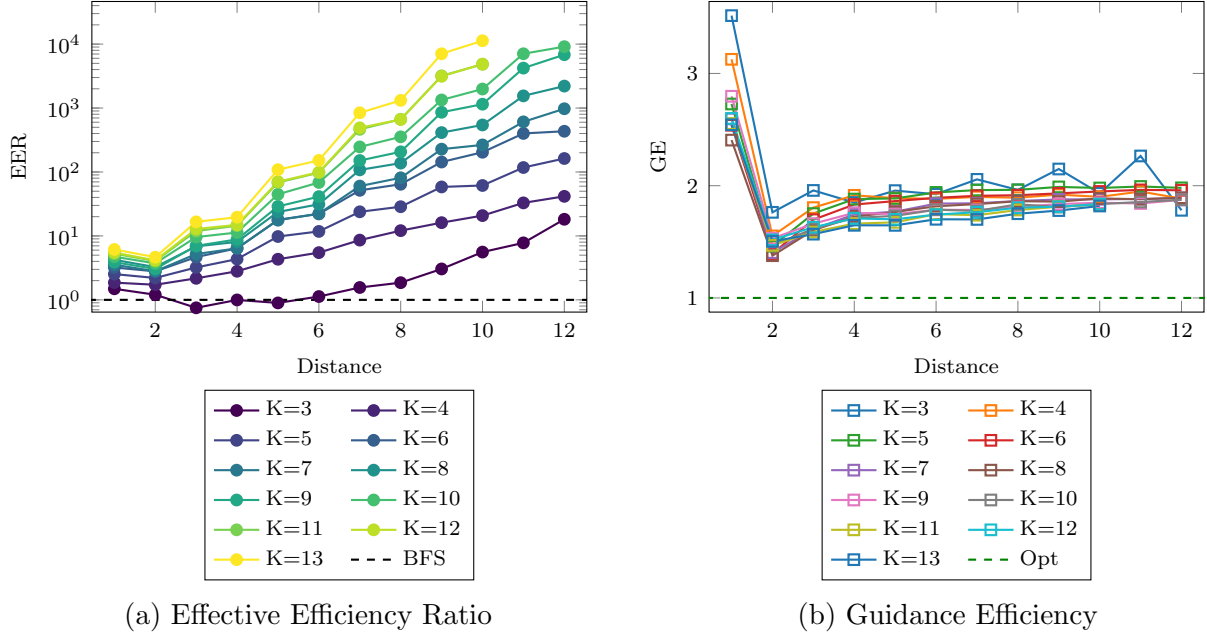


    \centering
    \begin{subfigure}[b]{0.49\textwidth}


\caption{Effective Efficiency Ratio}
        \label{sfig:EER_ar_ID}
    \end{subfigure}\hfill
    \begin{subfigure}[b]{0.49\textwidth}

%
\caption{Guidance Efficiency}
        \label{sfig:GE_ar_ID}
    \end{subfigure}

    \caption{Global performance of the AGNN pathfinder evaluated on theories in the testing set of AGNN.}
    \label{fig:ar_global_ID}
\end{figure}

The AGNN pathfinder is the only pathfinder in this work that is unidirectional. However, for consistency in the analysis with the other pathfinders, we still compare it with the bidirectional BFS baseline. Additionally, the beam search we performed set the beam width to $B=3$. This is the reason why, in Figure \ref{sfig:EER_ar_ID}, for quivers with $K=3$ and small distances, the performance of the AGNN pathfinder is worse than that of the BFS baseline since, by construction, the AGNN pathfinder explores all the nodes of the quiver at each step. However, for larger quivers, the AGNN pathfinder is an order of magnitude better than the bidirectional BFS baseline, despite its reduced SR. Furthermore, the efficiency of the pathfinder grows with the distance, even for small quivers. 

This sustained efficiency is directly explained by the GE in Figure \ref{sfig:GE_ar_ID}, which remains stable across all distances. The AGNN pathfinder resolves the large-distance breakdown seen in DGNN pathfinder for large distances because the pathfinder has a more definite criterion for choosing the direction of the path. However, this gain in efficiency comes at the expense of success rate 

Combining the AGNN and DGNN policies into a single pathfinder aims to recover the success rate while maintaining high GE.\footnote{Recall that, by definition in \eqref{eq:GEbidir}, the GE is better the closer it is to unity.}

\subsubsection{Hybrid Pathfinder Performance}
\label{sec:HybridPathPerformance}

\begin{figure}[!htp]
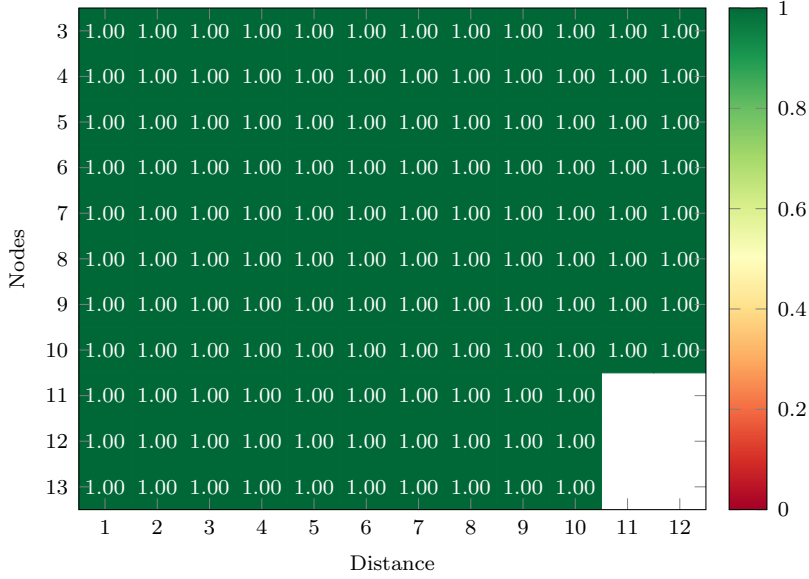

    \centering

\caption{Success Heatmap}
        \label{sfig:SH_hybrid_ID}
\end{figure}

We now evaluate the Hybrid pathfinder, designed to integrate the complementary guidance policies of the AGNN and DGNN to address their limitations. The evaluation is once again performed on the same test set. The most immediate result is the recovery of a $100\%$ SR across all tested duality distances, as shown in \cref{sfig:SH_hybrid_ID}.

\begin{figure}[!htp]
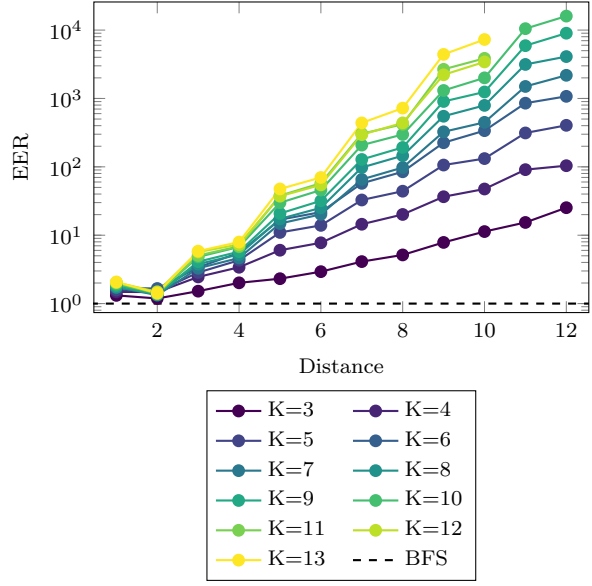
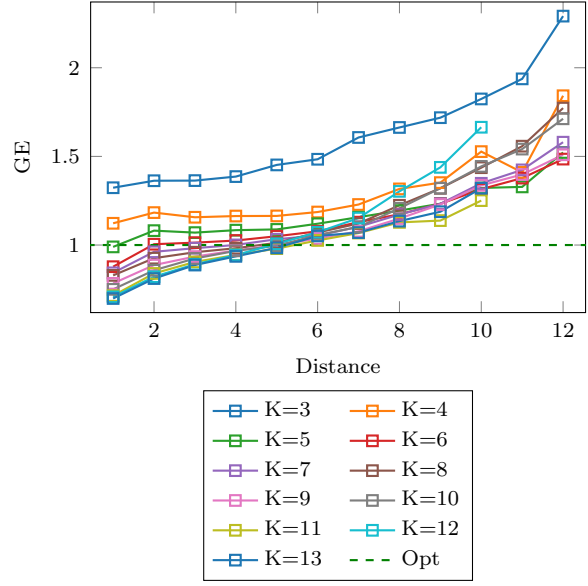


    \centering
   
    \begin{subfigure}[b]{0.49\textwidth}


\caption{Efficiency Ratio}
        \label{sfig:ER_hybrid_ID}
    \end{subfigure}\hfill
    \begin{subfigure}[b]{0.49\textwidth}

%
\caption{Guidance Efficiency}
        \label{sfig:GE_hybrid_ID}
    \end{subfigure}
    \caption{Global performance of the Hybrid pathfinder tested on theories in the testing set of DGNN and AGNN.}
    \label{fig:hybrid_global_ID}
\end{figure}

Figure \ref{sfig:ER_hybrid_ID} confirms that the ER outperforms the unguided BFS baseline across all distances, becoming more efficient also for small quivers. We also see that the GE in Figure \ref{sfig:GE_hybrid_ID} is closer, in order of magnitude, to the one observed for the AGNN pathfinder, starting to diverge only at large distances. The reason for this behavior is that the Hybrid pathfinder keeps the whole frontier of nodes to dualize, as does the DGNN pathfinder, but each node has a cost function given by the probability distribution predicted by the AGNN model.\footnote{In Figure \ref{sfig:GE_hybrid_ID}, we can see that the GE can become smaller than $1$. This happens when, given a starting quiver $Q_A$, multiple nodes exist that can be dualized to reach $Q_B$, while \eqref{eq:GEbidir} assumes that there is only one path connecting each pair.} 

For ID theories, the Hybrid pathfinder resolves the guidance inefficiency of DGNN and the success-rate drop of the AGNN pathfinder, giving the so-far most efficient NN-guided pathfinder with a 100\% SR in finding a path (if the path exists).

\subsubsection{LCA Pathfinder Performance}
\label{sec:LCA-performance}

\begin{figure}[!htp]
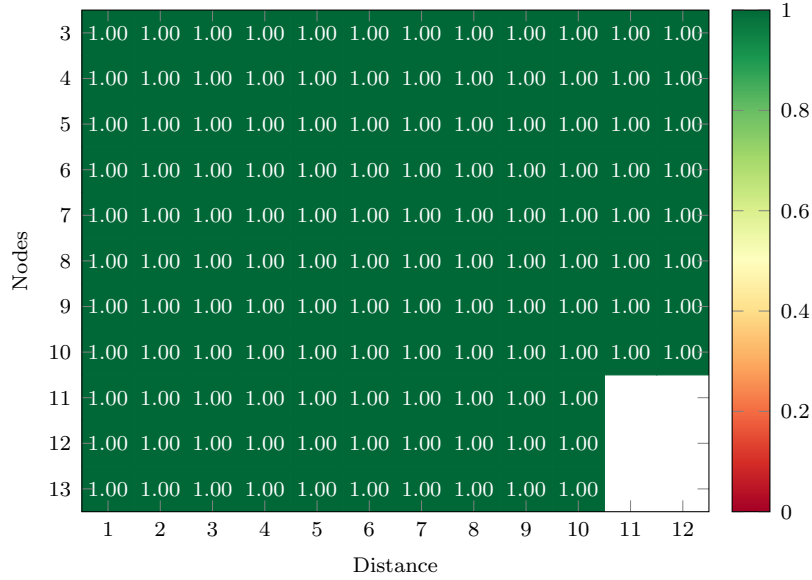

    \centering


        \caption{Success Heatmap}
        \label{sfig:SH_deterministic_ID}
\end{figure}

In this section, we discuss the performance of the LCA pathfinder over the BFS baseline. Despite the fact that this pathfinder does not have a NN-guided policy, we test it on the same sample of theories that we used for the previous pathfinders to be able to compare its performance. As shown in \cref{sfig:SH_deterministic_ID}, the LCA pathfinder reaches an SR near $100\%$, so we can use the ER as a ratio of efficiency when we compare it with the BFS baseline.

\begin{figure}[!htp]
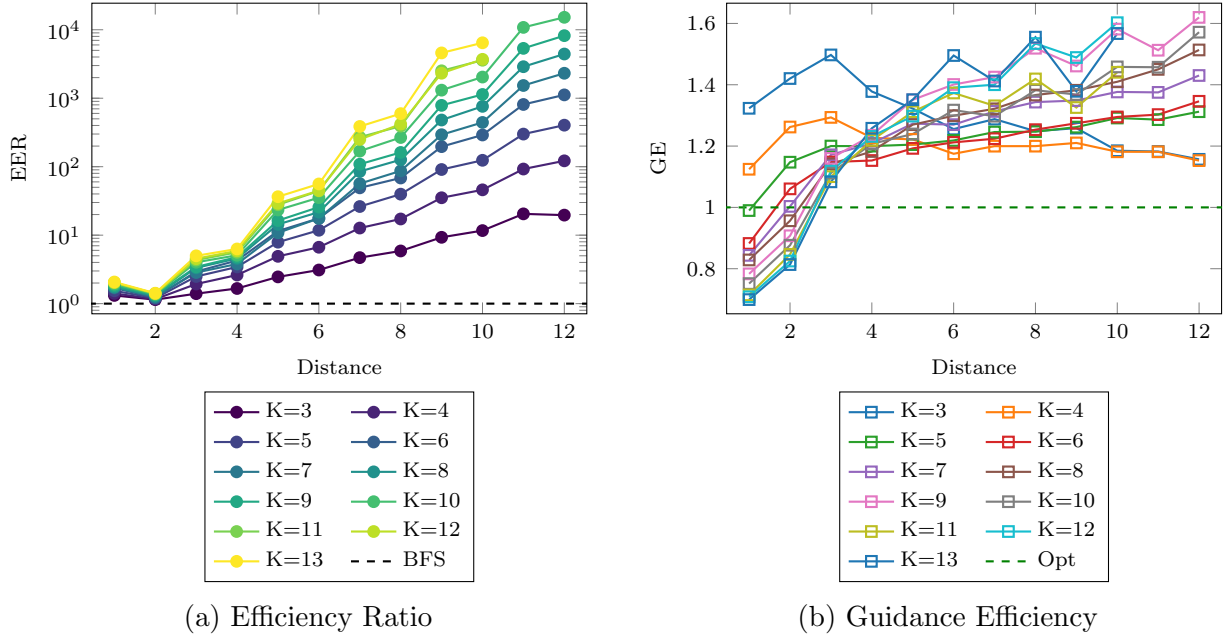


    \centering
    \begin{subfigure}[b]{0.49\textwidth}


\caption{Efficiency Ratio}
        \label{sfig:ER_deterministic_ID}
    \end{subfigure}\hfill
    \begin{subfigure}[b]{0.49\textwidth}

%
\caption{Guidance Efficiency}
        \label{sfig:GE_deterministic_ID}
    \end{subfigure}
    \caption{The global performance of the LCA pathfinder tested on the ID dataset.}
    \label{fig:deterministic_global_ID}
\end{figure}

We see from Figure \ref{sfig:ER_deterministic_ID} that the LCA pathfinder reaches levels of performance comparable to the Hybrid pathfinder in Figure \ref{sfig:ER_hybrid_ID}. However, the way in which each pathfinder is designed is different. The LCA pathfinder finds a common ancestor with lower ranks, while the Hybrid pathfinder has two NN-guided policies that try to reduce the distance between the quivers and predict the nodes to be dualized at each step. It is true that the NNs have been trained to minimize the differences between the quivers; however, the distances predicted or the probability distributions are generated on the resulting graphs processed by the Transformers. The fact that both pathfinders perform similarly is, in some sense, surprising because it seems that the NNs have learned some pattern to guide the path towards a common quiver. This raises the question of how much the Hybrid pathfinder improves upon the LCA baseline.

\subsubsection{Comparing Pathfinders with LCA Baseline}\label{sec:CompatingPathLCA}

Although the BFS baseline is the only objective comparison to judge the efficiency of a pathfinder, since it represents the blind search one would perform when given two graphs, the LCA pathfinder represents the algorithm that a physicist would write for the task of finding the chain of dualities connecting two theories; this should be, from now on, the baseline against which to compare our NN-guided pathfinders. 

\begin{figure}[!htp]
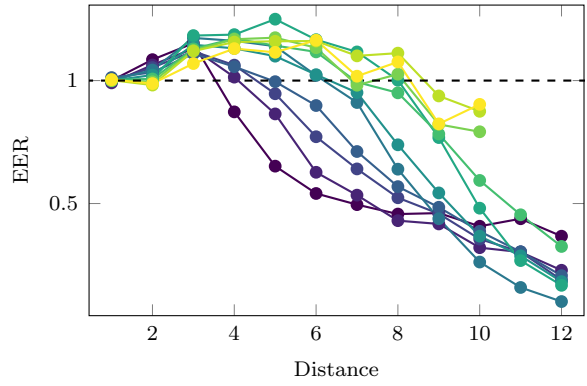
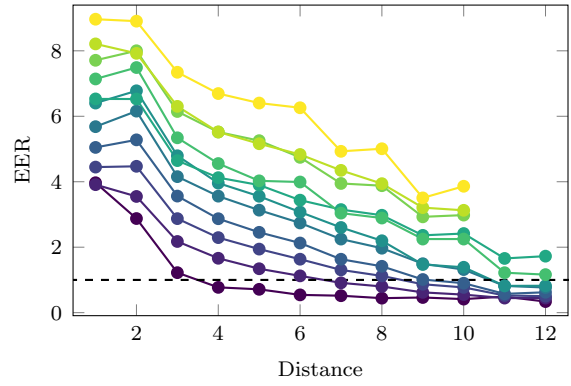
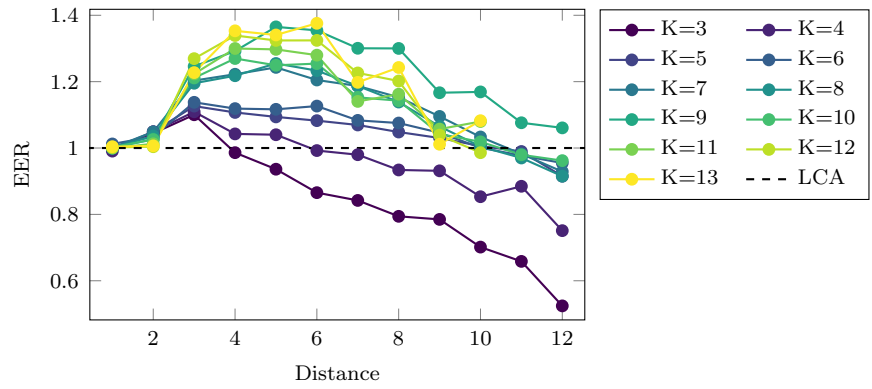

    \centering
    \begin{subfigure}[b]{0.49\textwidth}

\caption{DGNN Pathfinder}
        \label{sfig:eer_siamese_theories}
    \end{subfigure}\hfill
     \begin{subfigure}[b]{0.49\textwidth}
        %
\caption{AGNN Pathfinder}
        \label{sfig:eer_ar_theories}
    \end{subfigure}
    
    \vspace{0.2cm}
    \begin{subfigure}[b]{0.49\textwidth}
        %
\caption{Hybrid Pathfinder}
        \label{sfig:eer_hybrid_theories}
    \end{subfigure}
    \caption{EER for the three pathfinders using LCA baseline in the ID dataset.}
    \label{fig:eer_global_ID}
\end{figure}

In summary, DGNN achieves a high SR with degrading GE, whereas AGNN yields a lower SR with superior EER and GE. The Hybrid pathfinder reaches a 100\% SR, good performance, and stable GE. In \cref{fig:eer_global_ID}, we show how those same EERs compare when evaluated against the LCA pathfinder as the baseline. 

The first observation in Figure \ref{fig:eer_global_ID} is that the scale of the EER is now linear instead of logarithmic. Unlike the BFS comparisons, no model outperforms the LCA pathfinder by orders of magnitude; only the AGNN pathfinder in \cref{sfig:eer_ar_theories} has a $\sim 10\times$ speedup for small distances. 

\begin{figure}[!htp]
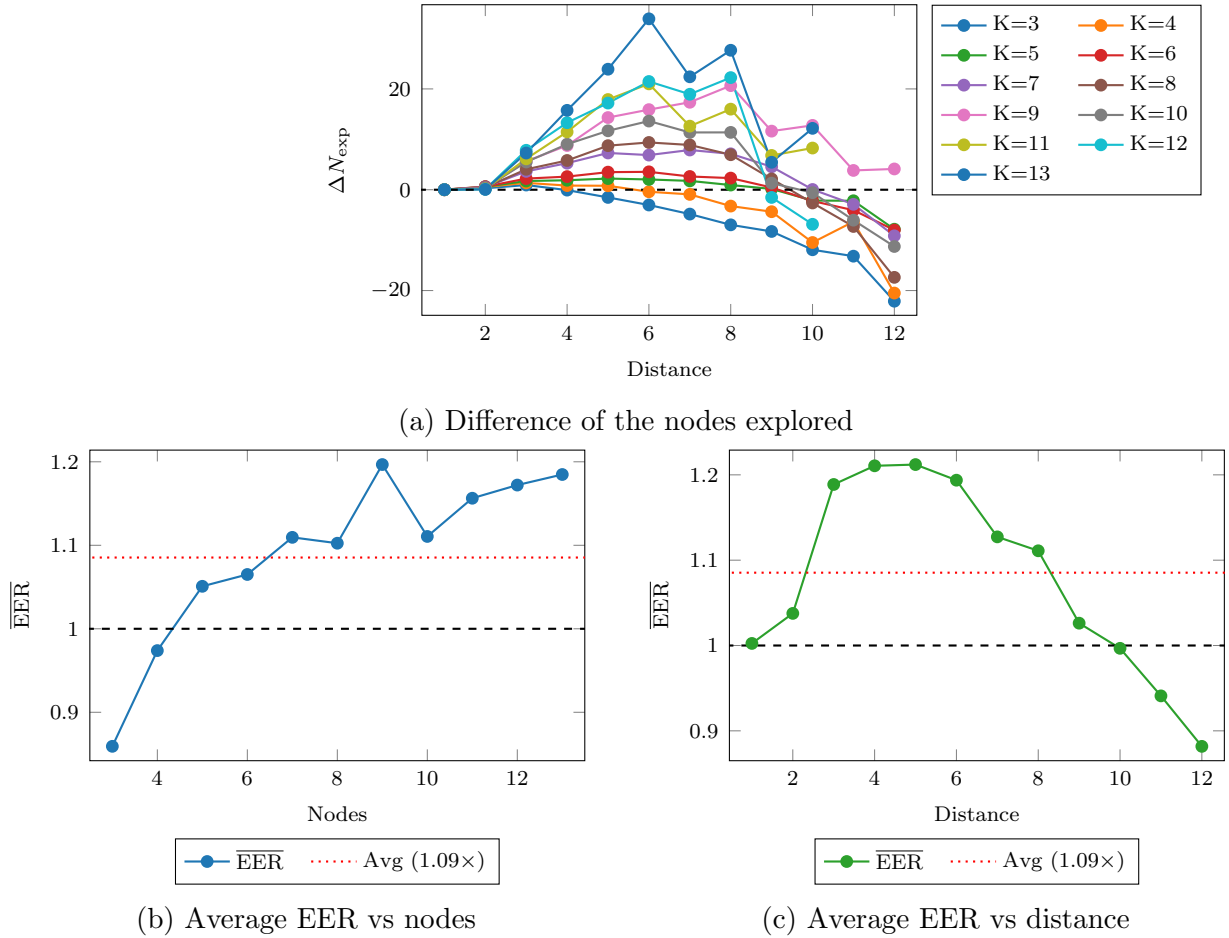

    \centering
      \begin{subfigure}[b]{0.49\textwidth}
        \definecolor{col1F77B4}{RGB}{31,119,180}
\definecolor{colFF7F0E}{RGB}{255,127,14}
\definecolor{col2CA02C}{RGB}{44,160,44}
\definecolor{colD62728}{RGB}{214,39,40}
\definecolor{col9467BD}{RGB}{148,103,189}
\definecolor{col8C564B}{RGB}{140,86,75}
\definecolor{colE377C2}{RGB}{227,119,194}
\definecolor{col7F7F7F}{RGB}{127,127,127}
\definecolor{colBCBD22}{RGB}{188,189,34}
\definecolor{col17BECF}{RGB}{23,190,207}
\definecolor{col000000}{RGB}{0,0,0}

        \caption{Difference of the nodes explored}
        \label{sfig:expsaving_ID}
          \end{subfigure}
          
  \begin{subfigure}[b]{0.49\textwidth}
\definecolor{col1F77B4}{RGB}{31,119,180}
\definecolor{col000000}{RGB}{0,0,0}
\definecolor{colFF0000}{RGB}{255,0,0}
%
        \caption{Average EER vs nodes}
        \label{sfig:avgEER_nodes_ID}
          \end{subfigure}\hfill  
\begin{subfigure}[b]{0.49\textwidth}
\definecolor{col2CA02C}{RGB}{44,160,44}
\definecolor{col000000}{RGB}{0,0,0}
\definecolor{colFF0000}{RGB}{255,0,0}
%
        \caption{Average EER vs distance}
        \label{sfig:avgEER_distance_ID}
          \end{subfigure}
        \caption{Nodes explored and average EER per nodes and distances of the Hybrid pathfinder compared with the LCA pathfinder in the ID dataset}
        \label{fig:exp+avgEER_ID}
\end{figure}

The Hybrid pathfinder achieves a global average speedup of $1.09\times$ over the LCA baseline, but its performance depends on the quiver size and the distance:
\begin{itemize}
    \item For small quivers ($K < 5$), the LCA pathfinder is more efficient, while Figure \ref{sfig:avgEER_nodes_ID} shows that the Hybrid pathfinder outperforms LCA on larger graphs ($K \ge 5$), with its average EER scaling from $1.05\times$ up to $1.18\times$ for $K \ge 10$.
    \item Figure \ref{sfig:avgEER_distance_ID} shows an immediate advantage for the Hybrid model across distances $d \in [2, 8]$, peaking at $\sim 1.21\times$ for $d \in [4, 5]$. The EER drops below unity only for small quivers at very high distances, $d>10$.
\end{itemize}
This is corroborated by the absolute node savings ($\Delta N_\text{exp} = N_\text{LCA} - N_\text{Hyb}$) in Figure \ref{sfig:expsaving_ID}, confirming that NN-guidance becomes  important as soon as the size of the quiver increases.

\begin{figure}[!htp]
     \centering
     \pgfplotstableread{
Label Shorter Equal Longer
3 0.00 100.00 0.00
4 1.30 96.53 2.16
5 3.36 89.31 7.33
6 3.98 87.13 8.89
7 4.71 83.12 12.16
8 5.87 81.80 12.33
9 4.04 79.33 16.63
10 5.40 78.64 15.96
11 1.93 75.64 22.42
12 5.31 80.06 14.63
13 1.68 72.58 25.74
}\dataAX

\begin{tikzpicture}[baseline=0,font=\footnotesize]
    \definecolor{col1}{HTML}{1F77B4}
    \definecolor{col3}{HTML}{D62728}
    \begin{axis}[
            ybar stacked,
            bar width=10pt,
            width=10cm, 
            height=8cm,
            xlabel={Nodes},
            ylabel={Percentage (\%)},
            ymin=0, ymax=30,
            xtick=data,
            xticklabels from table={\dataAX}{Label},
            ymajorgrids=true,
            grid style=dashed,
            legend style={at={(0.5,-0.2)}, anchor=north, legend columns=-1},
            enlarge x limits=0.12,
    ]
        \addplot [fill=col1, draw=black!80] table [y=Shorter, x expr=\coordindex] {\dataAX};
        \addplot [fill=col3, draw=black!80] table [y=Longer, x expr=\coordindex] {\dataAX};
        \legend{Shorter, Longer}
    \end{axis}
\end{tikzpicture}
        \caption{Total percentage of path deviations (Shorter and Longer) found in ID dataset by Hybrid pathfinder compared to the LCA pathfinder.}
        \label{fig:comp_paths_ID}
 \end{figure}
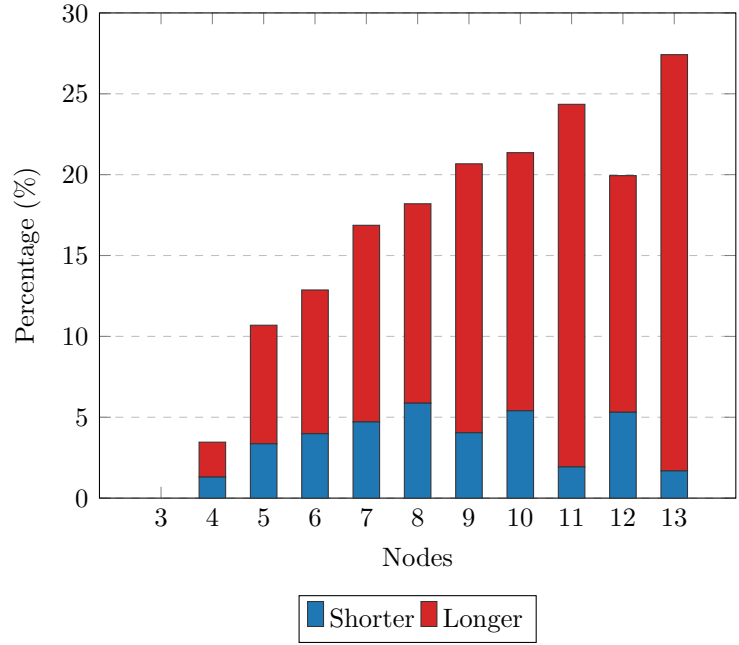

We must determine whether the Hybrid model's advantage over LCA stems from finding shorter paths or simply exploring fewer nodes. It could then be that the better performance (within limits) of the Hybrid pathfinder over the LCA that we are seeing in Figure \ref{fig:exp+avgEER_ID} is only due to the pathfinder being able to find shortcuts instead of finding the path to the common ancestor. For this reason, in Figure \ref{fig:comp_paths_ID}, we computed the percentage of times that the Hybrid pathfinder found a shorter or longer path than the LCA pathfinder.\footnote{By the length of a path, we mean the number of mutations connecting the pair of theories.} On average, in the ID dataset, the Hybrid pathfinder found a shorter path only $\sim 3.4\%$ of the time and found a longer path (due to the confusion of the NN policies) $\sim 12.6\%$ of the time. This means that $\sim 84\%$ of the time, the Hybrid pathfinder found a path of the same length as the LCA pathfinder, and its speedup is purely due to the exploration of fewer nodes during each step, as shown in Figure \ref{sfig:expsaving_ID}. Therefore, despite the different logic of guidance between the pathfinders, the Hybrid pathfinder behaves similarly to LCA $84\%$ of the time, while it is also able to find shortcuts when possible.\footnote{The presence of longer paths can be due to symmetries of the quivers that allow for multiple paths connecting the pair or the non-perfect guidance of the NN policies. Nevertheless, it is interesting to note that despite the path being longer, the number of nodes explored by the Hybrid pathfinder is generally smaller than those explored by the LCA pathfinder.} 

\subsubsection{Hybrid LCA Pathfinder Performance}
\label{sec:hyb_det_performance}

\begin{figure}[!htp]
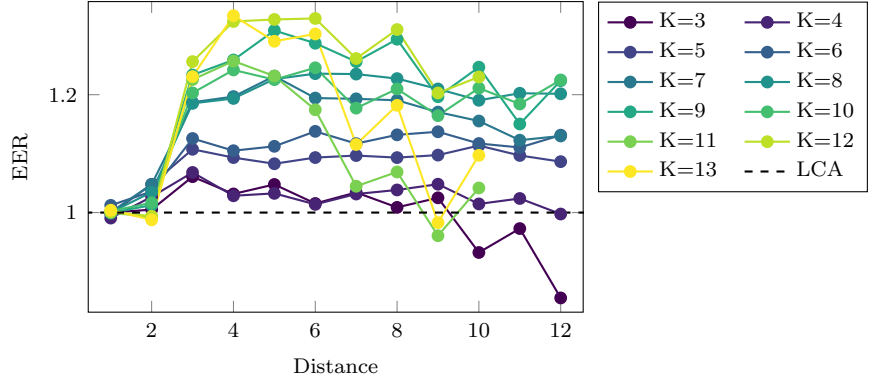

    \centering

    \caption{EER for the Hybrid LCA pathfinder on the ID dataset}
    \label{fig:eer_on_dataset_ID}
\end{figure}

The purpose of the Hybrid LCA pathfinder is to find a balance between the physics-informed policy defined by the LCA and the NN-guided policies of the Hybrid pathfinder. In this way, similarly to when we discussed the Hybrid pathfinder, we can try to improve the performance of the pathfinders: when the Hybrid pathfinder is struggling, i.e., at larger distances or smaller quivers, the LCA pathfinder can dominate the contribution to the heuristic. On the other hand, when the LCA pathfinder explores too many nodes in order to find the common ancestor, the NN can select a shorter and better path to improve overall efficiency. Figure \ref{fig:eer_on_dataset_ID} shows that the Hybrid LCA pathfinder improves upon, with the parameters discussed in the previous section, is an improvement compared to the corresponding performance of the Hybrid (DGNN--AGNN only) pathfinder, which we can see in \cref{sfig:eer_hybrid_theories,sfig:eer_hybrid_qtheories,sfig:eer_hybrid_newtheories}. 

\begin{figure}[!htp]
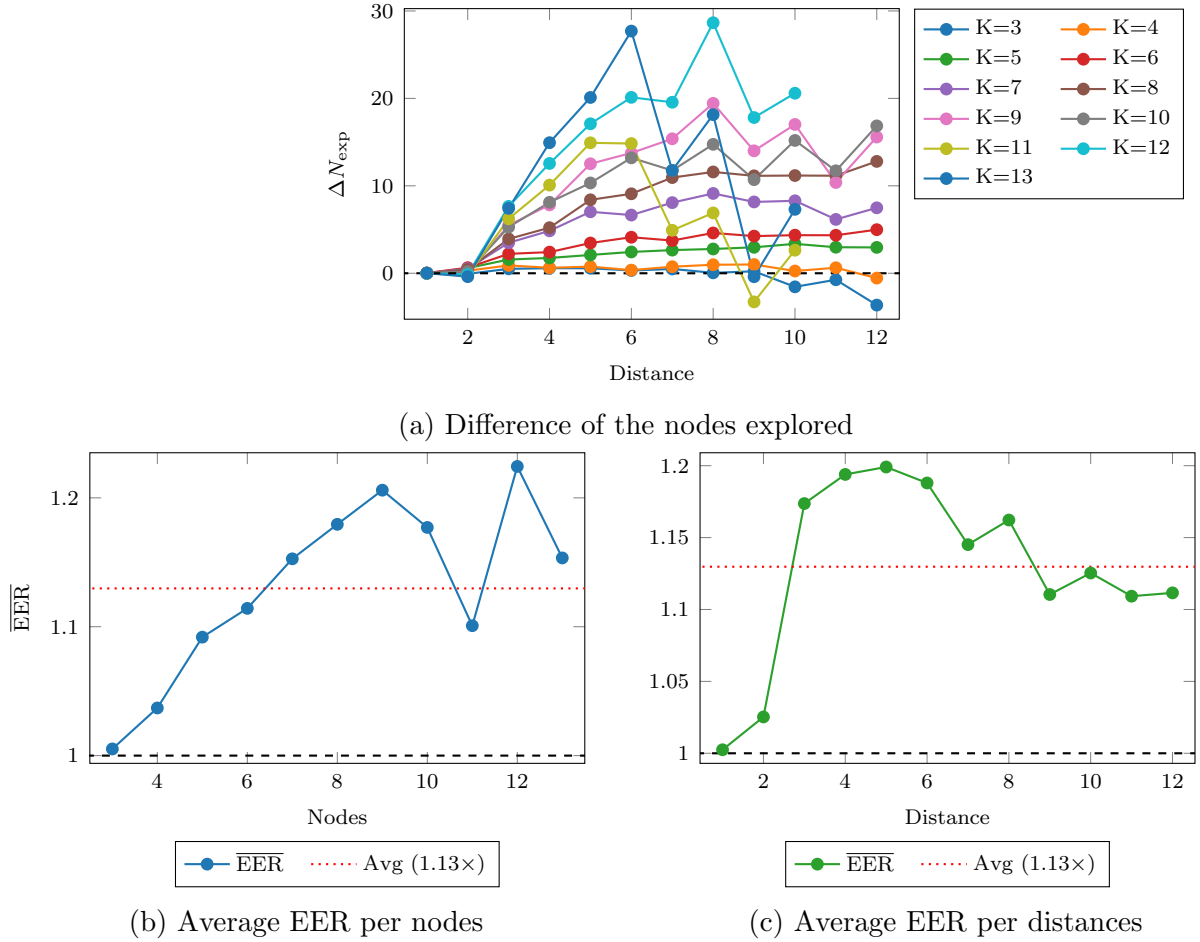

    \centering
      \begin{subfigure}[b]{0.49\textwidth}
\definecolor{col1F77B4}{RGB}{31,119,180}
\definecolor{colFF7F0E}{RGB}{255,127,14}
\definecolor{col2CA02C}{RGB}{44,160,44}
\definecolor{colD62728}{RGB}{214,39,40}
\definecolor{col9467BD}{RGB}{148,103,189}
\definecolor{col8C564B}{RGB}{140,86,75}
\definecolor{colE377C2}{RGB}{227,119,194}
\definecolor{col7F7F7F}{RGB}{127,127,127}
\definecolor{colBCBD22}{RGB}{188,189,34}
\definecolor{col17BECF}{RGB}{23,190,207}
\definecolor{col000000}{RGB}{0,0,0}

        \caption{Difference of the nodes explored}
        \label{sfig:hyb_det_expsaving_ID}
          \end{subfigure}
          
  \begin{subfigure}[b]{0.49\textwidth}
  \definecolor{col1F77B4}{RGB}{31,119,180}
\definecolor{col000000}{RGB}{0,0,0}
\definecolor{colFF0000}{RGB}{255,0,0}
%
        \caption{Average EER per nodes}
        \label{sfig:hyb_det_avgEER_nodes_ID}
          \end{subfigure}\hfill 
\begin{subfigure}[b]{0.49\textwidth}
            \definecolor{col2CA02C}{RGB}{44,160,44}
\definecolor{col000000}{RGB}{0,0,0}
\definecolor{colFF0000}{RGB}{255,0,0}
%
        \caption{Average EER per distances}
        \label{sfig:hyb_det_avgEER_distance_ID}
          \end{subfigure}
        \caption{Nodes explored and average EER per nodes and distances of the Hybrid LCA pathfinder compared with the LCA pathfinder in the ID dataset}
        \label{fig:hyb_det_exp+avgEER_ID}
\end{figure}

By inspection of Figure \ref{fig:hyb_det_exp+avgEER_ID}, the curves now show a positive trend in average EER for both nodes and distances, surpassing the pure LCA pathfinder as soon as the pairs of quivers are large and sufficiently separated. Despite the average EER being only slightly improved compared to what is shown in Figure \ref{fig:exp+avgEER_ID}, the trends seem to have a slower decay, guaranteeing that it is better than pure LCA for longer distances than the Hybrid pathfinder was.

\begin{figure}[!htp]
     \centering
     \pgfplotstableread{
Label Shorter Equal Longer
3 0.00 100.00 0.00
4 0.93 95.91 3.16
5 2.23 88.26 9.51
6 2.84 86.31 10.85
7 3.30 82.63 14.06
8 4.20 82.09 13.70
9 2.63 79.14 18.23
10 3.52 79.77 16.71
11 1.15 76.00 22.85
12 3.90 83.87 12.23
13 1.19 76.78 22.03
}\dataAX

\begin{tikzpicture}[baseline=0,font=\footnotesize]
    \definecolor{col1}{HTML}{1F77B4}
    \definecolor{col3}{HTML}{D62728}
    \begin{axis}[
            ybar stacked,
            bar width=10pt,
            width=10cm, 
            height=8cm,
            xlabel={Nodes},
            ylabel={Percentage (\%)},
            ymin=0, ymax=30,
            xtick=data,
            xticklabels from table={\dataAX}{Label},
            ymajorgrids=true,
            grid style=dashed,
            legend style={at={(0.5,-0.2)}, anchor=north, legend columns=-1},
            enlarge x limits=0.12,
    ]
        \addplot [fill=col1, draw=black!80] table [y=Shorter, x expr=\coordindex] {\dataAX};
        \addplot [fill=col3, draw=black!80] table [y=Longer, x expr=\coordindex] {\dataAX};
        \legend{Shorter, Longer}
    \end{axis}
\end{tikzpicture}
        \caption{Comparison of paths found in ID dataset by Hybrid LCA pathfinder when compared to the LCA pathfinder.}
        \label{fig:hyb_det_comp_paths_ID}
 \end{figure}
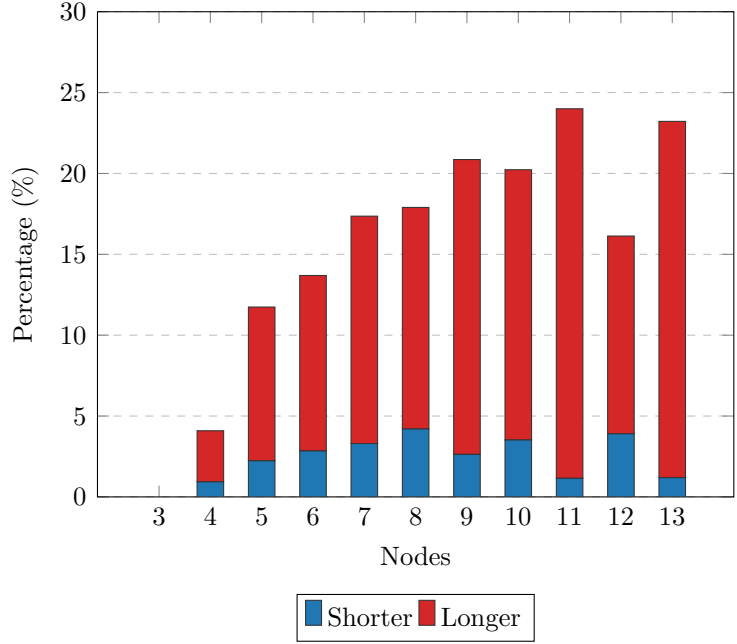

Finally, testing reveals that there is no significant competition between the Hybrid and LCA policies in the search for a path, leading to the preference for finding a path over finding the shortest path. We see in Figure \ref{fig:hyb_det_comp_paths_ID} that there is no significant increase in longer paths over the pure Hybrid pathfinder statistics shown in Figure \ref{fig:comp_paths_ID}. We believe that this is one of the signals that the optimization of the parameters has been successful and that the two policies are assisting each other in finding the path (rather than interfering with each other). 

 \subsection{Out-of-Distribution Theories Analysis}
 \label{sec:Performance-OODTh}

To probe the generalization capabilities of our NNs, we test them on the out-of-distribution (OOD) quiver families detailed in Appendix \ref{sec:OODTheories}, comparing our neural pathfinders directly against the BFS and LCA baselines. This OOD dataset encompasses both finite- and infinite-mutation type quivers, i.e., quivers with finite or infinitely many mutants, as well as quivers that violate gauge anomaly cancellation conditions. Although such anomalous configurations lack a physical realization as 4D $\mathcal{N}=1$ supersymmetric gauge theories, mutations are fundamentally well-defined combinatorial operations on directed graphs. Consequently, these ``unphysical'' examples provide a rigorous mathematical stress-test to evaluate the topological robustness of our pathfinder algorithms beyond familiar physical constraints.\footnote{Although these anomalous theories are inconsistent when interpreted as 4D gauge theories, they are consistent as quiver quantum mechanics theory, and appear, for example, as the BPS quivers for various 4D $\mathcal{N} = 2$ QFTs. In that context, mutations correspond to moving from one BPS chamber to another (see e.g., \cite{Alim:2011ae}).}

\subsubsection{Success Rates on OOD Theories}

\begin{figure}[!htp]
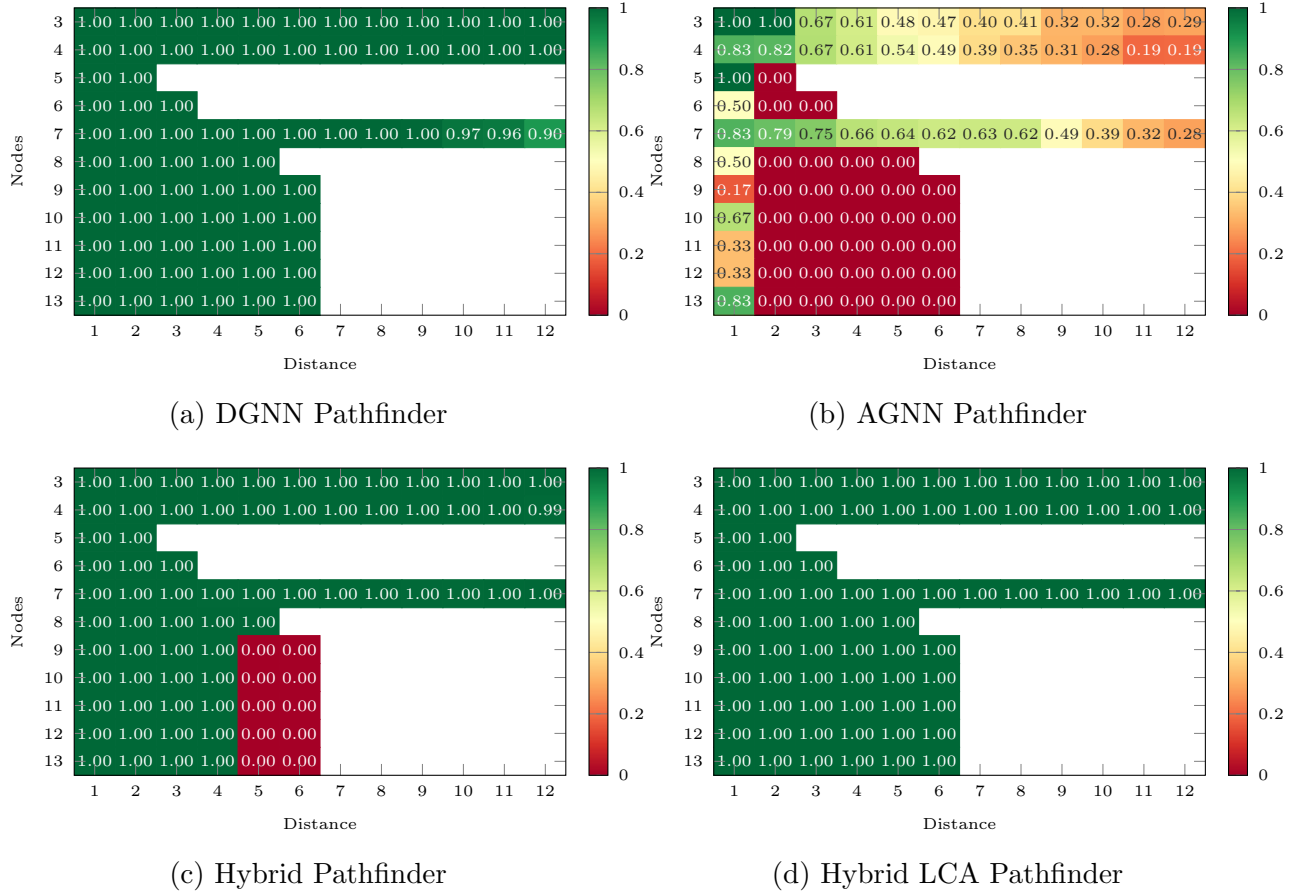

    \centering
    \begin{subfigure}[b]{0.49\textwidth}

\caption{DGNN Pathfinder}
        \label{sfig:SH_siamese_OOD_Q}
    \end{subfigure}\hfill
    \begin{subfigure}[b]{0.49\textwidth}
  %
\caption{AGNN Pathfinder}
        \label{sfig:SH_ar_OOD_Q}
    \end{subfigure}

    \vspace{0.2cm}
    \begin{subfigure}[b]{0.49\textwidth}
 %
\caption{Hybrid Pathfinder}
        \label{sfig:SH_hybrid_OOD_Q}
    \end{subfigure}\hfill
    \begin{subfigure}[b]{0.49\textwidth}
  %
\caption{Hybrid LCA Pathfinder}
        \label{sfig:SH_hybrid_det_OOD_Q}
    \end{subfigure}
    \caption{Success Heatmap for Theories in Figure \ref{fig:Baoquivers} (OOD dataset). The blank white regions are a consequence of these 
    theories being of finite mutation type, i.e., there is an upper bound on the distance between any pair of theories.}
    \label{fig:SH_global_OOD_Bao}
\end{figure}

We now analyze the performance of the various NN-guided pathfinders when tested on OOD theories. The heatmaps for the SRs for the theories in Figure \ref{fig:Baoquivers} are shown in Figure \ref{fig:SH_global_OOD_Bao}. We see that the DGNN pathfinder has an SR over $97\%$ in Figure \ref{sfig:SH_siamese_OOD_Q}, proving the ability of DGNN to guide the pathfinder regardless of whether that theory was part of the training set. This, however, does not apply to the AGNN pathfinder, which once again struggles to keep the SR up for large distances. For special sorts of OOD theories, we see in Figure \ref{sfig:SH_ar_OOD_Q} that the AGNN pathfinder fails completely due to the special nature of theories such as $\mathbf{Q1}$, $\mathbf{Q2}$, $\mathbf{Q3}$, $\mathbf{Q13}$, and $\mathbf{Q15}$ in Figure \ref{fig:Baoquivers} that either do not admit any non-anomalous realizations or admit only a finite number of mutations. The struggle of the AGNN model to guide the pathfinder is also reflected at the level of the Hybrid pathfinder, which failed to find the path for quivers with a large number of nodes, despite the DGNN pathfinder managing to do so alone. The success rate of $100\%$ is, however, achieved when the Hybrid pathfinder is augmented with the LCA policy, as shown in \cref{sfig:SH_hybrid_det_OOD_Q}.  

\begin{figure}[!htp]
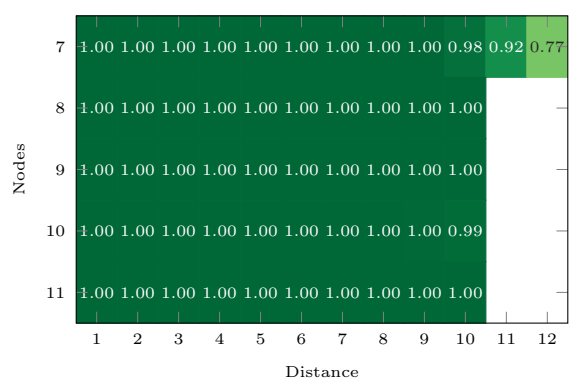
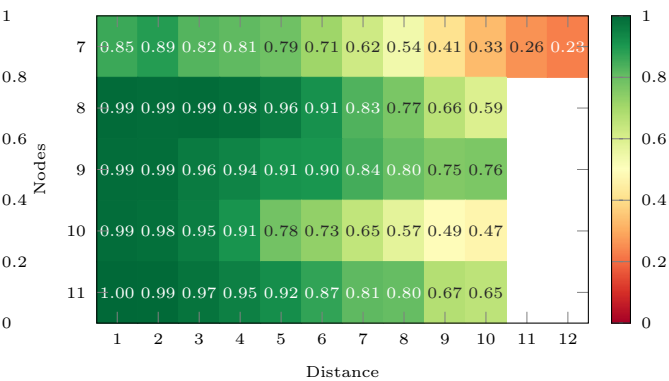
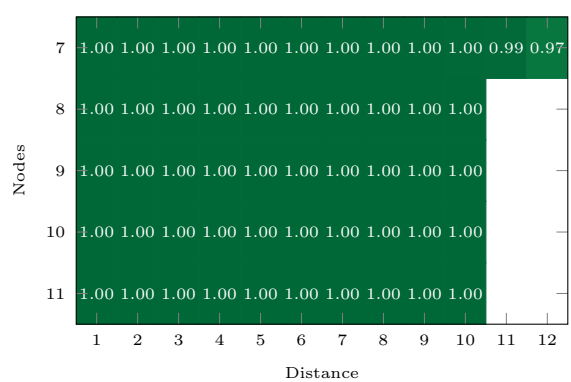
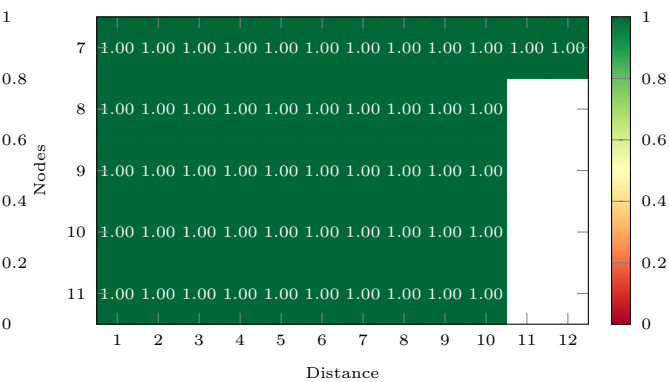

    \centering
    \begin{subfigure}[b]{0.49\textwidth}

\caption{DGNN Pathfinder}
        \label{sfig:SH_siamese_OOD_New}
    \end{subfigure}\hfill
    \begin{subfigure}[b]{0.49\textwidth}
  %
\caption{AGNN Pathfinder}
        \label{sfig:SH_ar_OOD_New}
    \end{subfigure}
    
    \vspace{0.2cm}
    \begin{subfigure}[b]{0.49\textwidth}
 %
\caption{Hybrid Pathfinder}
        \label{sfig:SH_hybrid_OOD_New}
    \end{subfigure}\hfill
    \begin{subfigure}[b]{0.49\textwidth}
  %
\caption{Hybrid LCA Pathfinder}
        \label{sfig:SH_hybrid_det_OOD_New}
    \end{subfigure}
    \caption{Success Heatmap for Theories in Figure \ref{fig:newquivers} (OOD dataset)}
    \label{fig:SH_global_OOD_New}
\end{figure}

The OOD theories in Figure \ref{fig:newquivers} admit infinitely many mutations, so they are closer to the theories considered during the training, although they were never seen by the NNs during the training. However, we see that the pathfinders behave similarly to the analysis done for each pathfinder in Section \ref{sec:Performance-IDTh}: the DGNN pathfinder has a SR of over $90\%$, while the AGNN pathfinder struggles to find a path at large distances, as we see in \cref{sfig:SH_siamese_OOD_New,sfig:SH_ar_OOD_New}. The Hybrid pathfinder has almost perfect SR due to the AGNN policy failing at large distances, but the perfect score is obtained by the Hybric LCA pathfinder, as we show in \cref{sfig:SH_hybrid_OOD_New,sfig:SH_hybrid_det_OOD_New}.

\subsubsection{Efficiency on OOD Theories}

\begin{figure}[!htp]
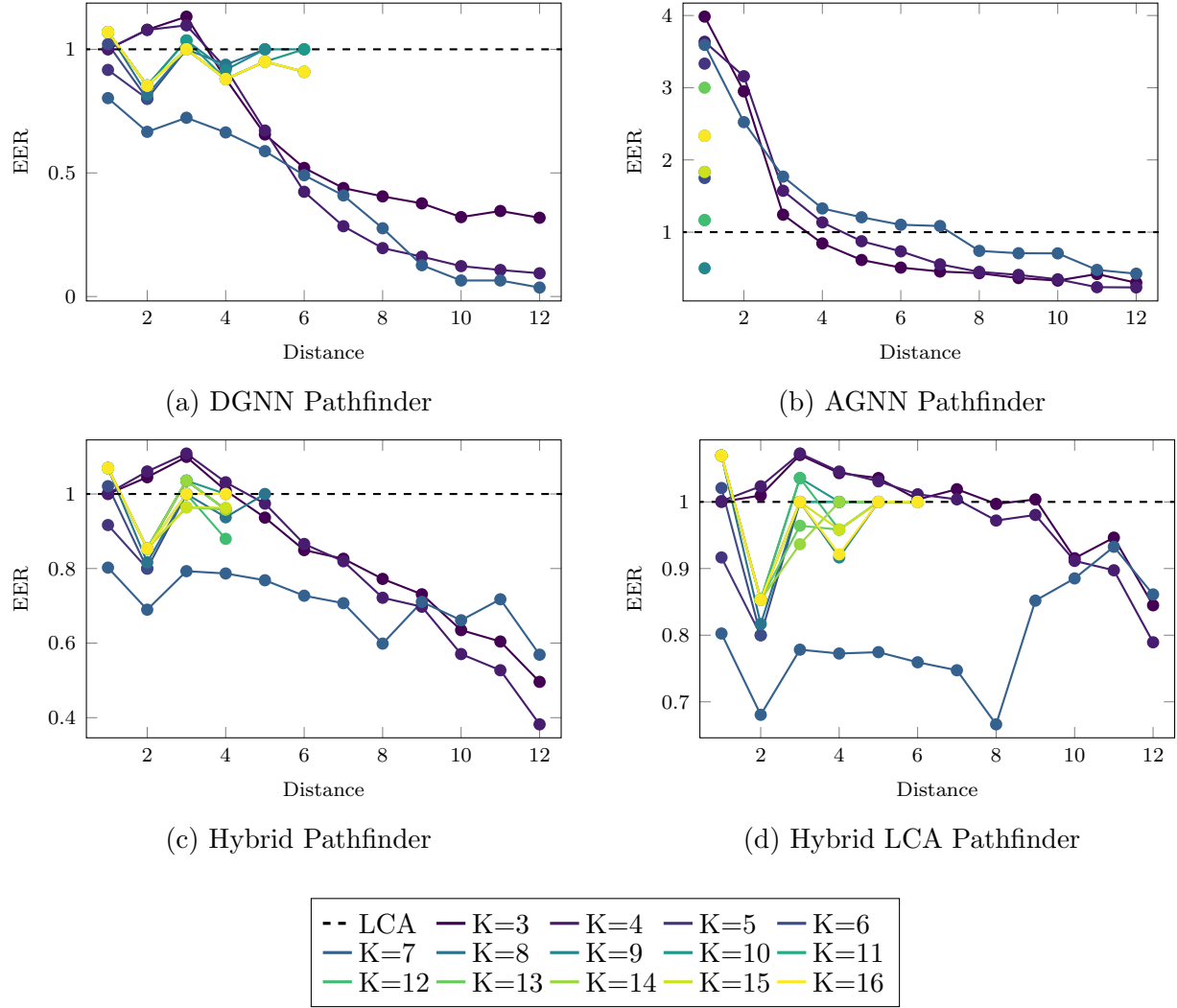

    \centering
    \begin{subfigure}[b]{0.49\textwidth}

\caption{DGNN Pathfinder}
        \label{sfig:eer_siamese_qtheories}
    \end{subfigure}\hfill
    \begin{subfigure}[b]{0.49\textwidth}
        %
\caption{AGNN Pathfinder}
        \label{sfig:eer_ar_qtheories}
    \end{subfigure}

    \vspace{0.2cm}
    \begin{subfigure}[b]{0.49\textwidth}
        %
\caption{Hybrid Pathfinder}
        \label{sfig:eer_hybrid_qtheories}
    \end{subfigure}\hfill
    \begin{subfigure}[b]{0.49\textwidth}
        %
\caption{Hybrid LCA Pathfinder}
        \label{sfig:eer_hybrid_det_qtheories}
    \end{subfigure}
    
    \vspace{0.5cm}
    %
    
    \caption{EER for the pathfinders using LCA baseline in the OOD dataset of Figure \ref{fig:Baoquivers}.}
    \label{fig:eer_global_OOD_Bao}
\end{figure}

The efficiencies of the NN-guided pathfinders on the OOD datasets, when compared to the LCA baseline, show interesting results. We can see in Figure \ref{fig:eer_global_OOD_Bao} that when the number of mutations of the quivers is finite or the theories are anomalous, all pathfinders perform worse than the LCA pathfinder in general ground. However, the worse performance of the Hybrid LCA is around the LCA performance. We believe that this is a sign that if the pairs of theories are not known to admit infinitely many mutations, the use of the Hybrid LCA pathfinder could still be more efficient, since it will only be outperformed by the LCA when the theories are anomalous. Moreover, beating the LCA pathfinder on the datasets we generated, as mentioned already in the previous sections, is the most difficult test for our pathfinders because the LCA pathfinder follows the same criterion (but opposite) that we used to generate the dataset in the first place, as explained in Section \ref{sec:BFS}. Therefore, the datasets, in particular when the number of allowed mutations is finite, are biased towards pairs that share a ``lowest common ancestor", by construction. 

If, instead, we consider theories that admit an infinite number of mutations, such as those in Figure \ref{fig:newquivers}, the NN-guided pathfinders behave similarly to how they performed on the ID datasets. We see in Figure \ref{fig:eer_global_OOD_New} that, in particular, the Hybrid LCA pathfinder has an LCA baseline that is usually below the efficiency of the NN-guided pathfinders. The overall performance is also consistent with the results found in the ID dataset, with a peak efficiency around $1.6\times$ faster than the LCA pathfinder, showing symptoms of degradation for large distances. Overall, however, we can conclude that the NNs guidance is not specific to the theories that were in the dataset; instead, it is based solely on the number of nodes and distances that were observed during the training of the NNs. For this reason, in Section \ref{sec:breakpoint}, we study the limitations of the pathfinders, i.e., where they start to exhibit failure modes, to understand how long we can achieve better performance than the LCA pathfinder given a certain training of the NNs.

\begin{figure}[!htp]
    \centering
    \begin{subfigure}[b]{0.49\textwidth}

\caption{DGNN Pathfinder}
        \label{sfig:eer_siamese_newtheories}
    \end{subfigure}\hfill
    \begin{subfigure}[b]{0.49\textwidth}
        %
\caption{AGNN Pathfinder}
        \label{sfig:eer_ar_newtheories}
    \end{subfigure}
    
    \vspace{0.2cm}
    \begin{subfigure}[b]{0.49\textwidth}
        %
\caption{Hybrid Pathfinder}
        \label{sfig:eer_hybrid_newtheories}
    \end{subfigure}\hfill
    \begin{subfigure}[b]{0.49\textwidth}
        %
\caption{Hybrid LCA Pathfinder}
        \label{sfig:eer_hybrid_det_newtheories}
    \end{subfigure}
    
    \vspace{0.5cm}
    %
    
    \caption{EER for the pathfinders using LCA baseline in the OOD dataset of Figure \ref{fig:newquivers}.}
    \label{fig:eer_global_OOD_New}
\end{figure}

\section{Failure Modes}
\label{sec:breakpoint}

So far, we have tested the performance of the different pathfinders against each other, analyzing the pros and cons of each strategy and trying to maximize efficiency while maintaining the highest possible success rate. In general, we see that all pathfinders, even those that are NN-guided, do not distinguish much between ID and OOD theories, maintaining similar efficiencies when compared to a BFS or LCA baseline. From the discussion in Section \ref{sec:RESULTS}, the Hybrid LCA pathfinder appears to be the most consistent, outperforming pure LCA every time when the theories admit an infinite number of mutations. However, even for these theories, the EER starts to decrease when small quivers are very far apart in terms of the number of mutations. The purpose of this section is to understand what the common failure modes and limitations (i.e., ``the breaking point'') of the pathfinders we propose are, given the training of the NNs over theories with at most $K_{\text{train}}$ nodes and at a distance $D_{\text{train}}$ from each other. In particular, we want to find the turning point when the efficiency ratio decreases and the maximum complexity after which the pathfinders degrade worse than either BFS or LCA baselines.  

For this analysis, we considered only theories with an infinite number of mutations, as found in the ID dataset and Figure \ref{fig:newquivers} of the OOD dataset. We introduce two new parameters:
\begin{equation}
    C = d_\text{db} \log_{10} K \coma \text{W} = \frac{N_\text{exp}}{d_\text{db}+1} \quad \left(\text{EW} = \frac{N_\text{exp}}{d_\text{db}+1}\frac{1}{\text{SR}}\right)\fstop
\end{equation}
The first is the logarithm of the complexity $K^{d_\text{db}}$ for a pair of quivers with $K$ nodes at a distance $d_\text{db}$ from each other, as saved from the database using BFS. The distance $d_\text{db}$ is not the distance found by the pathfinders; rather, it is the reference distance we save when generating the database, as explained in Section \ref{sec:BFS}. The parameter W (resp. EW) is the ``wandering" (resp. ``effective wandering"), which measures how many nodes the model actually explored compared to the minimum required if it made zero mistakes.

Throughout this work, and with the checkpoints available on the \href{https://github.com/alexmininno/GNN-Pathfinders}{GitHub repository}, the NNs have been trained up to distance $12$ for theories in the ID dataset with up to $10$ nodes, and up to distance $10$ for theories in the ID dataset with up to $13$ nodes. The complexities associated with these extremal values of training are
\begin{equation}
    C_1 = 12 \log_{10} 10 = 12 \coma C_2 = 10 \log_{10} 13 = 11.1394\fstop
\end{equation}
Therefore, despite being very close, we consider $C_1 = C_\text{train}$ as the reference complexity associated to the pathfinders for the moment. We can then plot in Figure \ref{fig:orig_complexity-plot} how the SR, EER (with respect to the LCA baseline), and EW differ across complexity.

\begin{figure}[!htp]
    \centering
    
    \begin{subfigure}[b]{0.49\textwidth}
        \centering

        \caption{Success Rate}
        \label{sfig:orig_complexity-SR}
    \end{subfigure}\hfill
    \begin{subfigure}[b]{0.49\textwidth}
        \centering
        %
        \caption{Effective Efficiency Ratio}
        \label{sfig:orig_complexity-EER}
    \end{subfigure}
    
    \vspace{0.5cm}

    \begin{subfigure}[b]{0.49\textwidth}
        \centering
        %

        \caption{Effective Wandering}
        \label{sfig:orig_complexity-EW}
    \end{subfigure}\hfill
    \begin{subfigure}[b]{0.49\textwidth}
        \centering
        %

        \caption{Effective Wandering without DGNN}
        \label{sfig:orig_complexity-EW_noS}
    \end{subfigure}
    
    \vspace{0.5cm}

    %
    
    \caption{Success Rate, Effective Efficiency Ratio (with respect to LCA baseline) and Effective Wandering for theories when considering NNs trained up to distance $12$ and $13$ nodes.}
    \label{fig:orig_complexity-plot}
\end{figure}

Interestingly, we see in Figure \ref{sfig:orig_complexity-SR} that the model, which started degrading even at small complexities, is the AGNN pathfinder. On the other hand, Figure \ref{sfig:orig_complexity-EER} presents it as the best model. This is consistent with the discussion that the efficiency of the AGNN is unmatched, but the lack of backtracking makes it unreliable for large complexities. However, the most interesting plot is Figure \ref{sfig:orig_complexity-EW}, where we see that the pathfinder that ``wandered" more is actually the DGNN, while all other models remained consistent, as shown in Figure \ref{sfig:orig_complexity-EW_noS}.\footnote{By definition, W and EW are ``better'' the closer they are to $1$.} In fact, the EER and EW of all models except for DGNN are better than the LCA baseline. The models show no signs of degradation even at the largest complexities used for training, i.e., $C_\text{train}$. 

\begin{figure}[!htp]
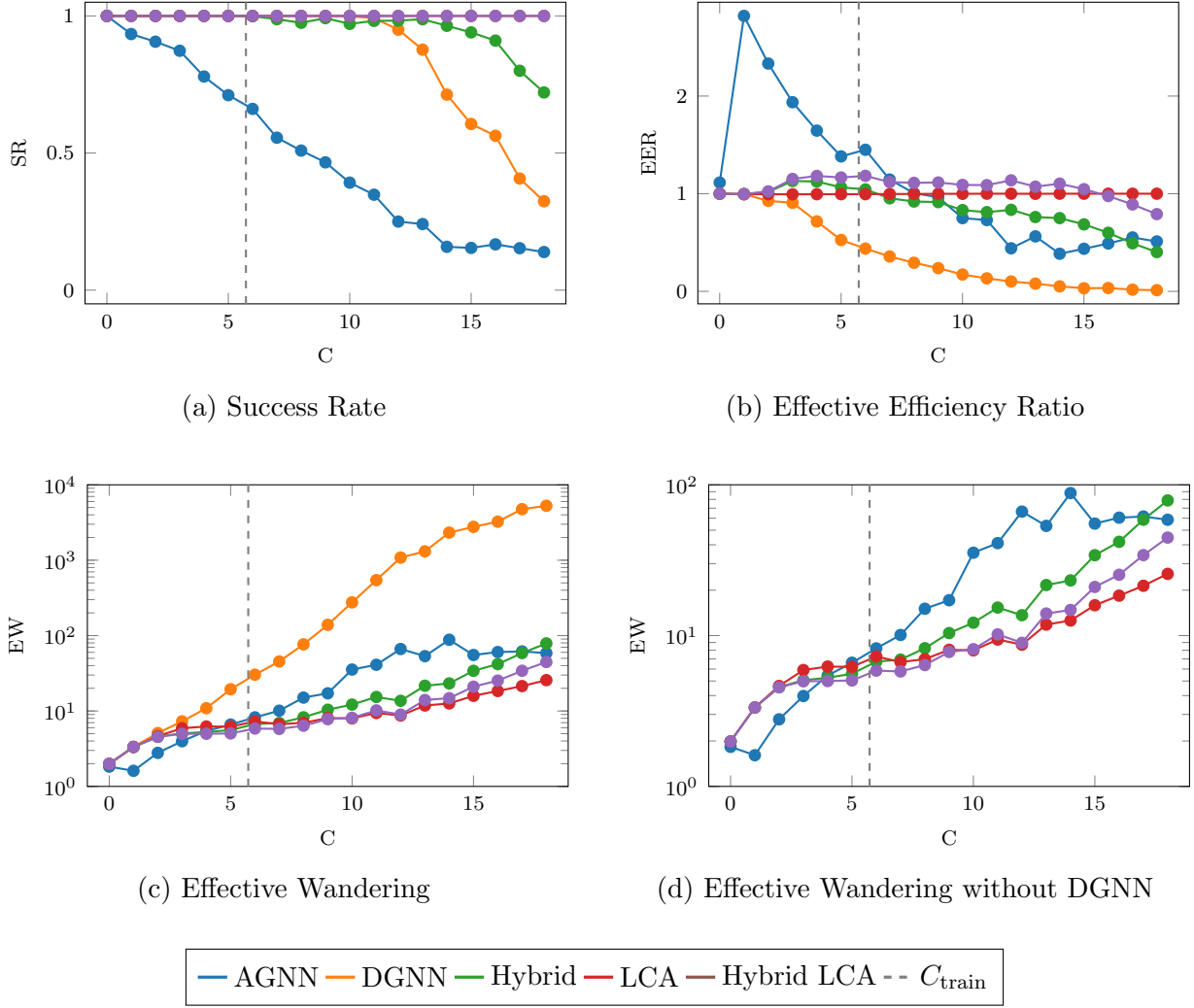

    \centering
    
    \begin{subfigure}[b]{0.49\textwidth}
        \centering

        \caption{Success Rate}
        \label{sfig:new_complexity-SR}
    \end{subfigure}\hfill
    \begin{subfigure}[b]{0.49\textwidth}
        \centering
        %
        \caption{Effective Efficiency Ratio}
        \label{sfig:new_complexity-EER}
    \end{subfigure}
    
    \vspace{0.5cm}

    \begin{subfigure}[b]{0.49\textwidth}
        \centering
        %
        \caption{Effective Wandering}
        \label{sfig:new_complexity-EW}
    \end{subfigure}\hfill
    \begin{subfigure}[b]{0.49\textwidth}
        \centering
        %
        \caption{Effective Wandering without DGNN}
        \label{sfig:new_complexity-EW_noS}
    \end{subfigure}
    
    \vspace{0.5cm}

    %
    \caption{Success Rate, Effective Efficiency Ratio (with respect to LCA baseline) and Effective Wandering for theories when considering the NNs trained up to distance $6$ on theories with up to $K=9$.}
    \label{fig:new_complexity-plot}
\end{figure}

In order to find the breaking point at which the NNs' guidance causes the pathfinder to degrade and become worse than an unguided LCA, we decided to artificially generate a situation where we could push the pathfinders to large complexities while training the NN on smaller ones. We therefore re-trained the NNs to a maximum complexity $C_\text{train} = 6 \log_{10} 9 \simeq 5.73$ and tested the pathfinders up to a complexity $C = 18.34$, obtained by considering theories in the ID dataset with $K=14$ at distance $16$. The results are shown in Figure \ref{fig:new_complexity-plot}.

This time, we clearly see that the pathfinders start to degrade when the $C_\text{train}$ threshold is passed. In fact, by focusing on the EER and EW for each model, we can predict the complexity value, $C_\text{pred}$, after which the LCA pathfinder will perform better than the NN-guided pathfinders:
\begin{equation}
\renewcommand{\arraystretch}{1.2}
\begin{tabular}{l|c|c|c|c|c|c}
 & \multicolumn{2}{c|}{{$C_\text{peak}$ ($\text{SR}_\text{peak}$)}} & \multicolumn{2}{c|}{${C_\text{pred}}$ ($\text{SR}_\text{pred}$)} & \multicolumn{2}{c}{{$C_\text{pred}/C_{\text{train}}$}} \\
\cline{2-7}
 & {EER} & {EW} & {EER} & {EW} & {EER} & {EW} \\
\hhline{=|=|=|=|=|=|=}
DGNN & 3.24 (1.00) & 3.24 (1.00) & 3.32 (1.00) & 3.34 (1.00) & 0.58 & 0.58 \\
AGNN & 1.11 (1.00) & 1.11 (1.00) & 8.14 (0.50) & 4.70 (0.73) & 1.42 & 0.82 \\
Hybrid & 5.57 (1.00) & 4.46 (1.00) & 6.48 (0.99) & 6.69 (0.99) & 1.13 & 1.17 \\
Hybrid LCA & 14.31 (1.00) & 5.57 (1.00) & 15.66 (1.00) & 9.67 (1.00) & 2.73 & 1.69 \\
\end{tabular}
\end{equation}
We see that the DGNN pathfinder is the one degrading even before reaching $C_\text{train}$, while the AGNN pathfinder appears to be the most efficient for larger complexities, but the predicted SR is rapidly dropping to zero. The most interesting pathfinders are the Hybrid and Hybrid LCA pathfinders, which perform better than LCA for complexities $1.13$ to $2.73\times$ larger than the training complexity, respectively. Also, note that the breaking point of the Hybrid and Hybrid LCA occurs before their success rates drop below $100\%$, meaning that they will be reliable for the entire duration they perform better than LCA. We believe this to be a good estimate of how long the NN-guided pathfinders can outperform the LCA pathfinder.

\section{Conclusions} \label{sec:CONC}

Dualities between quantum field theories provide important access to deep non-perturbative phenomena. It is therefore natural to ask: given two QFTs, could it be that they are actually dual? If so, what is the sequence of simple duality operations that connects the two theories? In practice, this is a challenging question and is amenable to machine learning techniques. Given the encouraging success of Transformers in tackling complex problems in string theory \cite{Yip:2025hon,Walden:2025cpf,Arnal:2026zyo,Yip:2026jhw} and conformal field theories \cite{Cao:2026nsz}, it is natural to ask whether this architecture can be similarly incorporated into the duality pathfinder we aim to develop. In this paper, we have constructed a dataset of dual QFTs with a seed theory given by a D3-brane probing a toric singularity. We used this dataset to train different graph neural networks.\footnote{For other but related uses of GNNs in physics see, e.g., \cite{Lawrie:2024kiw}.} Combining this with pathfinder algorithms, we have studied the computational complexity of connecting Seiberg dual theories and established that a hybrid approach tends to be both more efficient and more accurate. In the remainder of this section, we discuss some potential avenues for future investigation.

In this work, we have investigated the computational complexity of dualities between quiver gauge theories. From a holographic standpoint, such sequences of dualities prominently appear in warped throats of a string compactification, where motion down the throat triggers a specific sequence of duality moves. It would be interesting to investigate whether this notion of computational complexity matches more precisely with other bulk proposals of complexity (see, e.g., \cite{Brown:2015bva, Carmi:2016wjl, Balasubramanian:2019wgd}).\footnote{It would also be interesting to study various generalizations 
of complexity distance measures of the sort introduced in \cite{Heckman:2026beg}.}

An ambitious aim for the future would be to classify all physically distinct quiver gauge theories. To accomplish this, one would also need to have a sharp understanding of how many quiver gauge theories are, in fact, dual (i.e., to identify the equivalence classes). We anticipate using this work to tackle this classification problem.

We have primarily focused on 4D quiver gauge theories and their Seiberg duals, but we anticipate that the same techniques apply more broadly. It would be interesting to study the computational complexity of dualities in other well-motivated settings, including dualities of string, M-, or F-theory compactifications, as well as quantum field theories in other spacetime dimensions.

At a practical level, it would be interesting to increase both the size of the training dataset and the complexity of the network architectures used to establish Seiberg dual pairs. In particular, studying possible scaling laws in the degree of inference as a function of dataset and network size would be instructive.

Given recent advancements in large language models (LLMs), it is natural to ask whether the custom GNN architecture used here is really necessary. Said differently, can one simply provide two dualized quivers and ask an LLM to find the most efficient and accurate path between these theories? This provides a natural class of benchmark problems for studying the capabilities of state of the art inference models, but at the same time, it is difficult to provide a completely ``fair comparison'' since it can be challenging to properly compare the computational resources of an LLM with the specific (quite modest) computational budget used in the present work.\footnote{Can one truly put a price tag on finding all Seiberg dual quiver gauge theories? We think not.}

\subsubsection*{Acknowledgements}

We thank F. Carta, M. Danese, S. De, V. Jejjala, C. Lawrie, H. Lee, and D.S. Park for helpful comments and discussions. The work of JJH and SNM is supported by DOE (HEP) Award DE-SC0013528, BSF grant 2022100, and a University Research Foundation grant at the University of Pennsylvania. 
AM and GS are supported in part by DOE (HEP) Awards DE-SC0017647 and DE-SC0023719. GS is also supported in part by a WARF Named Professorship, provided by the University of Wisconsin--Madison Office of the Vice Chancellor for Research with funding from the Wisconsin Alumni Research Foundation.
This work was performed in part while JJH, AM, and GS were at the Aspen Center for Physics, which is supported by National Science Foundation grant PHY-2210452. JJH and GS especially thank the organizers of the Aspen workshop ``Theoretical Physics for Artificial Intelligence'' for providing a stimulating atmosphere that directly led to the start of this project. SNM and AM thank the Simons Center for Geometry and Physics for the hospitality during the ``23rd Simons Physics Summer Workshop: Theory, Experiment and the Emerging New Physics". The code produced for this work was written with the help of \texttt{Gemini Pro 3.1}.

\appendix

\section{Seed Theories for Dataset Generation}
\label{sec:basictheories}

In this appendix, we describe in more detail the seed theories used for dataset generation. We focus primarily on D3-branes probing toric CY singularities. See \cite{Franco:2017jeo} and references therein for additional discussion of the construction and study of these theories.

\subsection{In-Distribution Theories}
\label{sec:IDTheories}

We trained our NNs on five families of 4d $\mathcal{N}=1$ SCFTs that arise from D3-branes probing toric CY threefolds:
\begin{enumerate}
    \item The 4d $\mathcal{N}=1$ SCFTs on D3-branes probing $\CC^3/\ZZ^{(1,1,-2)}_n$, with $n$ being odd and orbifold action $(1,1,-2)$ on the $\CC^3$ coordinates. These are orbifolds of  $\mathcal{N}=4$ $\SU(N)$ SYM, usually characterized by three chiral fields $X$,$Y$,$Z$ in the adjoint of $\SU(N)$, with a superpotential term
    \begin{equation}
        W = \tr\left(X[Y,Z]\right)\coma
    \end{equation}
    where the trace is over color indices. The action of the orbifold leads to a theory with $n$ gauge nodes $\SU(N_i)$, with chiral fields in the bifundamental representation of $\SU(N_i)\times \SU(N_j)$. Using the same notation of the parent theory, they can be grouped in 
    \begin{equation}
        X_{i|i-2\mod n}\,,\,Y_{i|i+1\mod n}\,,\,Z_{i|i+1\mod n}\coma
    \end{equation}
    while the superpotential becomes
    \begin{equation}\label{eq:C3ZnSup}
        W_{\CC^3/\ZZ_{n}} = \sum_{i=1}^n \tr\left(X_{i|i-2}Y_{i-2|i-1}Z_{i-1|i}-X_{i|i-2}Z_{i-2|i-1}Y_{i-1|i}\right)\fstop
    \end{equation}
    Figure \ref{fig:C3Zntheories} shows representative $\CC^3/\ZZ_n$ quivers. For NN training, we used odd orbifolds with $3\leq n\leq 13$.

    \begin{figure}[!htp]
    \centering
    \def\rad{1.5cm}
    \def\d{1.8cm}

    \begin{subfigure}[b]{0.3\textwidth}
        \centering

        \caption{$\mathbb{C}^3/\mathbb{Z}_3$}
        \label{sfig:C3Z3}
    \end{subfigure}
    \begin{subfigure}[b]{0.3\textwidth}
        \centering
        %
        \caption{$\mathbb{C}^3/\mathbb{Z}_5$}
        \label{sfig:C3Z5}
    \end{subfigure}
    \begin{subfigure}[b]{0.3\textwidth}
        \centering
        %
        \caption{$\mathbb{C}^3/\mathbb{Z}_7$}
        \label{sfig:C3Z7}
    \end{subfigure}
    \caption{Examples of $\CC^3/\ZZ_n$ quivers theories. The corresponding superpotential can be derived from \eqref{eq:C3ZnSup}.}
    \label{fig:C3Zntheories}
\end{figure}

    \item The 4d $\mathcal{N}=1$ SCFTs on D3-branes probing $C(dP_1)$, where by $C(.)$ we mean the complex cone over the $dP_1$ surface. This theory is characterized by four gauge groups and bifundamental fields
    \begin{equation}
        X_{14}\,,\,Y_{14}\,,\,X_{21}\,,\,X_{24}\,,\,X_{31}\,,\,X_{32}\,,\,Y_{32}\,,\,X_{43}\,,\,Y_{43}\,,\,Z_{43}\coma
    \end{equation}
    subjected to the superpotential
    \begin{equation}\label{eq:dP1Sup}
    \begin{split}
        W_{C(dP_1)} = &\,X_{14} Y_{43} Y_{32} X_{21} - Y_{14} Y_{43} X_{32} X_{21} + X_{24} Z_{43} X_{32} - X_{24} X_{43} Y_{32} \\
        &+ X_{31} Y_{14} X_{43} - X_{31} X_{14} Z_{43}\fstop
    \end{split}
    \end{equation}
    The quiver for this theory is shown in Figure \ref{sfig:dP1}.
    \item The 4d $\mathcal{N}=1$ SCFTs on D3-branes probing $C(dP_2)$. This theory is characterized by five gauge groups and bifundamental fields
    \begin{equation}
X_{15}\,,\,Y_{15}\,,\,X_{21}\,,\,X_{25}\,,\,X_{31}\,,\,X_{32}\,,\,X_{42}\,,\,X_{43}\,,\,X_{53}\,,\,X_{54}\,,\,Y_{54}\coma
    \end{equation}
    subjected to the superpotential
    \begin{equation}\label{eq:dP2Sup}
    \begin{split}
        W_{C(dP_2)} = &\, X_{15} X_{53}X_{31}  - X_{25} X_{53}X_{32}  + X_{25}  X_{54}X_{42} - X_{31} Y_{15} X_{54} X_{43}  \\
        &- X_{15} Y_{54} X_{42} X_{21}   + X_{21} Y_{15} Y_{54} X_{43} X_{32}   \fstop
    \end{split}
    \end{equation}
    The quiver for this theory is shown in Figure \ref{sfig:dP2}.
\item The 4d $\mathcal{N}=1$ SCFTs on D3-branes probing $C(dP_3)$. This theory is characterized by six gauge groups and bifundamental fields
    \begin{equation}
X_{15}\,,\,X_{16}\,,\,X_{21}\,,\,X_{26}\,,\,X_{31}\,,\,X_{32}\,,\,X_{42}\,,\,X_{43}\,,\,X_{53}\,,\,X_{54}\,,\,X_{64}\,,\,X_{65}\coma
    \end{equation}
    subjected to the superpotential
    \begin{equation}\label{eq:dP3Sup}
    \begin{split}
          W_{C(dP_3)} =&\, X_{15}  X_{53}X_{31} - X_{15} X_{54} X_{42}X_{21}   + X_{26} X_{64} X_{42}  - X_{16} X_{64} X_{43} X_{31} \\
          &- X_{26} X_{65} X_{53} X_{32}   + X_{16} X_{65} X_{54}  X_{43} X_{32}  X_{21}  \fstop
    \end{split}
    \end{equation}
    The quiver for this theory is shown in Figure \ref{sfig:dP3}.

    \begin{figure}[!htp]
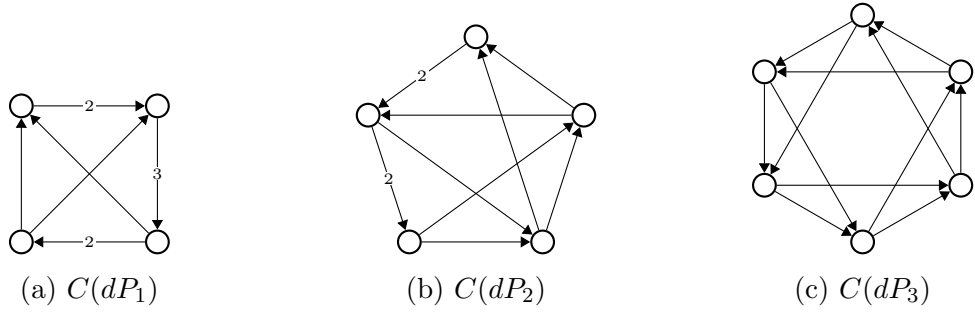

    \centering
    \def\rad{1.5cm}
    \def\d{1.8cm}
    \begin{subfigure}[b]{0.3\textwidth}
        \centering

        \caption{$C(dP_1)$}
        \label{sfig:dP1}
    \end{subfigure}
    \begin{subfigure}[b]{0.3\textwidth}
        \centering
        %
        \caption{$C(dP_2)$}
        \label{sfig:dP2}
    \end{subfigure}
    \begin{subfigure}[b]{0.3\textwidth}
        \centering
        %
        \caption{$C(dP_3)$}
        \label{sfig:dP3}
    \end{subfigure}
    \caption{Quivers of $C(dP_n)$ theories. The corresponding superpotentials are in \cref{eq:dP1Sup,eq:dP2Sup,eq:dP3Sup}.}
    \label{fig:dPntheories}
\end{figure}

    \item The 4d $\mathcal{N}=1$ SCFTs on D3-branes probing the $C(Y^{p,q})$ singularities. This family of theories is characterized by $2p$ gauge groups and bifundamental fields that can be categorized into $\SU(2)$ doublets $U^\alpha$, $V^\alpha$, and singlets $Z$, $Y$. With gauge node indices understood modulo $2p$, they can be grouped as
\begin{equation}
U^\alpha_{2k-1|2k} \quad (k=1 \dots p)\,,\, V^\alpha_{2k|2k+1} \quad (k=1 \dots q)\,,\, \quad Z_{2k|2k+1} \quad (k=q+1 \dots p)\coma
\end{equation}
\begin{equation}
Y_{2k+1|2k-1} \,,\, Y_{2k+2|2k} \quad (k=1 \dots q)\,,\, \quad Y_{2k+2|2k-1} \quad (k=q+1 \dots p)\coma
\end{equation}
where $\alpha=1,2$ is the doublet index. These fields enter to a superpotential as
\begin{equation}\label{eq:YpqSup}
\begin{split}
W_{C(Y^{p,q})} = &\,\sum_{k=1}^{q} \epsilon_{\alpha\beta} \left( U^\alpha_{2k-1|2k} V^\beta_{2k|2k+1} Y_{2k+1|2k-1} - V^\alpha_{2k|2k+1} U^\beta_{2k+1|2k+2} Y_{2k+2|2k} \right) \\
& + \sum_{k=q+1}^{p} (-1)^{k-q-1} \epsilon_{\alpha\beta}  U^\alpha_{2k-1|2k} Z_{2k|2k+1} U^\beta_{2k+1|2k+2} Y_{2k+2|2k-1} \fstop
\end{split}
\end{equation}
Some examples of $C(Y^{p,q})$ quivers we considered in our database are shown in Figure \ref{fig:Ypqtheories}. There are some known identities for the $C(Y^{p,q})$ theories, such as $C(Y^{2,1})\simeq C(dP_1)$, or $C(Y^{p,0})\simeq \mathcal{C}/\ZZ_p$, where $\mathcal{C}$ is the conifold theory. Moreover, $C(Y^{p,p})\simeq \CC^3/\ZZ_{2p}$. Since we want to avoid 2-loops in the quiver, we exclude the latter theories, and $C(Y^{1,0})\simeq \mathcal{C}$. For the training of the NNs we considered $2\leq p\leq 6$ and $1\leq q\leq 6$ with $q\neq p$, so that the total number of nodes in the quivers was up to $12$. 
\end{enumerate}

\begin{figure}[!htp]
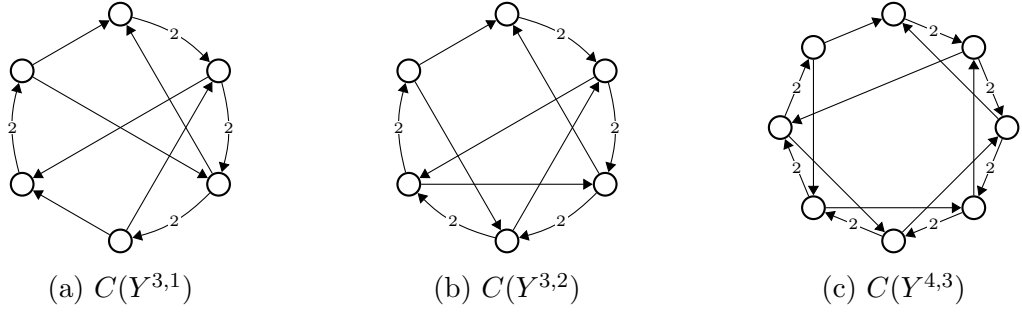

    \centering
    \def\rad{1.5cm}
    \def\d{1.8cm}

    \begin{subfigure}[b]{0.3\textwidth}
        \centering

        \caption{$C(Y^{3,1})$}
        \label{sfig:Y31}
    \end{subfigure}
    \begin{subfigure}[b]{0.3\textwidth}
        \centering
        %
        \caption{$C(Y^{3,2})$}
        \label{sfig:Y32}
    \end{subfigure}
    \begin{subfigure}[b]{0.3\textwidth}
        \centering
        %
        \caption{$C(Y^{4,3})$}
        \label{sfig:Y43}
    \end{subfigure}
    \caption{Examples of $C(Y^{p,q})$ quivers theories. The corresponding superpotential can be derived from \eqref{eq:YpqSup}.}
    \label{fig:Ypqtheories}
\end{figure}

\subsection{Out-of-Distribution Theories}
\label{sec:OODTheories}

\begin{figure}[!htp]
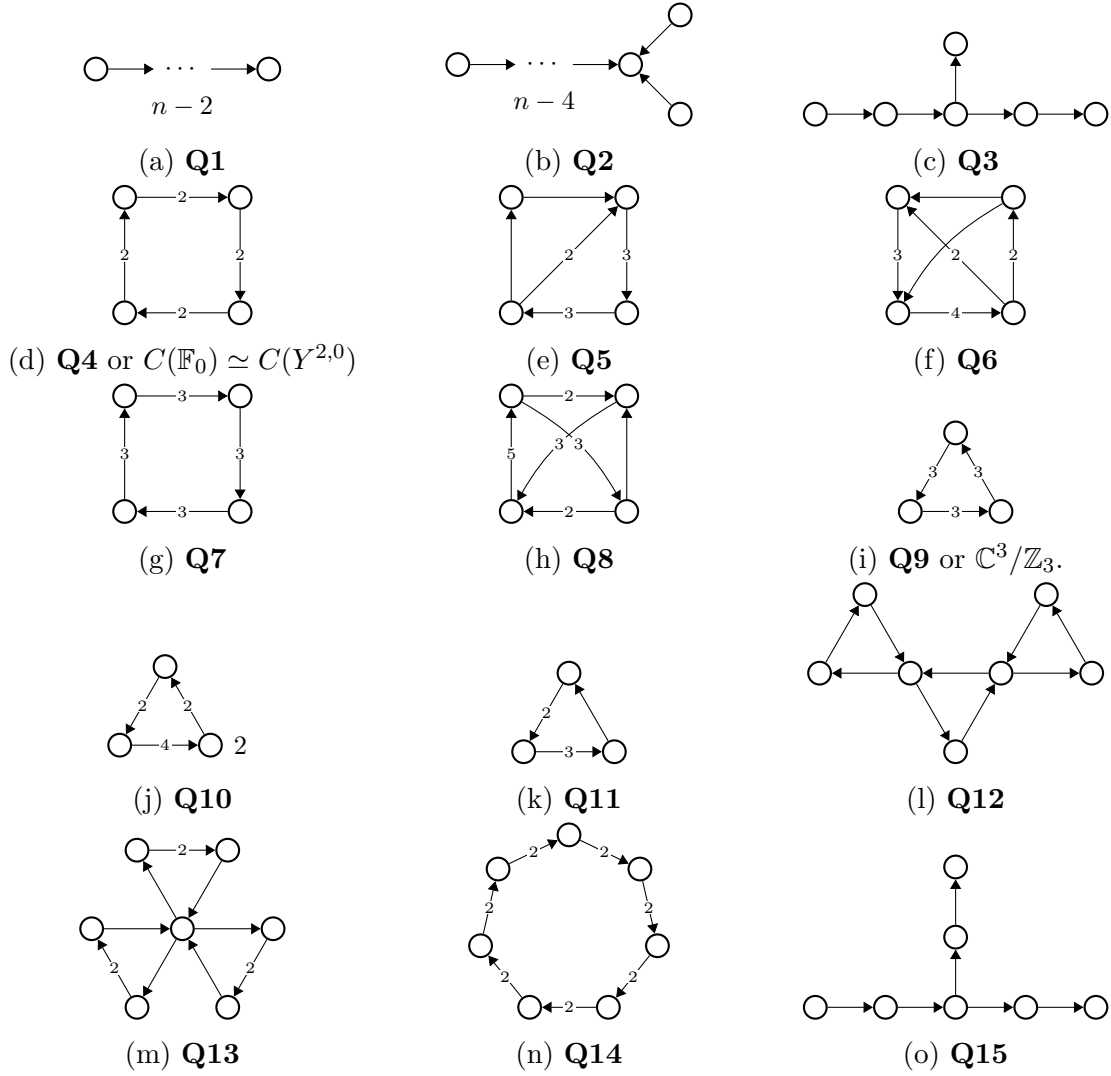

    \centering
    \begin{subfigure}[b]{0.3\textwidth}
        \centering

        \caption{$\mathbf{Q1}$}
        \label{sfig:Q1}
    \end{subfigure}
    \begin{subfigure}[b]{0.3\textwidth}
        \centering
        %
        \caption{$\mathbf{Q2}$}
        \label{sfig:Q2}
    \end{subfigure}
    \begin{subfigure}[b]{0.3\textwidth}
        \centering
        %
        \caption{$\mathbf{Q3}$}
        \label{sfig:Q3}
    \end{subfigure}
    \begin{subfigure}[b]{0.3\textwidth}
        \centering
        %
        \caption{$\mathbf{Q4}$ or $C(\mathbb{F}_0)\simeq C(Y^{2,0})$}
        \label{sfig:Q4}
    \end{subfigure}
    \begin{subfigure}[b]{0.3\textwidth}
        \centering
        %
        \caption{$\mathbf{Q9}$ or $\CC^3/\ZZ_3$.}
        \label{sfig:Q9}
    \end{subfigure}
    \begin{subfigure}[b]{0.3\textwidth}
        \centering
        %
        \caption{$\mathbf{Q15}$}
        \label{sfig:Q15}
    \end{subfigure}
    \caption{Quiver families considered in \cite{Bao:2020nbi}. We comment that some of these quivers are intrinsically anomalous when inerpreted as 4D $\mathcal{N} = 1$ QFTs (i.e., inconsistent), but are perfectly consistent when viewed as the BPS quivers of 4D $\mathcal{N} = 2$ QFTs. As such, they provide an interesting testing ground for studying mutations of quivers.}
    \label{fig:Baoquivers}
\end{figure}

\begin{figure}[!htp]
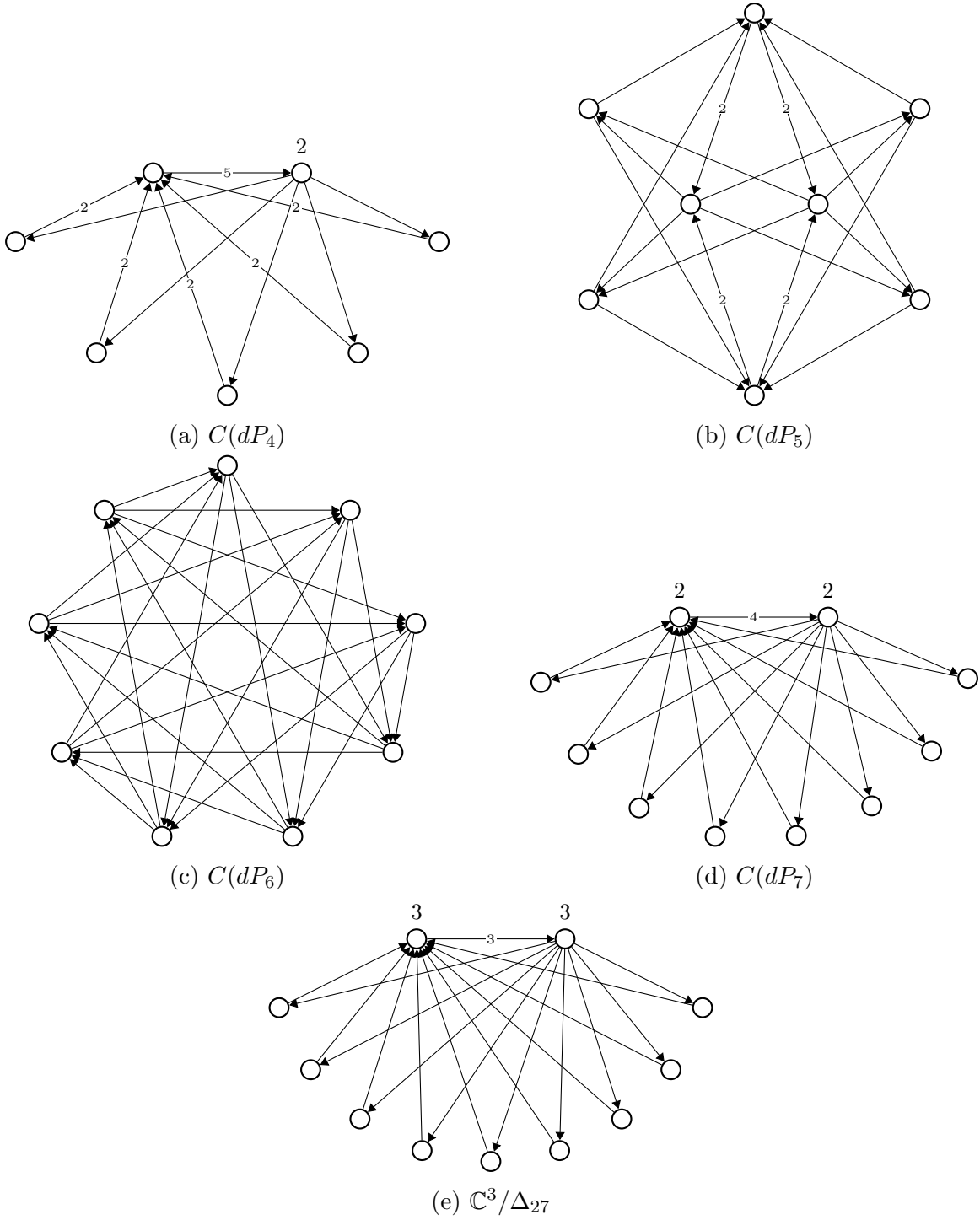

    \centering
    \def\x{20mm} 
    \def\d{1.8cm} 
    \def\r{3} 

    \begin{subfigure}[b]{0.49\textwidth}
        \centering

        \caption{$C(dP_4)$}
        \label{sfig:CdP4}
    \end{subfigure}
    \begin{subfigure}[b]{0.49\textwidth}
        \centering
       %
        \caption{$C(dP_5)$}
        \label{sfig:CdP5}
    \end{subfigure}
    \begin{subfigure}[b]{0.49\textwidth}
        \centering
        %
        \caption{$C(dP_6)$}
        \label{sfig:CdP6}
    \end{subfigure}
    \begin{subfigure}[b]{0.49\textwidth}
        \centering
        %
        \caption{$C(dP_7)$}
        \label{sfig:CdP7}
    \end{subfigure}

  \begin{subfigure}[b]{\textwidth}
        \centering
        %
        \caption{$\CC^3/\Delta_{27}$}
        \label{sfig:C3Delta27}
    \end{subfigure}
    \caption{Further quiver theories used to test the NN. Nodes with a rank other than one are specified whenever it is necessary to make the theory non-anomalous.}
    \label{fig:newquivers}
\end{figure}

To evaluate the neural network's generalization to unseen theories, we tested it on the family of quivers $\mathbf{Q1}$--$\mathbf{Q15}$ from \cite{Bao:2020nbi} (Figure \ref{fig:Baoquivers}), using non-anomalous rank assignments when possible. We also evaluated quivers from D3-branes probing non-toric CY threefolds \cite{Burrington:2007mj,Beaujard:2020sgs}, shown in Figure \ref{fig:newquivers}. 

\section{Overview of Graph Neural Networks}
\label{app:gnn_appendix}

Graph Neural Networks (GNNs) process data natively structured as graphs by assigning features to each node and iteratively updating these representations over successive layers. This refinement is regulated by a localized neighborhood aggregation mechanism known as \textit{message passing} \cite{Gilmer:2017}. In the following, we provide a brief review of this technique.

\subsection{Message Passing Paradigm}
\label{app:MessagePassing}

In a standard Message Passing Neural Network (MPNN) \cite{Gilmer:2017}, the network updates the hidden representation $h_i$ of node $i$ at layer $\ell$ by aggregating messages from its neighbors $j \in \mathcal{N}(i)$. The unactivated update rule commonly takes the form:
\begin{equation}
    m_i^{(\ell+1)} = W_{\text{self}}^{(\ell)} h_i^{(\ell)} + \sum_{j \in \mathcal{N}(i)} W_{\text{msg}}^{(\ell)} h_j^{(\ell)}\coma
\end{equation}
where $W_{\text{self}}$ and $W_{\text{msg}}$ are learnable weight matrices. After aggregating the messages, the network normalizes the features. A frequent choice is Layer Normalization \cite{LeiBa:2016}, which standardizes a vector $m_i \in \mathbb{R}^H$ using its mean $\mu_i$ and variance $\sigma_i^2$, and shifts it with learnable parameters $\gamma, \beta$:
\begin{equation}\label{eq:LayerNorm}
\text{LayerNorm}(m_i) = \gamma \odot \frac{m_i - \mu_i}{\sqrt{\sigma_i^2 + \epsilon}} + \beta\coma
\end{equation}
where $\odot$ is the Hadamard product, i.e., element-wise multiplication.

The normalized vector is then passed through an activation function. Since Layer Normalization can shift feature values to be negative, the Leaky Rectified Linear Unit (LeakyReLU) \cite{Maas:2013} is often used to prevent ``dead" neurons:
\begin{equation}\label{eq:LeakyRelu}
    \text{LeakyReLU}(x) = \begin{cases} 
      x & \text{if } x \geq 0\coma \\
      \alpha x & \text{if } x < 0 \fstop
   \end{cases}
\end{equation}
The parameter $\alpha$ is a small positive constant. Stacking several message passing layers allows a node to gather information from nodes multiple hops away in the graph.

\subsection{Graph Transformers and Global Context}

While standard MPNNs effectively capture local graph topology, their ability to gather long-range information is limited. In particular, the phenomenon of oversmoothing \cite{Wu:2023}, where node representations become indistinguishable after too many layers, remains one of the primary architectural bottlenecks of deep MPNNs. To solve this, architectures like GraphGPS \cite{Rampavek:2022} and Graph Transformers combine local message passing with global attention mechanisms \cite{Vaswani:2017}.

In a Graph Transformer layer, the sequence of node embeddings $X \in \mathbb{R}^{K \times H}$ is projected into Queries ($Q$), Keys ($K$), and Values ($V$) using learnable linear transformations:
\begin{equation}
    Q = XW^Q\coma  K = XW^K\coma V = XW^V\fstop
\end{equation}
The attention weights determine how much focus each node places on every other node in the graph:
\begin{equation}
    \text{Attention}(Q, K, V) = \text{softmax}\left( \frac{QK^T}{\sqrt{d_k}} \right) V\fstop
\end{equation}
To evaluate different types of relationships simultaneously, the network splits the embeddings into multiple independent heads using Multi-Head Attention (MHA). The outputs are concatenated and projected back to the original dimension:
\begin{equation}
    \text{MHA}(X) = \text{Concat}(\text{head}_1, \dots, \text{head}_h) W^O\fstop
\end{equation}
A single Transformer encoder layer typically wraps this operation with position-wise Feed-Forward Networks (FFN) and residual connections:
\begin{equation}
\begin{split}
    \tilde{X} &= \text{LayerNorm}(X + \text{MHA}(X))\coma \\
    X_{\text{out}} &= \text{LayerNorm}(\tilde{X} + \text{FFN}(\tilde{X}))\fstop
    \end{split}
\end{equation}

\section{Overview of Search Algorithms}
\label{app:pathfindersgeneralities}

In this Appendix, we review some generalities about A$^*$ Search and Beam Search pathfinders.

\subsection{\texorpdfstring{A$^*$}{A*} Search Algorithms}

A$^*$ is an informed search algorithm optimized for weighted graphs: starting from a specific node, the algorithm seeks a path to a given final node by minimizing a specified cost criterion. The algorithm identifies the trajectory by maintaining a tree of paths originating from the starting point, extending them until the target node is reached. 
In contrast to uninformed search strategies like Dijkstra's algorithm, A$^*$ determines which path to extend based on an evaluation function $f(n)$ that it minimizes. This function is defined as
\begin{equation}
    f(n) = g(n) + h(n)\coma
\end{equation}
where $n$ designates the next node on the path, $g(n)$ denotes the accumulated cost of the path from the starting node to $n$, and $h(n)$ represents the heuristic function estimating the cost of the cheapest remaining path from $n$ to the goal. 
Typically, the A$^*$ implementation relies on a priority queue known as the \textit{frontier}. At each iteration, the node with the lowest $f(\cdot)$ value is selected (dequeued) from the frontier, compared against the target, and, if it differs, all its neighbors are added to the queue. Crucially, these new states supplement rather than overwrite the existing frontier. Preserving previously explored branches is essential to allow the algorithm to backtrack if a promising path reaches a dead-end. This cycle of selecting the lowest-cost nodes and expanding the frontier repeats continuously until the dequeued node coincides with the target.

We now see how bidirectional A$^*$ search (illustrated schematically in Figure \ref{sfig:schematicbidirA*search} and detailed in Algorithm \ref{alg:bidirectional_astar}) is implemented to explore the space of quiver mutations in Section \ref{sec:branching_factor}. The search expands two frontiers simultaneously: a forward frontier ($\mathcal{Q}_{\text{fwd}}$) originating from $Q_A$ and a backward frontier ($\mathcal{Q}_{\text{bwd}}$) originating from $Q_B$. 

For any intermediate state $Q_n$ reached from $Q_A$ or $Q_m$ reached from $Q_B$, the forward and backward evaluation functions are defined as:
\begin{equation}
    f_{\text{fwd}}(Q_n) = g_{\text{fwd}}(Q_n) + h_{\text{fwd}}(Q_n)\coma \quad f_{\text{bwd}}(Q_m) = g_{\text{bwd}}(Q_m) + h_{\text{bwd}}(Q_m)\coma
\end{equation}
where $g(\cdot)$ and $h(\cdot)$ depend on the specific pathfinder considered in Section \ref{sec:PathforSeibergDualities}.

At each iteration, the algorithm prioritizes the expansion of whichever frontier currently holds the minimum estimated path cost $\min(f)$. When a quiver is selected for expansion, its valid Seiberg mutants are generated and compared against the states already visited by the opposing frontier.

\begin{algorithm}[!htp]
\caption{Bidirectional A$^*$ Search for Seiberg Dualities}
\label{alg:bidirectional_astar}
\footnotesize
\begin{algorithmic}[1]
\State \textbf{Input:} Initial quiver $Q_A$, target quiver $Q_B$
\State \textbf{Output:} Shortest path between $Q_A$ and $Q_B$, or \text{Failure}
\State Initialize $\mathcal{Q}_{\text{fwd}}$ and $\mathcal{Q}_{\text{bwd}}$ as empty priority queues
\State $\mathcal{V}_{\text{fwd}} \gets \{ Q_A: 0 \}$, $\mathcal{V}_{\text{bwd}} \gets \{ Q_B: 0 \}$ \Comment{Track states and accumulated costs}
\State Push $(g_{\text{fwd}}(Q_A) + h_{\text{fwd}}(Q_A), Q_A)$ to $\mathcal{Q}_{\text{fwd}}$
\State Push $(g_{\text{bwd}}(Q_B) + h_{\text{bwd}}(Q_B), Q_B)$ to $\mathcal{Q}_{\text{bwd}}$
\While{$\mathcal{Q}_{\text{fwd}}$ is not empty \textbf{and} $\mathcal{Q}_{\text{bwd}}$ is not empty}
    \If{$\min(f_{\text{fwd}}) \leq \min(f_{\text{bwd}})$}
        \State Pop $Q_n$ from $\mathcal{Q}_{\text{fwd}}$
        \For{$k \gets 1$ to $K$}
            \State $Q_{n'} \gets D_k\,Q_n$
            \If{$Q_{n'} \neq \emptyset$}
                \If{$\exists \, Q_m \in \mathcal{V}_{\text{bwd}} \text{ s.t. } Q_{n'} \cong Q_m$}
                    \State \Return ReconstructPath($\mathcal{V}_{\text{fwd}}, \mathcal{V}_{\text{bwd}}, Q_{n'}, Q_m$) \Comment{Intersection found up to isomorphism}
                \EndIf
                \State $g_{\text{new}} \gets \mathcal{V}_{\text{fwd}}[Q_n] + \text{step\_cost}$
                \If{$Q_{n'} \notin \mathcal{V}_{\text{fwd}}$ \textbf{or} $g_{\text{new}} < \mathcal{V}_{\text{fwd}}[Q_{n'}]$}
                    \State $\mathcal{V}_{\text{fwd}}[Q_{n'}] \gets g_{\text{new}}$
                    \State Push $(g_{\text{new}} + h_{\text{fwd}}(Q_{n'}), Q_{n'})$ to $\mathcal{Q}_{\text{fwd}}$
                \EndIf
            \EndIf
        \EndFor
    \Else
        \State \textit{// Perform expansion for the backward frontier ($\mathcal{Q}_{\text{bwd}}$)}
    \EndIf
\EndWhile
\State \Return \text{Failure}
\end{algorithmic}
\end{algorithm}

\subsection{Beam Search Algorithms}

While A$^*$ employs a selection strategy, prioritizing the state with the lowest $f$-score at each iteration, it preserves completeness by maintaining the entire frontier of unexplored branches. This memory allows the algorithm to backtrack from dead-ends, guaranteeing that a valid path will be found in a finite number of steps if one exists.

In contrast, we now introduce a Beam Search pathfinder, which is a purely greedy pathfinder that retains only a subset of candidates at each step, discarding all alternative branches. While this approach reduces memory overhead and execution time, the absence of a backtracking mechanism makes it vulnerable to local minima: a single misleading evaluation by the heuristic can permanently trap the search in a dead-end.

Formally, for a $K$-node quiver, we fix a ``beam width" $B \leq K$ and define the ``beam'' $\mathcal{B}_t$ as the set of candidate quivers retained in memory at depth $t$, with $|\mathcal{B}_t| \leq B$. At each exploration step, the algorithm generates the pool of all valid Seiberg mutants originating from the current beam:
\begin{equation}
    \mathcal{C}_{t+1} = \bigcup_{Q \in \mathcal{B}_t} \left\{ D_k\,Q \;\middle|\; D_k\,Q \neq \emptyset \right\}\fstop
\end{equation}
Rather than maintaining an exhaustive frontier, the new beam $\mathcal{B}_{t+1}$ is formed by selecting the $B$ quivers from $\mathcal{C}_{t+1}$ that minimize a guiding cost function $g(Q)$:
\begin{equation}
    \mathcal{B}_{t+1} = \left\{ Q \in \mathcal{C}_{t+1} \;\middle|\; \left| \left\{ Q' \in \mathcal{C}_{t+1} \setminus \{Q\} \;\middle|\; g(Q') < g(Q) \right\} \right| < B \right\}\coma
\end{equation}
where ties at the boundary $|\mathcal{B}_{t+1}| = B$ are broken deterministically by graph lexicographical ordering. Morally, the Beam Search is therefore an A$^*$ Search with $f(n) = g(n)$ and $h(n) = 0$, and a limited frontier. In Section \ref{sec:PathforSeibergDualities}, $g(\cdot)$ is estimated using the output of the AGNN model.

During expansion, every generated candidate $Q'$ is checked against the target $Q_B$. A schematic representation of the beam search is shown in Figure \ref{sfig:schematicbeamsearch}, with the corresponding pseudocode provided in Algorithm \ref{alg:beam_search}.

\begin{algorithm}[!htp]
\caption{Beam Search Algorithm}
\label{alg:beam_search}
\footnotesize
\begin{algorithmic}[1]
\State \textbf{Input:} Initial quiver $Q_A$, target quiver $Q_B$, beam width $B$, max depth $\mathfrak{d}_\text{max}$
\State \textbf{Output:} Path to target state $Q_B$, or \text{Failure}
\State $\mathcal{B}_0 \gets \{ Q_A \}$
\State $\mathcal{V} \gets \{ Q_A \}$
\For{$t \gets 0$ to $\mathfrak{d}_\text{max} - 1$}
    \State $\mathcal{C}_{t+1} \gets \emptyset$ \Comment{Candidate set}
    \For{$Q \in \mathcal{B}_t$}
        \For{$k \gets 1$ to $K$}
            \State $Q' \gets D_k \,Q$
            \If{$Q' \cong Q_B$}
                \State \Return Path($Q_A \to \dots \to Q'$) \Comment{Success up to isomorphism}
            \EndIf
            \If{$Q' \neq \emptyset$ \textbf{and} $Q' \notin \mathcal{V}$}
                \State Compute cost $g(Q') \gets g(Q) + \text{cost\_function}(Q')$
                \State $\mathcal{C}_{t+1} \gets \mathcal{C}_{t+1} \cup \{ (Q', g(Q')) \}$
            \EndIf
        \EndFor
    \EndFor
    \If{$\mathcal{C}_{t+1} = \emptyset$}
        \State \Return \text{Failure} \Comment{dead-end}
    \EndIf
    \State $\mathcal{B}_{t+1} \gets \underset{s \in \mathcal{C}_{t+1}}{\text{arg min-}B} \ g(s)$ \Comment{Retain top $B$ candidates}
    \State $\mathcal{V} \gets \mathcal{V} \cup \mathcal{B}_{t+1}$
\EndFor
\State \Return \text{Failure}
\end{algorithmic}
\end{algorithm}

\newpage

\bibliographystyle{JHEP}
\bibliography{references}

\end{document}